\documentclass[aps,prl,twocolumn,notitlepage,amsmath,amssymb,superscriptaddress]{revtex4-2}
\usepackage{dcolumn}
\usepackage{amsthm}
\usepackage{bm}
\usepackage{graphicx,graphics,times,bm,amsthm,mathrsfs,MnSymbol}
\usepackage{gensymb}
\usepackage{amsfonts}
\pdfoutput=1
\usepackage[pdfstartview=FitH]{hyperref}
\hypersetup{
    colorlinks=true,       
    linkcolor=red,          
    citecolor=magenta,        
    filecolor=magenta,      
    urlcolor=cyan,           
    runcolor=cyan}
\usepackage[pdftex]{color}
\usepackage{braket}
\newcommand{\po}[2]{\hat{\sigma}_{#1}^{#2}}
\usepackage{enumerate}
\usepackage[usenames,dvipsnames]{xcolor}
\usepackage{multirow}
\usepackage{mathtools}
\usepackage{bbold}

\newcommand{\ignore}[1]{}

\usepackage[defaultcolor=blue,final]{changes}

\begin{document}
\title{Exploring Hilbert-Space Fragmentation on a Superconducting Processor}

\author{Yong-Yi Wang}
\thanks{These authors contributed equally to this work.}
\affiliation{Institute of Physics, Chinese Academy of Sciences, Beijing 100190, China}
\affiliation{School of Physical Sciences, University of Chinese Academy of Sciences, Beijing 100049, China}

\author{Yun-Hao Shi}
\thanks{These authors contributed equally to this work.}
\affiliation{Institute of Physics, Chinese Academy of Sciences, Beijing 100190, China}
\affiliation{School of Physical Sciences, University of Chinese Academy of Sciences, Beijing 100049, China}
\affiliation{Beijing Academy of Quantum Information Sciences, Beijing 100193, China}

\author{Zheng-Hang Sun}
\thanks{These authors contributed equally to this work.}
\affiliation{Theoretical Physics III, Center for Electronic Correlations and Magnetism, Institute of Physics, University of Augsburg, D-86135 Augsburg, Germany}

\author{Chi-Tong Chen}
\affiliation{Quantum Science Center of Guangdong-Hong Kong-Macao Greater Bay Area, Shenzhen, Guangdong 518045, China}

\author{Zheng-An Wang}
\affiliation{Beijing Academy of Quantum Information Sciences, Beijing 100193, China}
\affiliation{Hefei National Laboratory, Hefei 230088, China}

\author{Kui Zhao}
\affiliation{Beijing Academy of Quantum Information Sciences, Beijing 100193, China}

\author{Hao-Tian Liu}
\affiliation{Institute of Physics, Chinese Academy of Sciences, Beijing 100190, China}
\affiliation{School of Physical Sciences, University of Chinese Academy of Sciences, Beijing 100049, China}

\author{Wei-Guo Ma}
\affiliation{Institute of Physics, Chinese Academy of Sciences, Beijing 100190, China}
\affiliation{School of Physical Sciences, University of Chinese Academy of Sciences, Beijing 100049, China}

\author{Ziting Wang}
\affiliation{Beijing Academy of Quantum Information Sciences, Beijing 100193, China}

\author{Hao Li}
\affiliation{Beijing Academy of Quantum Information Sciences, Beijing 100193, China}

\author{Jia-Chi Zhang}
\affiliation{Institute of Physics, Chinese Academy of Sciences, Beijing 100190, China}
\affiliation{School of Physical Sciences, University of Chinese Academy of Sciences, Beijing 100049, China}

\author{Yu Liu}
\affiliation{Institute of Physics, Chinese Academy of Sciences, Beijing 100190, China}
\affiliation{School of Physical Sciences, University of Chinese Academy of Sciences, Beijing 100049, China}

\author{Cheng-Lin Deng}
\affiliation{Institute of Physics, Chinese Academy of Sciences, Beijing 100190, China}
\affiliation{School of Physical Sciences, University of Chinese Academy of Sciences, Beijing 100049, China}

\author{Tian-Ming Li}
\affiliation{Institute of Physics, Chinese Academy of Sciences, Beijing 100190, China}
\affiliation{School of Physical Sciences, University of Chinese Academy of Sciences, Beijing 100049, China}

\author{Yang He}
\affiliation{Institute of Physics, Chinese Academy of Sciences, Beijing 100190, China}
\affiliation{School of Physical Sciences, University of Chinese Academy of Sciences, Beijing 100049, China}

\author{Zheng-He Liu}
\affiliation{Institute of Physics, Chinese Academy of Sciences, Beijing 100190, China}
\affiliation{School of Physical Sciences, University of Chinese Academy of Sciences, Beijing 100049, China}

\author{Zhen-Yu Peng}
\affiliation{Institute of Physics, Chinese Academy of Sciences, Beijing 100190, China}
\affiliation{School of Physical Sciences, University of Chinese Academy of Sciences, Beijing 100049, China}

\author{Xiaohui Song}
\affiliation{Institute of Physics, Chinese Academy of Sciences, Beijing 100190, China}
\affiliation{School of Physical Sciences, University of Chinese Academy of Sciences, Beijing 100049, China}

\author{Guangming Xue}
\affiliation{Beijing Academy of Quantum Information Sciences, Beijing 100193, China}

\author{Haifeng Yu}
\affiliation{Beijing Academy of Quantum Information Sciences, Beijing 100193, China}

\author{Kaixuan Huang}
\email{huangkx@baqis.ac.cn}
\affiliation{Beijing Academy of Quantum Information Sciences, Beijing 100193, China}

\author{Zhongcheng Xiang}
\email{zcxiang@iphy.ac.cn}
\affiliation{Institute of Physics, Chinese Academy of Sciences, Beijing 100190, China}
\affiliation{School of Physical Sciences, University of Chinese Academy of Sciences, Beijing 100049, China}

\author{Dongning Zheng}
\affiliation{Institute of Physics, Chinese Academy of Sciences, Beijing 100190, China}
\affiliation{School of Physical Sciences, University of Chinese Academy of Sciences, Beijing 100049, China}
\affiliation{Songshan Lake Materials Laboratory, Dongguan, Guangdong 523808, China}
\affiliation{CAS Center for Excellence in Topological Quantum Computation, UCAS, Beijing 100190, China, and Mozi Laboratory, Zhengzhou 450001, China}

\author{Kai Xu}
\email{kaixu@iphy.ac.cn}
\affiliation{Institute of Physics, Chinese Academy of Sciences, Beijing 100190, China}
\affiliation{School of Physical Sciences, University of Chinese Academy of Sciences, Beijing 100049, China}
\affiliation{Beijing Academy of Quantum Information Sciences, Beijing 100193, China}
\affiliation{Songshan Lake Materials Laboratory, Dongguan, Guangdong 523808, China}
\affiliation{CAS Center for Excellence in Topological Quantum Computation, UCAS, Beijing 100190, China, and Mozi Laboratory, Zhengzhou 450001, China}

\author{Heng Fan}
\email{hfan@iphy.ac.cn}
\affiliation{Institute of Physics, Chinese Academy of Sciences, Beijing 100190, China}
\affiliation{School of Physical Sciences, University of Chinese Academy of Sciences, Beijing 100049, China}
\affiliation{Beijing Academy of Quantum Information Sciences, Beijing 100193, China}
\affiliation{Songshan Lake Materials Laboratory, Dongguan, Guangdong 523808, China}
\affiliation{CAS Center for Excellence in Topological Quantum Computation, UCAS, Beijing 100190, China, and Mozi Laboratory, Zhengzhou 450001, China}

\begin{abstract}
    Isolated interacting quantum systems generally thermalize, yet there are several examples for the breakdown of ergodicity, such as many-body localization and quantum scars. Recently, ergodicity breaking has been observed in systems subjected to linear potentials, termed Stark many-body localization. This phenomenon is closely associated with Hilbert-space fragmentation, characterized by a strong dependence of dynamics on initial conditions. Here, we explore initial-state dependent dynamics using a ladder-type superconducting processor with up to 24 qubits, which enables precise control of the qubit frequency and initial state preparation. In systems with linear potentials, we experimentally observe distinct non-equilibrium dynamics for initial states with the same quantum numbers and energy, but with varying domain wall numbers. Accompanied by the numerical simulation for systems with larger sizes, we reveal that this distinction becomes increasingly pronounced as the system size grows, in contrast with weakly disordered interacting systems. Our results provide convincing experimental evidence of the fragmentation in Stark systems, enriching our understanding of the weak breakdown of ergodicity.
\end{abstract}

\maketitle

\begin{figure*}[t]
	\includegraphics[width=0.85\linewidth]{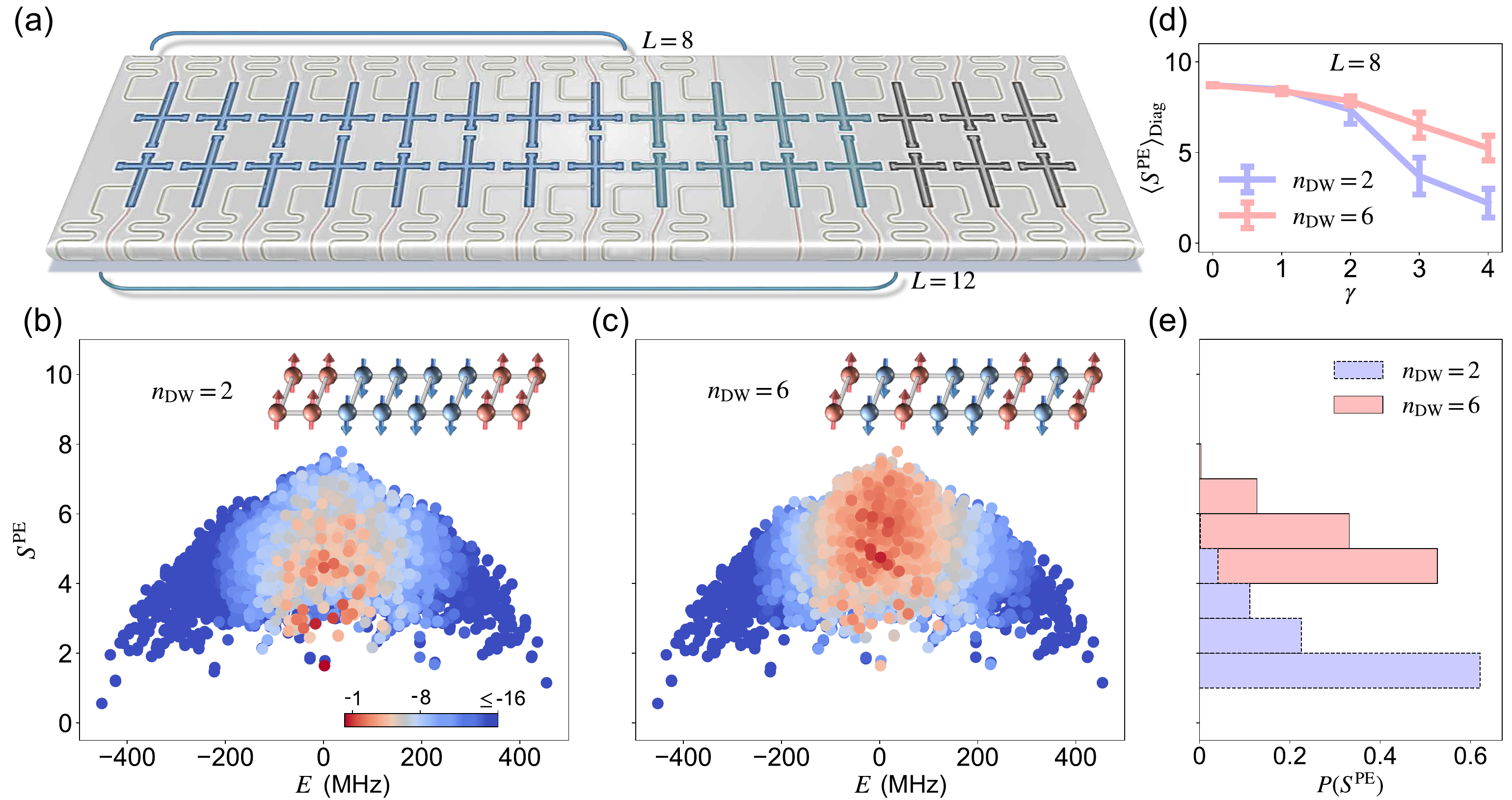}\\
	\caption{Experimental setup. (a) Schematic representation of the ladder-type superconducting processor. For the experiment, $L (8\ \&\ 12) \times 2$ qubits of the 30-qubit ladder are utilized, with nearest-neighboring $\bar{J}_{\text{NN}}/2\pi \simeq 7$ MHz and next-nearest-neighboring couplings $\bar{J}_{\text{NNN}} \lesssim \bar{J}_{\text{NN}}/6$. (b), (c) Participation entropy of eigenstates for Hamiltonian [Eq.~(\ref{Ham1})] with $\gamma=4$, colored by the logarithm of the eigenstate occupation numbers $\log_{10}{\left| {\langle \psi_0|E_n\rangle } \right|^2}$ for the initial state (b) $\ket{\psi_{n_{\text{DW}}=2}}=\ket{\mathbb{11000011}}$, and (c) $\ket{\psi_{n_{\text{DW}}=6}}=\ket{\mathbb{10100101}}$, with $\ket{\mathbb{0(1)}}$ denotes the ground (excited) state $\ket{0(1)}$ on both legs of each ladder rung. (d) Diagonal ensemble average of the PE for different gradients $\gamma$, with the errorbar presenting the standard deviation. (e)  Histogram of the participation entropy with $\gamma=4$ in the diagonal ensembles corresponding to the initial states $\ket{\psi_{n_{\text{DW}}=2}}$ and $\ket{\psi_{n_{\text{DW}}=6}}$.}
    \label{fig1}
\end{figure*}

\begin{figure*}[htbp]
	\includegraphics[width=0.9\linewidth]{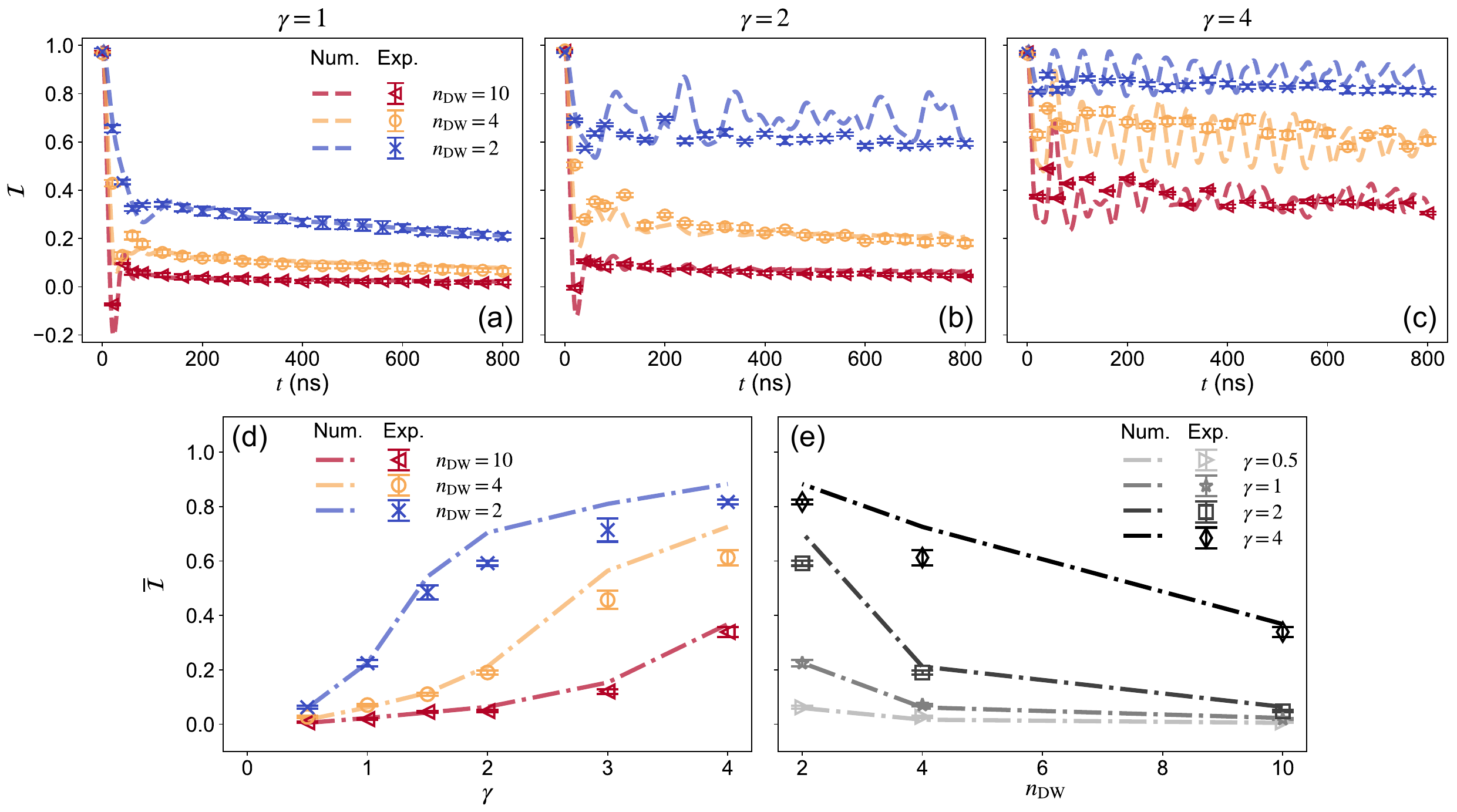}\\
	\caption{Initial-state dependent dynamics of imbalance in the $12 \times 2$ ladder. (a--c) The time evolution of the imbalance $\mathcal{I}(t)$ for different initial states $\ket{\psi_{n_{\text{DW}}=2}}=\ket{\mathbb{111000000111}}$, $\ket{\psi_{n_{\text{DW}}=4}}=\ket{\mathbb{110001100011}}$, and $\ket{\psi_{n_{\text{DW}}=10}}=\ket{\mathbb{101010010101}}$ at gradients $\gamma = 1, 2$ and $4$, respectively. Points are experimental data, each averaged over $5000$ repetitions, and the dashed lines are numerical simulation. (d) The late-time averaged imbalance $\bar{\mathcal{I}}$ as a function of the gradient $\gamma$ for initial states with different domain wall numbers $n_{\text{DW}}$. (e) The late-time averaged imbalance $\bar{\mathcal{I}}$ as a function of $n_{\text{DW}}$ for different $\gamma$. Averages are taken over a time window from $600$ to $800$ ns. Error bars represent the standard deviation.} 
    \label{fig2}
\end{figure*}

\begin{figure}[b]
	\includegraphics[width=1\linewidth]{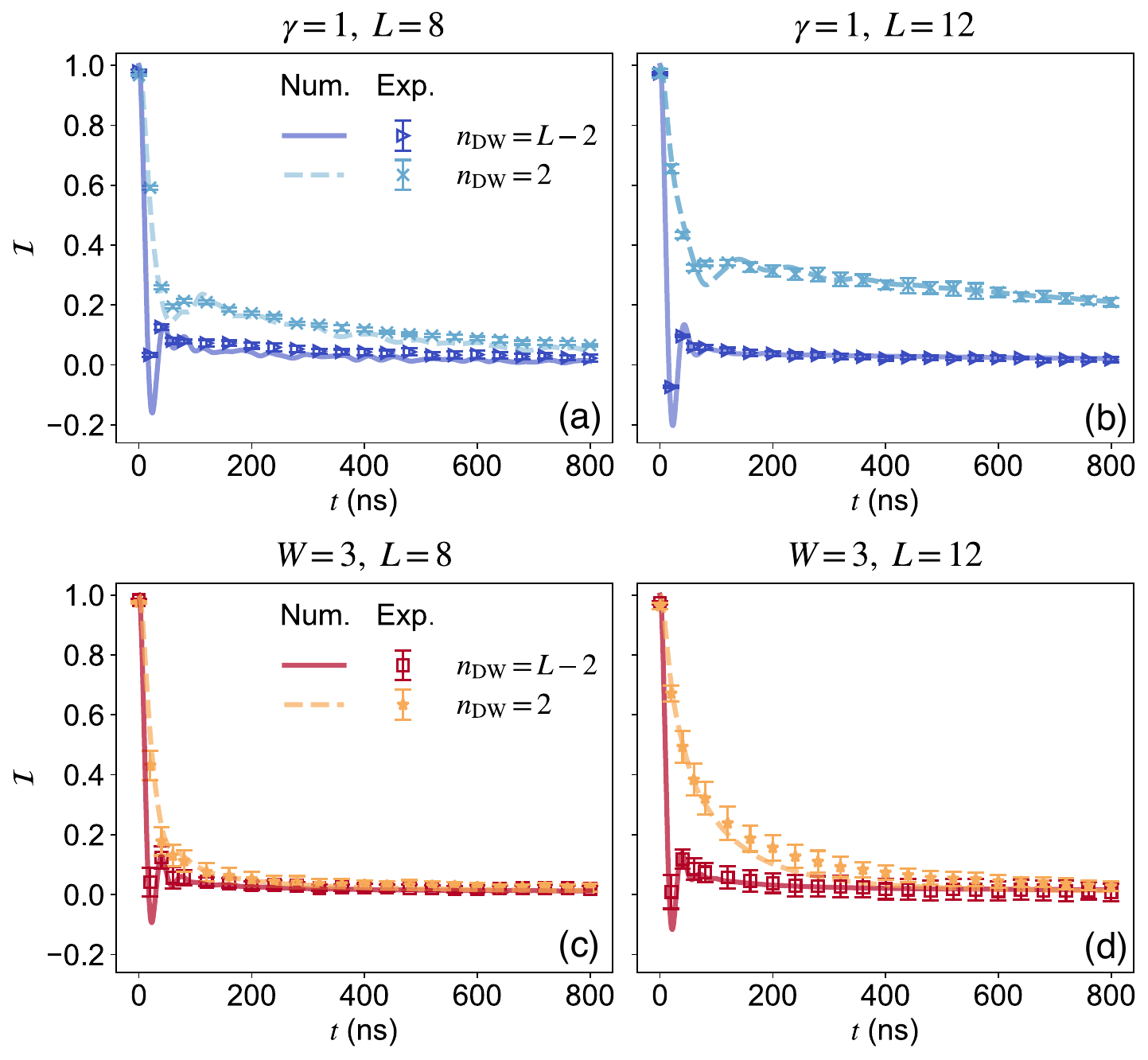}\\
	\caption{Time evolution of imbalance for initial states $\ket{\psi_0}=\ket{\psi_{n_{\text{DW}}=2}}$ and $\ket{\psi_{n_{\text{DW}}=L-2}}$ in different systems: (a) Stark system with the gradient $\gamma=1$ for $L=8$, and (b) Stark system with $\gamma=1$ for $L=12$; (c) disordered system with the disorder strength $W=3$ for $L=8$, and (d) disordered system with $W=3$ for $L=12$. In Stark systems (a, b), points are experimental data, each averaged over $5000$ repetitions with error bars indicating the standard deviation. In disordered systems (c, d), the experimental data points are the average of 20 disorder realizations with $5000$ repetitions for each realization, and the error bar shows the standard error of the mean. Lines represent numerical simulation.}
    \label{fig3}
\end{figure}

\begin{figure*}[htbp]
	\includegraphics[width=0.95\linewidth]{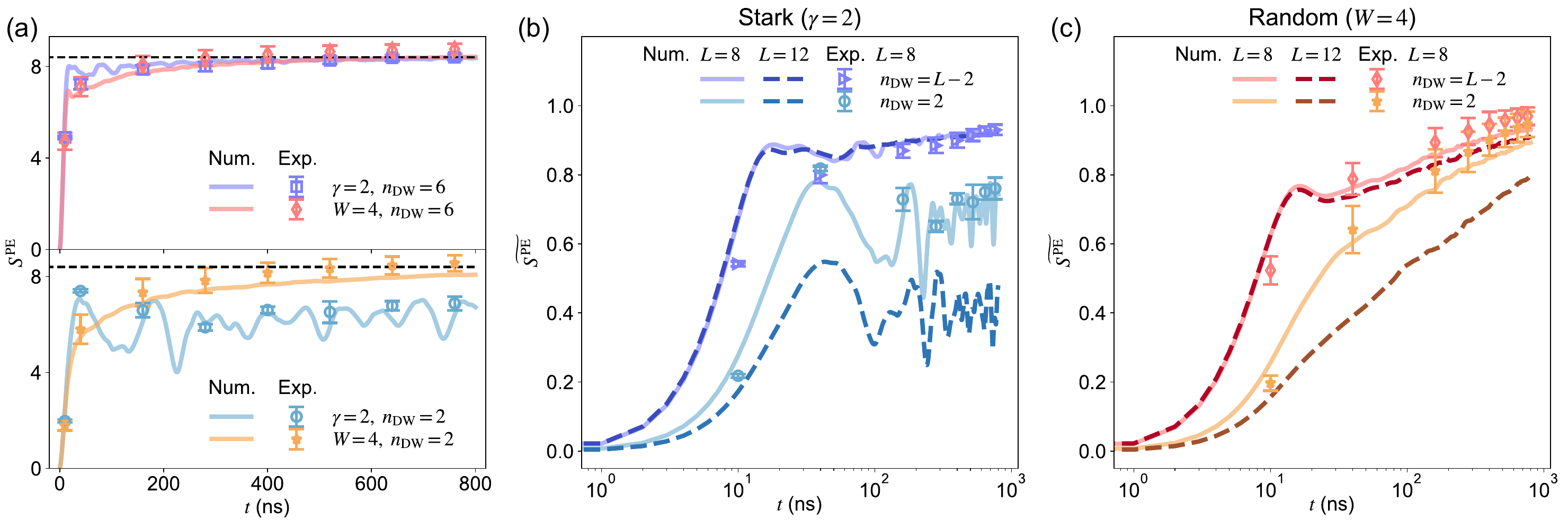}\\
	\caption{Time evolution of participation entropy. (a) Dynamical participation entropy in the Stark system with $\gamma=2$ and the disordered system with $W=4$ for different initial states $\ket{\psi_0}=\ket{\psi_{n_{\text{DW}}=6}}$ (upper panel) and $\ket{\psi_{n_{\text{DW}}=2}}$ (lower panel) with the system length $L=8$. For $\ket{\psi_0}=\ket{\psi_{n_{\text{DW}}=6}}$, the numerical results of the late-time $S^{\text{PE}}$ is around $8.4$ in both systems, marked by the horizontal dashed lines in both panels. (b, c) Normalized PE $\widetilde{S^{\text{PE}}}=S^{\text{PE}}/S^{\text{PE}}_{\text{T}}$ with the same experimental data as (a) against the logarithm of time, in (b) the Stark system with $\gamma=2$ and (c) the disordered system with $W=4$. Numerical results are depicted by lines for $L=8$ and $12$. The experimental data for $S^{\text{PE}}$ in Stark systems are obtained from five sets of independent experiments, each consisting of $5 \times 10^5$ repeated single-shot measurements. For disordered systems, data are from $20$ disorder realizations, each with $5 \times 10^5$ measurements. Error bars indicate the standard deviation.}
    \label{fig4}
\end{figure*}

\textit{Introduction.}---%
In the past decades, the development of artificial quantum systems has motivated considerable studies of quantum statistical mechanics and non-equilibrium dynamics in many-body systems~\cite{Georgescu2014}. As a general framework for quantum thermalization, the eigenstate thermalization hypothesis (ETH) proposes that generic closed interacting systems thermalize under their own dynamics~\cite{Rigol2008,DAlessio2016,Deutsch2018}, while the breakdown of ergodicity can typically occur in interacting systems with strong disorder, known as many-body localization (MBL)~\cite{Nandkishore2015,Altman2018,Abanin2019}. Recent advances reveal intermediate behavior between the two extreme limits, referred to as weak ergodicity breaking, including phenomena such as quantum many-body scars~\cite{Bernien2017,Turner2018,Serbyn2021,Zhang2023,Su2023} and Hilbert-space fragmentation (HSF)~\cite{Sala2020,Khemani2020,Herviou2021,Yang2020,Hahn2021,Moudgalya2021,Moudgalya2022}. In weak ergodicity breaking systems, a small fraction of nonthermal eigenstates coexist within the majority of the thermal spectrum, allowing for halted thermalization under certain initial conditions~\cite{Moudgalya2022_2}.

As a prominent example of weak ergodicity breaking, HSF has been extensively explored theoretically in dipole-moment conserving systems, where the Hilbert space fragments into exponentially many disconnected Krylov subspaces due to the interplay between charge and dipole conservation, leading to the strong dependence of the dynamics on the initial conditions~\cite{Pai2019,Sala2020,Khemani2020,Herviou2021}. On the other hand, systems subjected to a linear potential (hereinafter referred to as Stark systems) can also exhibit MBL-like behavior, such as long-lived initial state memory and multifractality, termed as Stark many-body localization (SMBL)~\cite{Schulz2019, VanNieuwenburg2019,Taylor2020,Yao2020,Wang2021,Zisling2022}. Experimental studies of SMBL have been conducted using various experimental platforms ~\cite{Guardado-sanchez2020, Morong2021,Guo2021,Scherg2021,Kohlert2023}. In these systems subjected to a linear potential, the emergence of dipole moment has raised questions and debates about its relationship with HSF~\cite{Taylor2020,Khemani2020,Doggen2021,Yao2021,Doggen2022}. A previous work showed a significant dependence of the dynamics on initial doublon fraction in the tilted one-dimensional Fermi-Hubbard model, but the initial conditions considered do not possess the same quantum numbers and energy~\cite{Kohlert2023}. Actually, the dynamics is profoundly related to the energy of chosen initial states even in disordered interacting systems, as a consequence of many-body mobility edge~\cite{Guo2021_2,Luitz2015}. Therefore, a direct demonstration of HSF in Stark systems requires a more thorough investigation into the dynamics for the initial states with the same quantum numbers and energy, excluding the potential influences stemming from different total charge and emergent dipole moment, as well as many-body mobility edges.

In this work, leveraging the precise control and flexibility of our ladder-type superconducting processor, we engineer the Hamiltonian and prepare various initial states with the same quantum numbers (charge and dipole moment) and energy for systems with up to 24 qubits. Applying site-dependent frequency deviations to individual qubits precisely, we observe distinct dynamical behavior in the Stark systems for initial states featuring varying domain wall numbers, even at a very small gradient. 

Moreover, recent work~\cite{Prasad2022} suggests a dependence of the dynamics on the domain wall number in initial states in disordered systems. To distinguish this phenomenon in disordered systems from HSF, we also investigate the dependence on the domain wall number in weakly disordered systems, specifically in the ergodic phase. We demonstrate that, in weakly disordered systems, slow relaxation for initial states with small domain wall numbers is a finite-time effect. In contrast, in systems with a weak linear potential, the distinct dynamics associated with different domain wall numbers, persisting over long timescales, become increasingly pronounced as the system size grows.

In addition, the efficient single-shot simultaneous readout enables us to experimentally measure the dynamics of participation entropy (PE), which directly characterizes the available Hilbert space for a certain initial state~\cite{Luitz2014,Luitz2015,Mace2019}. The limited growth of PE observed in systems subjected to a linear potential reveals that an initial state with a small domain wall number can only spread within a limited fraction of the Hilbert space, providing direct evidence of HSF in Stark systems. For larger systems, where measuring the PE of the entire system becomes challenging, we propose a practical scheme to estimate the upper bound of PE by experimentally measuring subsystems of moderate length, from which we can also extract key evidence of HSF.

\textit{Model and set-up.}---%
Our experiments utilize a quantum processor equipped with a two-legged ladder structure consisting of 30 transmon qubits, as shown in Fig.~\ref{fig1}(a). The effective Hamiltonian of the qubit-ladder processor reads~\cite{Zhu2022,Xiang2023,Shi2023}
\begin{equation}
    \begin{aligned}
    \frac{\hat{H}}{\hbar}= &\sum_{<jm;j^\prime m^\prime>} J_{jm;j^\prime m^\prime}\left(\po{jm}{+} \po{j^\prime m^\prime}{-}+\po{jm}{-} \po{j^\prime m^\prime}{+}\right)\\
    &+\sum_{m=1,2}\sum_{j = 1}^L W_{jm} \po{jm}{+} \po{jm}{-},
    \end{aligned}
    \label{Ham1}
\end{equation}
where $\hbar = h/2\pi$ denotes the Planck constant, $L$ is the length of the ladder, $\po{jm}{+} (\po{jm}{-})$ is the two-level raising (lowering) operator for the qubit $Q_{jm}$. Here, the first summation runs over all nearest- and next-nearest-neighboring sites $(j, m)$ and $(j^\prime, m^\prime)$ with the averaged nearest-neighboring qubit-qubit coupling $\bar{J}_{\text{NN}}/2\pi \simeq 7$ MHz and next-nearest-neighboring $\bar{J}_{\text{NNN}} \lesssim \bar{J}_{\text{NN}}/6$ (see Supplementary Materials~\cite{supp} for coupling strengths in details). The on-site potential, denoted as $W_{jm}$, can be adjusted by applying fast flux bias to the Z control lines of qubits. This flexibility enables the tuning of $W_{jm}$ to both Stark (linear) potentials $W_{jm}/ \bar{J}_{\text{NN}}=-j\gamma$, as well as random potentials $W_{jm}/ \bar{J}_{\text{NN}}$ drawn from a uniform distribution in $[-W, W]$, facilitating the quantum simulation of both SMBL and conventional MBL. 

Utilizing a generalized form of Jordan-Wigner transformation~\cite{Strong1992,Azzouz1994}, the approximating ladder-$XX$ model can be mapped to an interacting spinless fermion model. As the gradient $\gamma$ (or the disorder strength $W$) increases, a transition from ergodicity to MBL exists in this typical nonintegrable model~\cite{supp}. One can expect that in the regime of a large gradient $\gamma$, the conservation of total $U(1)$ charge $\hat{Q} \equiv \sum_{j,m}{\po{jm}{+}\po{jm}{-}}$, together with the emergent conservation of dipole moment $\hat{P} \equiv \sum_{j,m} j \po{jm}{+}\po{jm}{-}$, gives rise to HSF, so that the system would exhibit distinct dynamics with different initial states $\ket{\psi_0}$.

For $\ket{\psi_0}$ considered in this work, the initial spin distribution is identical in the two rows of the ladder. These initial states share the same $Q=\langle \hat{Q} \rangle, P=\langle \hat{P} \rangle$, and energy $E = \bra{\psi_0} \hat{H} \ket{\psi_0}$, but vary in terms of total domain wall number, defined as $n_{\text{DW}}=\sum_{j=1}^{L-1} (1-\langle \hat{\bar{\sigma}}_{j}^{z} \hat{\bar{\sigma}}_{j+1}^{z} \rangle)/2$, with $\hat{\bar{\sigma}}_{j}^{z}=(\po{j,1}{z}+\po{j,2}{z})/2$. Specifically, we examine two initial states for the system length $L=8$: $\ket{\psi_{n_{\text{DW}}=2}}=\ket{\mathbb{11000011}}$ and $\ket{\psi_{n_{\text{DW}}=6}}=\ket{\mathbb{10100101}}$, where $\ket{\mathbb{0(1)}}$ denotes the ground (excited) state of the single qubit $\ket{0(1)}$ on both legs of each ladder rung (see the insets in Fig.~\ref{fig1}(b) and (c)).

To quantify the localization in Hilbert space of spin configurations, we can calculate the PE of all eigenstates with exact diagonalization~\cite{Luitz2014,Luitz2015,Mace2019}, defined as $S^{\text{PE}}(n) = -\sum_i^{\mathcal{N}} {{p_i(n)}\log {p_i(n)}}$, where $\mathcal{N}$ is the dimension of Hilbert space, and $p_i(n) = {\left| {\langle i|E_n\rangle } \right|^2}$ with $\{\ket{i}\}$ being spin configuration basis (i.e., $\{\sigma^z\}$). Thus, $S^{\text{PE}} = 0$ if the state is a single configuration, and $S^{\text{PE}}_{\text{GOE}}=\log{0.482\mathcal{N}}$ for Gaussian random states~\cite{supp,Torres-Herrera2016}. In Fig.~\ref{fig1}(b) and (c), the PE of eigenstates is displayed for the Hamiltonian [Eq.~(\ref{Ham1})] with a Stark potential at $\gamma = 4$. The colorbars represent the eigenstate occupation numbers $|C_n|^2\equiv\left| {\langle \psi_0|E_n\rangle } \right|^2$, indicating the weight of eigenstate $\ket{E_n}$ in the diagonal ensemble~\cite{Rigol2008,DAlessio2016}, with respect to $\ket{\psi_0}=\ket{\psi_{n_{\text{DW}}=2}}$ and $\ket{\psi_{n_{\text{DW}}=6}}$, respectively. Different from the ETH prediction, substantial fluctuations in PE are observed across the eigenstates within a narrow window around $E=0$. 

Notably, as depicted in Fig.~\ref{fig1}(e), $\ket{\psi_{n_{\text{DW}}=2}}$ tends to predominantly overlap with a small number of eigenstates with relatively low PE, whereas the weight distribution for $\ket{\psi_{n_{\text{DW}}=6}}$ spreads over a greater number of eigenstates with higher PE. We calculate the diagonal ensemble average of the PE $\langle S^{\rm{PE}}\rangle_\text{Diag} \equiv \sum_n |C_n|^2 S^{\rm{PE}}(n)$, and the variance $\sigma^2 \equiv \sum_n |C_n|^2 \left[S^{\rm{PE}}(n)-\langle S^{\rm{PE}}\rangle_\text{Diag}\right]^2$, to characterize these two distributions for various $\gamma$. We observe that as $\gamma$ increases, the averages between the two distributions of the PE increasingly diverge, with the standard deviation being smaller than the difference between their averages (see Fig.~\ref{fig1}(d)), suggesting a decrease in the overlap of the distributions. At $\gamma = 4$, the diagonal-ensemble predictions of PE is $\langle S^{\rm{PE}}\rangle_\text{Diag}=5.24 \pm 0.68 $ and $2.20 \pm 0.79$ corresponding to $\ket{\psi_0}=\ket{\psi_{n_{\rm{DW}}=6}}$ and $\ket{\psi_{n_{\rm{DW}}=2}}$, respectively. This indicates that these two states respectively reside within nearly disconnected fragments, with $\ket{\psi_{n_{\text{DW}}=2}}$ more localized within a smaller fragment in the Hilbert space than $\ket{\psi_{n_{\text{DW}}=6}}$. The results for different $\gamma$, similar behavior in entanglement entropy, and the relation between the PE and domain wall numbers can be found in Supplementary Materials~\cite{supp}.

\textit{Initial-state dependent dynamics.}---%
In contrast to ergodic dynamics, where memory of initial conditions is hidden in global operators, in MBL systems, memory of initial conditions can be preserved in local observables for generic high-energy initial states at long times after sudden quench~\cite{Nandkishore2015,Altman2018,Abanin2019}. To quantify the preservation of the information encoded in the initial state, we consider the imbalance generalized for any initial product state in the $\{\sigma^z\}$ ~\cite{Schreiber2015,Guo2021,Guo2021_2}, defined as
\begin{equation}
    \mathcal{I}=\frac{1}{2L} \sum_{jm}\bra{\psi_0}\po{jm}{z}(t) \po{jm}{z}(0)\ket{\psi_0}.
\end{equation}
The generalized imbalance will relax from initial value $\mathcal{I}=1$ to $0$ if the system thermalizes, while it will maintain a finite value at long times if the system fails to thermalize.

To observe the initial-state dependent dynamics due to HSF, we consider three typical initial states for the system length $L=12$, including 
$\ket{\psi_{n_{\text{DW}}=2}}=\ket{\mathbb{111000000111}}$, 
$\ket{\psi_{n_{\text{DW}}=4}}=\ket{\mathbb{110001100011}}$, and
$\ket{\psi_{n_{\text{DW}}=10}}=\ket{\mathbb{101010010101}}$.
We prepare these initial states with the same quantum numbers and energy but different domain wall numbers, by simultaneously applying $\pi$ pulses to half of the qubits. Calibrating the rectangular Z pulses for all qubits to create Stark potentials, we measure the generalized imbalance $\mathcal{I}$ for different initial states following evolution over time $t$. The experimental data at $\gamma=1, 2$ and $4$ are depicted in Fig.~\ref{fig2}(a--c), along with numerical simulation~\cite{supp}. We note that for $\gamma=1, 2$, the average ratio of adjacent level spacings $\langle r \rangle\gtrsim 0.5$, close to prediction of Gaussian orthogonal ensemble, while $\langle r \rangle\lesssim 0.4$ for $\gamma=4$, exhibiting Poisson statistics~\cite{supp,Luitz2015,Atas2013,Roushan2017}. Therefore, at $\gamma=4$, the imbalance oscillates around a finite value after a fast drop at short times for all three states, signifying a complete breakdown of ergodicity.
At both $\gamma=1$ and $2$, the imbalance for $\ket{\psi_0}=\ket{\psi_{n_{\text{DW}}=10}}$ relaxes to zero, in accordance with ergodic dynamics. However, the relaxation gradually slows down as $n_{\text{DW}}$ decreases. For $\gamma=2$, the imbalance for $\ket{\psi_0}=\ket{\psi_{n_{\text{DW}}=2}}$ has already exhibited pronounced oscillations around a value significantly above zero with negligible decay.

We then average the late-time imbalance over a time window from $600$ to $800$ ns, which is plotted against $\gamma$ and $n_{\text{DW}}$ in Fig.~\ref{fig2}(d) and (e), respectively. On the one hand, the late-time imbalance for all initial states exhibits a clear upward trend as $\gamma$ increases. On the other hand, we observe a monotonic decrease in the late-time imbalance as $n_{\text{DW}}$ decreases. For initial states with smaller $n_{\text{DW}}$, the late-time imbalance displays a quicker rise as $\gamma$ increases. By selecting initial states with identical quantum numbers and energy, the above experimental results demonstrate the initial-state dependent dynamics concerning domain wall numbers in Stark systems, which is a significant feature of HSF and weak breakdown of ergodicity.

\textit{Comparison with weakly disordered systems.}---%
According to the ETH, a fully delocalized regime exists in the disordered interacting systems with weak disorder, where the choice of the initial state is of no significance at long times~\cite{Hauschild2016,Prasad2022}. However, in terms of the ETH in its strong sense, numerical evidence has been put forward for the absence of an ETH-MBL transition in Stark systems due to HSF~\cite{Doggen2021}. The question of whether HSF will lead to distinctions between Stark systems and disordered systems is a crucial aspect to explore. Here, we focus on Hamiltonian [Eq.~(\ref{Ham1})] with the gradient $\gamma=1$, and the disorder strengths $W=3$. For these parameters, $\langle r \rangle$ remains close to $0.531$, often viewed as a signature of ergodicity~\cite{supp}. In other word, we focus on systems with weak Stark and disordered potentials, which exhibit ETH-like level statistics.

For both models, we measure the generalized imbalance for $\ket{\psi_0}=\ket{\psi_{n_{\text{DW}}=2}}$ and $\ket{\psi_{n_{\text{DW}}=L-2}}$ with system lengths $L=8$ and $12$. The experimental data and numerical results are presented in Fig.~\ref{fig3}. We can see that the imbalance for $\ket{\psi_0}=\ket{\psi_{n_{\text{DW}}=L-2}}$ relaxes to zero in a short time for both Stark systems and disordered systems, irrespective of system size.

However, in Stark systems, an evident slowdown in relaxation can be observed for $\ket{\psi_0}=\ket{\psi_{n_{\text{DW}}=2}}$, which becomes more evident with increasing system size (see Fig.~\ref{fig3}(a) and (b)). A significantly higher late-time imbalance for $L=12$ compared to $L=8$ suggests a nonvanishing imbalance for systems with a sufficiently large size. This phenomenon persists even for a smaller gradient $\gamma=0.5$~\cite{supp}.

In contrast, in weakly disordered systems, we can merely observe a slightly slower thermalization for $\ket{\psi_0}=\ket{\psi_{n_{\text{DW}}=2}}$ compared to $\ket{\psi_{n_{\text{DW}}=L-2}}$, with the disorder-average imbalance approaching zero at long times for both initial states. This behavior holds true for both system lengths $L=8$ and $12$, as shown in Fig.~\ref{fig3}(c) and (d), which starkly contrasts with the observation in Stark systems. A more quantitative analysis of the imbalance decay in the two models is provided in the Supplementary Materials~\cite{supp}.

To elucidate the finite-size effect in the two systems~\cite{Abanin2021,Sierant2022}, in Supplementary Materials~\cite{supp}, we present numerical simulations of the dynamics with lengths up to $L=20$ using matrix product state (MPS) techniques. Moreover, we consider a longer evolution time up to $t=40$ $\mu$s ($J_\text{NN}t \sim 2 \times 10^3$) to provide a definitive observation of the initial-state dependent dynamics in Stark systems. We show that, in weakly disordered systems, although the dynamics for the two-domain-wall initial state suffers from a strong finite-time effect, the imbalance is expected to approach zero eventually in the thermodynamic limit. This behavior is consistent with ``strong'' ETH, where the long-time behavior of local observables does not depend on the initial states, provided they share the energy and quantum numbers. In contrast, the Stark system with a small linear potential exhibits a robust dynamical signature of ``weak'' ETH, where a small set of eigenstates may violate ETH, leading to the initial-state dependent dynamics.

\textit{Dynamical signature of HSF via participation entropy.}---%
As discussed above, the Hilbert space tends to fragment into many disconnected Krylov subspaces with different dimensions in Stark systems. A direct quantity reflecting the extent to which a time-evolved state $\ket{\psi(t)}$ spreads over the Hilbert space~\cite{Guo2021_2,Li2023} is the dynamical PE, defined as 
\begin{equation}
    S^{\text{PE}}(t) = -\sum_i^{\mathcal{N}} {{p_i(t)}\log {p_i(t)}},
    \label{PE}
\end{equation}
with the multiqubit probabilities $p_{i}(t) = |\langle \psi(t)| i\rangle|^{2}$. 
For ergodic dynamics, the dynamical PE increases from $S^{\text{PE}}(0)=0$ for any initial product state to the late-time value $S^{\text{PE}}_{\text{T}}=\log \mathcal{N} -1 + \gamma_e$ (with $\gamma_e$ Euler's constant)~\cite{supp,Boixo2018}. The dynamical PE determined by the multiqubit probabilities can be observed with our superconducting processor thanks to its efficient multiqubit simultaneous readout capability.
Applying a Stark potential with $\gamma=2$ and a random potential with $W=4$ to the system, we track the time evolution of the multiqubit probabilities with site-resolved simultaneous readout at different times $t$. For the calculation of the PE, the experimental data of the multiqubit probabilities in Fig.~\ref{fig4} are obtained from $5 \times 10^5$ repeated measurements and post-selected within the $Q = L$ sector. As shown in Fig.~\ref{fig4}(a), the initial product states, as Fock states, gradually propagate and cover the available Hilbert space at long times, reflected by the increase of PE. For both models, dynamics for $\ket{\psi_0}=\ket{n_{\text{DW}}=6}$ approaches similar values of PE at long times. However, in a random potential, the PE for $\ket{\psi_0}=\ket{\psi_{n_{\text{DW}}=2}}$ increases and eventually approaches the late-time value for $\ket{\psi_{n_{\text{DW}}=6}}$ (marked by dashed lines in Fig.~\ref{fig4}(a)), while in a Stark potential, it oscillates at a lower value, suggesting that $\ket{\psi_{n_{\text{DW}}=2}}$ can only propagate in a restricted fraction of the Hilbert space.

In Fig.~\ref{fig4}(b) and (c), we plot the experimental data of normalized PE $\widetilde{S^{\text{PE}}}=S^{\text{PE}}/S^{\text{PE}}_{\text{T}}$ against the logarithm of time, with numerical results for $L=12$, in systems with small Stark and random potentials, respectively. In Stark systems, we observe that the data of normalized PE for $\ket{\psi_{n_{\text{DW}}=L-2}}$ nearly coincide for different system sizes, while those for $\ket{\psi_{n_{\text{DW}}=2}}$ decreases with the increasing system size. This indicates that $\ket{\psi_{n_{\text{DW}}=L-2}}$ can spread in a fraction of Hilbert space whose dimension scales with the size of the entire Hilbert space, while the dimension of Krylov subspace corresponding to $\ket{\psi_{n_{\text{DW}}=2}}$ becomes vanishingly small in the thermodynamic limit~\cite{supp}. In contrast, in disordered systems, $\widetilde{S^{\text{PE}}}$ exhibits unceasing growth within the experimental time for both initial states (and approaches a similar value of $\widetilde{S^{\text{PE}}} \sim 1$ over a longer time~\cite{supp}), despite differences in relaxation. The comparison to disordered interacting systems further reveals a distinctive characteristic in Stark systems, where the initial-state dependent dynamics is closely associated with disconnected Krylov subspaces with different scaling behaviors.

\begin{figure}[t]
	\centering
	\includegraphics[width=1\linewidth]{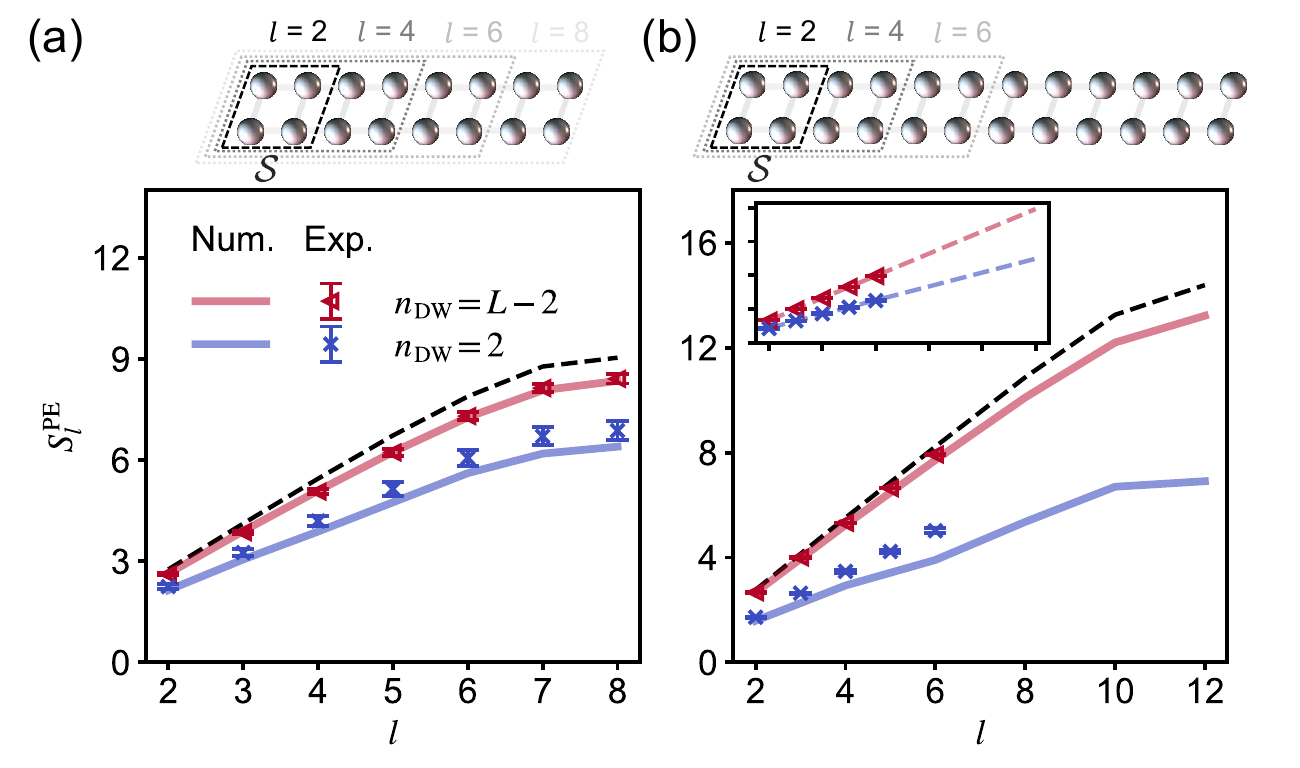}
	\caption{Late-time local participation entropy of subsystems. (a) The late-time local participation entropy in the Stark system with $\gamma=2$ for $L=8$. Lines denote numerical simulations, while points are experimental data averaged over five sets of independent experiments, each set consisting of $5 \times 10^5$ repeated single-shot measurements. The error bars denote the standard deviation. (b) Same as (a), but for $L=12$, with each set consisting of $1 \times 10^6$ repeated single-shot measurements. The data points are measured around $t \approx 800$ ns. The black dashed lines denote local participation entropies of random pure states for (a) $L=8$ and (b) $L=12$, respectively. Inset in (b): The dashed lines, colored according to the different initial states, denote the fit of the data from $l=2$ to $l=6$ with a linear function. The fitted functions are $f(l)=1.31 l+0.05$ for $\ket{\psi_0}=\ket{\psi_{n_{\text{DW}}=L-2}}$, and $f(l)=0.82 l+0.14$ for $\ket{\psi_0}=\ket{\psi_{n_{\text{DW}}=2}}$, respectively.}
    \label{fig5}
\end{figure}

For a system with the length of 12 or larger, it is a challenging experimental task to measure PE directly for the entire system. To address this challenge, we introduce the concept of local PE. Let us first divide the system into two parts: the subsystem $\mathcal{S}$ and its complement. As depicted in the upper part of Fig.~\ref{fig5}, the subsystem $\mathcal{S}$ includes consecutive $l \times 2$ qubits $\mathcal{S}(j;l)=\{Q_{j,m},Q_{j+1,m},\ldots,Q_{j+l-1,m}\}$ with $m=1,2$. The local PE can then be defined as the PE for the subsystem $\mathcal{S}$, which is given by
\begin{equation}
    S^{\mathrm{PE}}[\mathcal{S}]=\sum_{i}p_i^\mathcal{S}\log p_i^\mathcal{S},
\end{equation}
where $p_i^\mathcal{S}$ denotes the probabilities of a time-evolved state in spin configuation basis $\ket{i}^\mathcal{S}\in\{ \ket{\boldsymbol{i_{Q}}}|i_Q\in\{0,1\} ,Q \in \mathcal{S}\}$ for the subsystem $\mathcal{S}(j;l)$. When $l=L$, the local PE corresponds to the PE of the entire system. Regardless of the system length $L$, local PE can be efficiently measured in quantum processors with relatively small number of measurements for subsystems of moderate length $l$~\cite{supp}.

In Fig.~\ref{fig5}, we present both the numerical and experimental results of the late-time spatially averaged local PE $S^{\mathrm{PE}}_l = \langle S^{\mathrm{PE}}[\mathcal{S}(j; l)]\rangle_j$ ($\langle\cdots\rangle_j$ denotes average over all possible subsystems with consecutive $l \times 2$ qubits) for the Stark system with $\gamma=2$, around $t \approx 800$ ns, for system lengths $L=8$ (Fig.~\ref{fig5}(a)) and $L=12$ (Fig.~\ref{fig5}(b)). For these two system lengths $L$, we observe that the local PE increases approximately linearly with the subsystem length $l$ for both initial states $ \ket{\psi_0}=\ket{\psi_{n_{\text{DW}}=2}}$ and $\ket{\psi_{n_{\text{DW}}=L-2}}$, except for a downward bending as $l$ approaches the entire system length $L$. This downward bending is also observed for random pure states, i.e., the late-time states for fully ergodic systems (see black dashed lines in Fig.~\ref{fig5}). We argue that this downward bending is a result of the conservation of total $U(1)$ charge, and becomes more moderate with larger system sizes~\cite{supp}. This behavior suggests a practical and scalable approach for estimating the upper bound of PE by fitting the local PE measured in subsystems of moderate length $l$, and then extrapolating to $l=L$.

For $L=12$, as shown in the inset of Fig.~\ref{fig5}(b), we fit the experimental data for $l=2$ to $l=6$ with a linear function, and extrapolate to estimate the upper bound for the PE of the entire system, denoted as $S^{\mathrm{PE}}_{\mathrm{est}}$. The difference in estimated PE $\Delta S^{\mathrm{PE}}_{\mathrm{est}}\approx 6$ between the two initial states  reflects the significant discrepancy between the Krylov subspaces corresponding to the two initial states.

\textit{Conclusions and outlook.}---%
In summary, we have reported the experimental observation of distinctively initial-state dependent dynamics in Stark systems, by studying the dynamics of imbalance and PE for initial states with varying domain wall numbers. Our experimental results are primarily relevant for system sizes up to $12 \times 2$ and timescales of tunneling times up to $\bar{J}_\text{NN}t\sim 40$, qualitatively showcasing signatures of HSF in Stark systems. This experiment elucidates the crucial distinctions between the systems with weak linear and disordered potentials, demonstrating weak ETH due to HSF in the former and strong ETH in the latter. The distinctions are further numerically demonstrated for larger systems and longer timescales using MPS techniques in Supplementary Materials~\cite{supp}. Moreover, the close relation between the late-time imbalance and domain wall numbers of initial states indicates the presence of a more intricate landscape of the mobility edge in weak ergodicity-breaking systems like Stark systems~\cite{Zhang2021,Wei2022}, in contrast with disorder-driven MBL where mobility edge relies solely on energy~\cite{Guo2021_2}. 

The efficient readout of our superconducting processor enables the measurement of PE, offering direct evidence of HSF. However, it is a challenging experimental task for systems with a length of 12 or larger. To this end, we propose a scheme for estimating the upper bound of PE by measuring the local PE, which requires relatively fewer experimental measurements. As a direct measure of the Krylov subspace dimension, the PE, along with the associated local PE, merits systematic study in typical toy models of HSF, such as pair hopping models~\cite{Pai2019,Khemani2020,Sala2020,Moudgalya2021}. This approach will also enable further investigations of HSF in Stark systems or other higher-dimensional weak ergodicity-breaking systems~\cite{Naik2023, Will2023}.

\begin{acknowledgments}
    This work was supported by the National Natural Science Foundation of China (Grants Nos. 92265207, T2121001, 92365301, T2322030, 12122504, 12274142, 12475017, and 12204528), 
    the Innovation Program for Quantum Science and Technology (Grant No. 2021ZD0301800), 
    the Beijing Nova Program (No. 20220484121),
    the Natural Science Foundation of Guangdong Province (Grant No. 2024A1515010398),
    Scientific Instrument Developing Project of Chinese Academy of Sciences (Grant No. YJKYYQ20200041), 
    the Key-Area Research and Development Program of Guang-Dong Province (Grant Nos. 2018B030326001 and 2020B0303030001), 
    the State Key Development Program for Basic Research of China (Grant No. 2017YFA0304300),
    the China Postdoctoral Science Foundation (Grant No. GZB20240815),
    and the support from the Synergetic Extreme Condition User Facility (SECUF) in Huairou District, Beijing. Devices were made at the Nanofabrication Facilities at Institute of Physics, CAS in Beijing.
    \end{acknowledgments}

\bibliography{refe}
\clearpage    

\end{document}


\title{Supplementary Materials: \\Exploring Hilbert-Space Fragmentation on a Superconducting Processor}

\author{Yong-Yi Wang}
\thanks{These authors contributed equally to this work.}
\affiliation{Institute of Physics, Chinese Academy of Sciences, Beijing 100190, China}
\affiliation{School of Physical Sciences, University of Chinese Academy of Sciences, Beijing 100049, China}

\author{Yun-Hao Shi}
\thanks{These authors contributed equally to this work.}
\affiliation{Institute of Physics, Chinese Academy of Sciences, Beijing 100190, China}
\affiliation{School of Physical Sciences, University of Chinese Academy of Sciences, Beijing 100049, China}
\affiliation{Beijing Academy of Quantum Information Sciences, Beijing 100193, China}

\author{Zheng-Hang Sun}
\thanks{These authors contributed equally to this work.}
\affiliation{Theoretical Physics III, Center for Electronic Correlations and Magnetism, Institute of Physics, University of Augsburg, D-86135 Augsburg, Germany}

\author{Chi-Tong Chen}
\affiliation{Quantum Science Center of Guangdong-Hong Kong-Macao Greater Bay Area, Shenzhen, Guangdong 518045, China}

\author{Zheng-An Wang}
\affiliation{Beijing Academy of Quantum Information Sciences, Beijing 100193, China}
\affiliation{Hefei National Laboratory, Hefei 230088, China}

\author{Kui Zhao}
\affiliation{Beijing Academy of Quantum Information Sciences, Beijing 100193, China}

\author{Hao-Tian Liu}
\affiliation{Institute of Physics, Chinese Academy of Sciences, Beijing 100190, China}
\affiliation{School of Physical Sciences, University of Chinese Academy of Sciences, Beijing 100049, China}

\author{Wei-Guo Ma}
\affiliation{Institute of Physics, Chinese Academy of Sciences, Beijing 100190, China}
\affiliation{School of Physical Sciences, University of Chinese Academy of Sciences, Beijing 100049, China}

\author{Ziting Wang}
\affiliation{Beijing Academy of Quantum Information Sciences, Beijing 100193, China}

\author{Hao Li}
\affiliation{Beijing Academy of Quantum Information Sciences, Beijing 100193, China}

\author{Jia-Chi Zhang}
\affiliation{Institute of Physics, Chinese Academy of Sciences, Beijing 100190, China}
\affiliation{School of Physical Sciences, University of Chinese Academy of Sciences, Beijing 100049, China}

\author{Yu Liu}
\affiliation{Institute of Physics, Chinese Academy of Sciences, Beijing 100190, China}
\affiliation{School of Physical Sciences, University of Chinese Academy of Sciences, Beijing 100049, China}

\author{Cheng-Lin Deng}
\affiliation{Institute of Physics, Chinese Academy of Sciences, Beijing 100190, China}
\affiliation{School of Physical Sciences, University of Chinese Academy of Sciences, Beijing 100049, China}

\author{Tian-Ming Li}
\affiliation{Institute of Physics, Chinese Academy of Sciences, Beijing 100190, China}
\affiliation{School of Physical Sciences, University of Chinese Academy of Sciences, Beijing 100049, China}

\author{Yang He}
\affiliation{Institute of Physics, Chinese Academy of Sciences, Beijing 100190, China}
\affiliation{School of Physical Sciences, University of Chinese Academy of Sciences, Beijing 100049, China}

\author{Zheng-He Liu}
\affiliation{Institute of Physics, Chinese Academy of Sciences, Beijing 100190, China}
\affiliation{School of Physical Sciences, University of Chinese Academy of Sciences, Beijing 100049, China}

\author{Zhen-Yu Peng}
\affiliation{Institute of Physics, Chinese Academy of Sciences, Beijing 100190, China}
\affiliation{School of Physical Sciences, University of Chinese Academy of Sciences, Beijing 100049, China}

\author{Xiaohui Song}
\affiliation{Institute of Physics, Chinese Academy of Sciences, Beijing 100190, China}
\affiliation{School of Physical Sciences, University of Chinese Academy of Sciences, Beijing 100049, China}

\author{Guangming Xue}
\affiliation{Beijing Academy of Quantum Information Sciences, Beijing 100193, China}

\author{Haifeng Yu}
\affiliation{Beijing Academy of Quantum Information Sciences, Beijing 100193, China}

\author{Kaixuan Huang}
\email{huangkx@baqis.ac.cn}
\affiliation{Beijing Academy of Quantum Information Sciences, Beijing 100193, China}

\author{Zhongcheng Xiang}
\email{zcxiang@iphy.ac.cn}
\affiliation{Institute of Physics, Chinese Academy of Sciences, Beijing 100190, China}
\affiliation{School of Physical Sciences, University of Chinese Academy of Sciences, Beijing 100049, China}

\author{Dongning Zheng}
\affiliation{Institute of Physics, Chinese Academy of Sciences, Beijing 100190, China}
\affiliation{School of Physical Sciences, University of Chinese Academy of Sciences, Beijing 100049, China}
\affiliation{Songshan Lake Materials Laboratory, Dongguan, Guangdong 523808, China}
\affiliation{CAS Center for Excellence in Topological Quantum Computation, UCAS, Beijing 100190, China, and Mozi Laboratory, Zhengzhou 450001, China}

\author{Kai Xu}
\email{kaixu@iphy.ac.cn}
\affiliation{Institute of Physics, Chinese Academy of Sciences, Beijing 100190, China}
\affiliation{School of Physical Sciences, University of Chinese Academy of Sciences, Beijing 100049, China}
\affiliation{Beijing Academy of Quantum Information Sciences, Beijing 100193, China}
\affiliation{Songshan Lake Materials Laboratory, Dongguan, Guangdong 523808, China}
\affiliation{CAS Center for Excellence in Topological Quantum Computation, UCAS, Beijing 100190, China, and Mozi Laboratory, Zhengzhou 450001, China}

\author{Heng Fan}
\email{hfan@iphy.ac.cn}
\affiliation{Institute of Physics, Chinese Academy of Sciences, Beijing 100190, China}
\affiliation{School of Physical Sciences, University of Chinese Academy of Sciences, Beijing 100049, China}
\affiliation{Beijing Academy of Quantum Information Sciences, Beijing 100193, China}
\affiliation{Songshan Lake Materials Laboratory, Dongguan, Guangdong 523808, China}
\affiliation{CAS Center for Excellence in Topological Quantum Computation, UCAS, Beijing 100190, China, and Mozi Laboratory, Zhengzhou 450001, China}

\maketitle
\beginsupplement
\onecolumngrid
\tableofcontents
\begin{figure*}[htbp]
	\centering
	\includegraphics[width=0.96\textwidth]{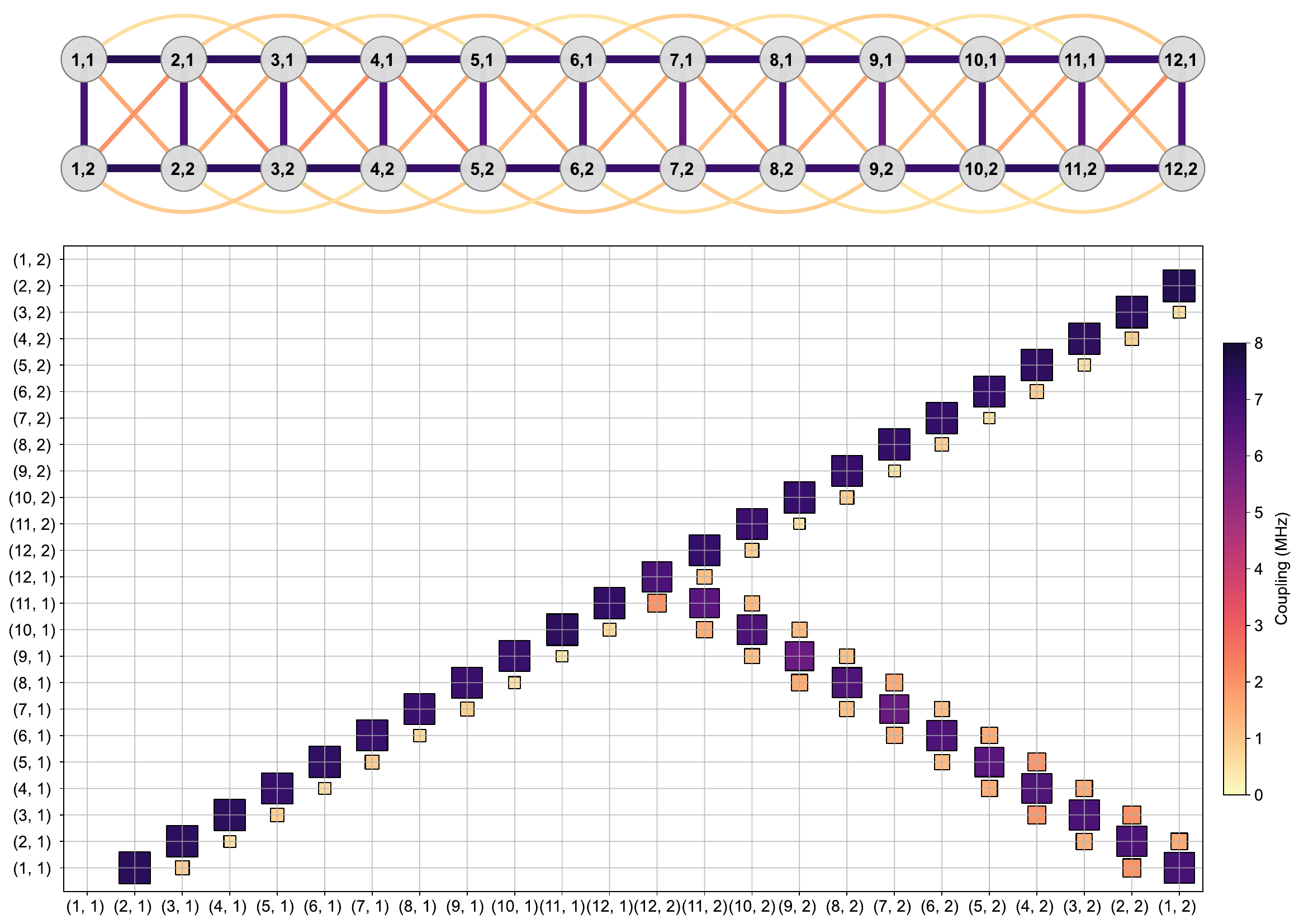}
	\caption{\textbf{Coupling strengths of the 24 qubits used in the ladder-type chip.}}
	\label{fig:coupling}
\end{figure*}
\clearpage
\twocolumngrid

\section{Frequency Arrangement}

Before multi-qubit experiments, we also need to arrange the idle frequency of each single qubit to ensure a long coherence time and a low level of XY crosstalk. We optimize the frequency arrangement by considering the two-level system (TLS), the overlap of energy levels, and the residual coupling between qubits. 
\begin{itemize}
	\item To avoid TLS, we can measure the energy relaxation time $T_1$ near the idle point. If the TLS is coupled to the qubit near the idle point, then the curve of $T_1$ will drop sharply or oscillate. Besides, the dephasing time $T_2$ also needs to be considered.
	
	\item If the qubit at the idle point has the transition frequency of $\omega_{\text{10}}$ from $\ket{0}$ to $\ket{1}$ and the transition frequency of $\omega_{\text{21}}$ from $\ket{1}$ to $\ket{2}$, we should avoid overlapping $\omega_{\text{10}}$, $\omega_{\text{21}}$, and the two-photons frequency $(\omega_{\text{10}}+\omega_{\text{21}})/2$. In addition, the leakage frequencies when we excite the qubit cannot overlap with these frequencies. Specifically, the leakage frequencies include $\omega_{\text{c}}$~(intrinsic leakage) and $2\omega_{\text{c}}-\omega_{\text{10}}$~(mirror leakage), where $\omega_{\text{c}}$ denotes the carrier frequency. 

	\item The XY crosstalk includes classical microwave crosstalk and quantum crosstalk. The classical microwave crosstalk can be effectively suppressed by appropriately increasing the duration of the excitation microwave pulse (reducing the broadening in the frequency domain). In our experiment, we set the duration of $\pi$ pulse to $120~\mathrm{ns}$. However, the quantum crosstalk resulting from the residual coupling between qubits can only be reduced by changing the frequencies and increasing the qubit frequency difference.
\end{itemize}
\begin{table}[htbp]
		\centering
            \scalebox{0.95}
		{\begin{tabular}{lcccc}
			Parameter & Median & Mean & Stdev. & Units\\[.05cm] \hline
			\\[-.2cm]
			Qubit maximum frequency & $5.025$ & $5.032$ & $0.240$ & GHz \\[.09cm]
			Qubit idle frequency & $4.723$ & $4.728$ & $0.346$ & GHz \\[.09cm]
			Qubit anharmonicity & $-0.222$ & $-0.222$ & $0.022$ & GHz\\[.09cm]
			Readout frequency & $6.715$ & $6.714$ & $0.061$ & GHz\\[.09cm]
			Mean energy relaxation time $\overline{T}_1$ & $33.2$ & $32.1$ & $7.5$ & $\mu$s \\[.09cm]
			Pure dephasing time at idle frequency $T_2^{\ast}$ & $1.0$ & $2.4$ & $4.2$ & $\mu$s \\[.09cm]
		\end{tabular}}
	\caption{\textbf{List of device parameters.} Reproduced from Ref.~\cite{Shi2023}.}\label{tab:paras}
\end{table}

\begin{figure}[b]
	\centering
	\includegraphics[width=0.47\textwidth]{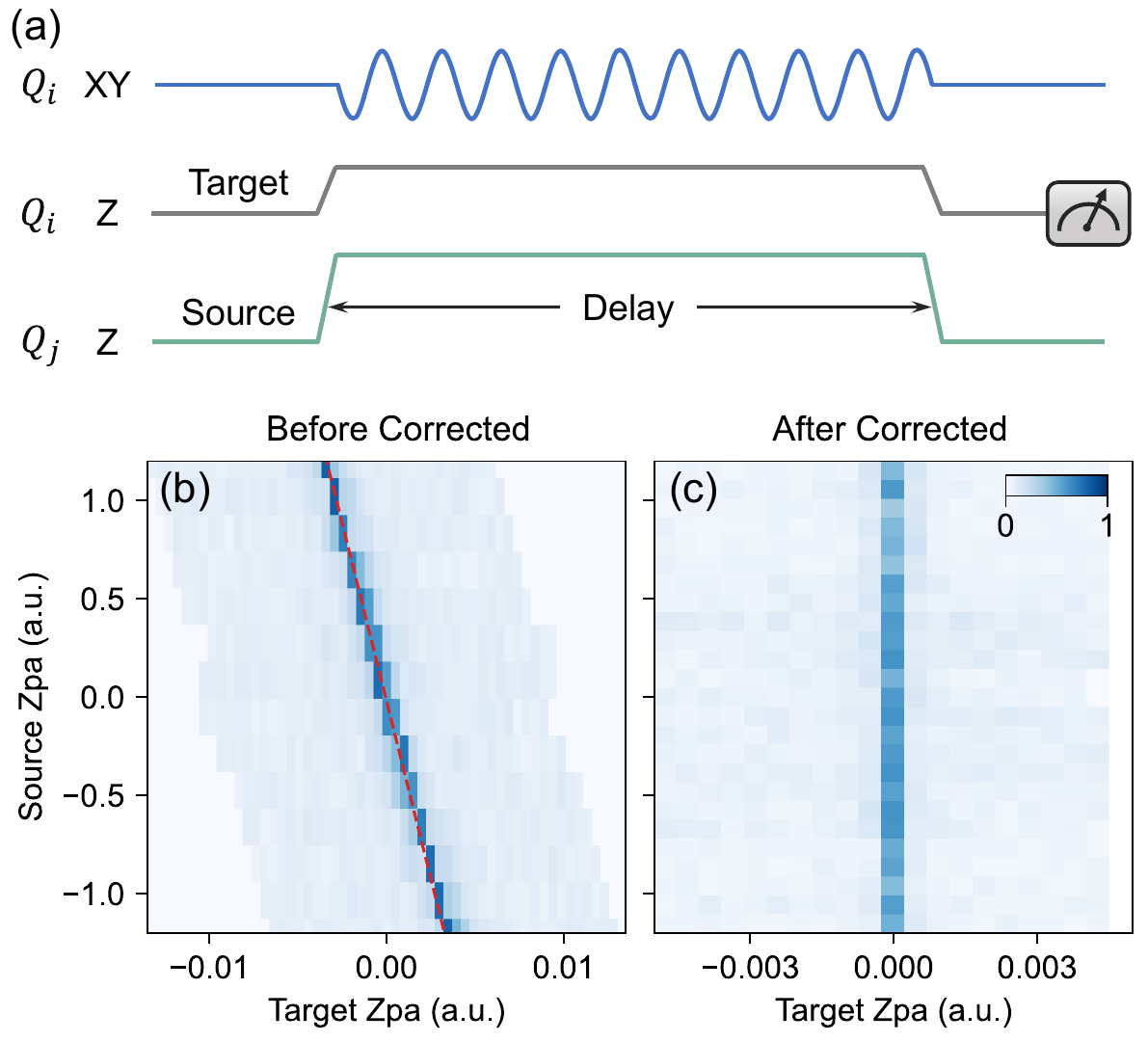}
	\caption{\textbf{Typical experimental data of measuring Z crosstalk.} (a) Pulse sequences for measuring the Z crosstalk coefficient of $Q_j$ to $Q_i$, i.e., $M_{i,j}$. Here we fix the delay at $1.5~\mu\mathrm{s}$ to suppress the broadening of data. (b) Heatmap showing the probability of $Q_i$ in $\ket{1}$ before and after the Z crosstalk correction. The red dash line is the result of linear fitting to obtain $M_{i,j}$.}
	\label{fig:zxtalk_spec}
\end{figure}

\begin{figure*}[htbp]
	\centering
	\includegraphics[width=0.96\textwidth]{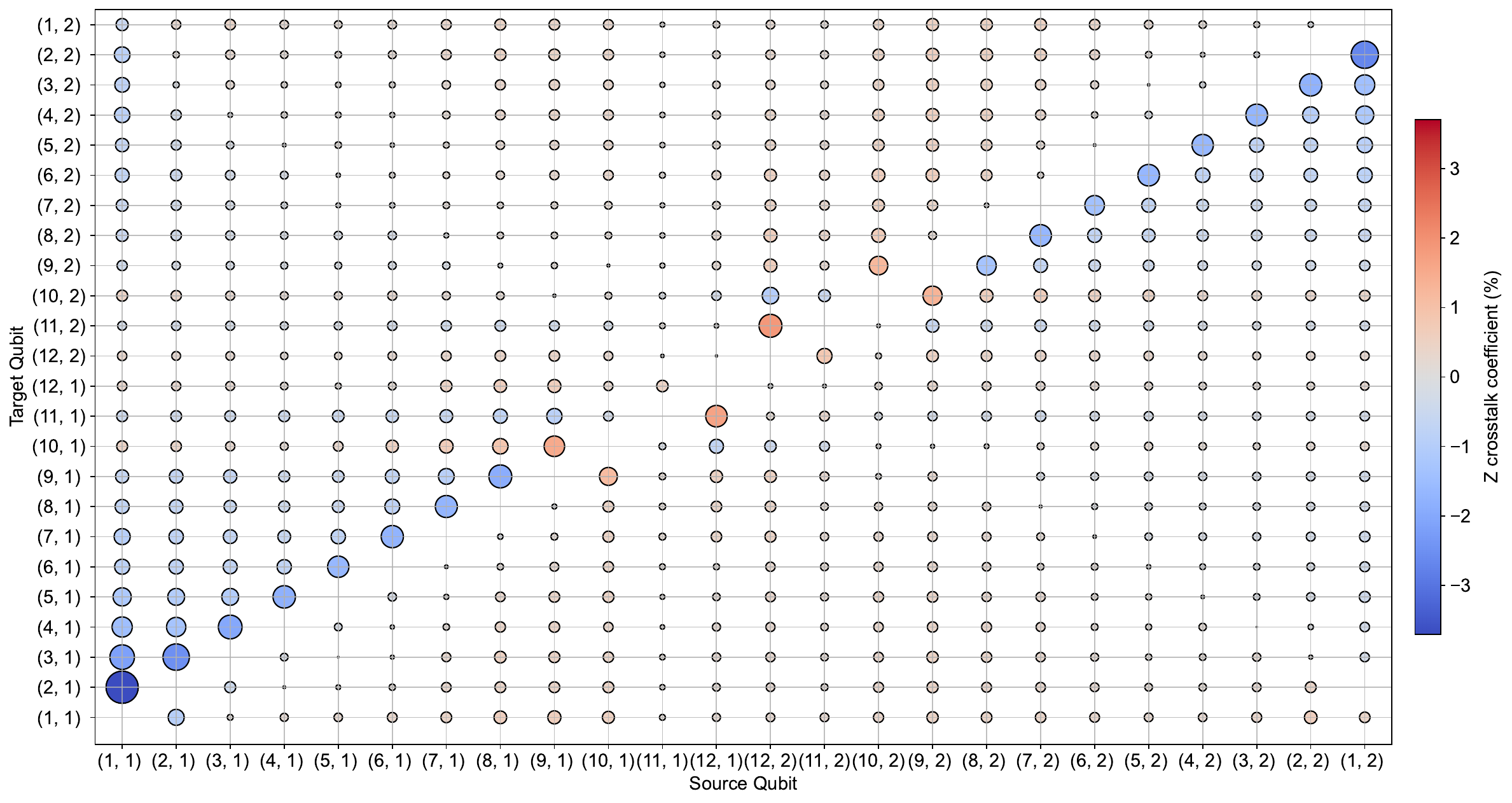}
	\caption{\textbf{Off-diagonal Z crosstalk matrix.} The heatmap shows the Z crosstalk coefficient $M_{i,j}$ ($i\neq{j}$). The size of the colored bubble represents $|M_{i,j}|$. Here we show the crosstalk coefficients of the 24 qubits used in the ladder-type chip.}
	\label{fig:zxtalk_mat}
\end{figure*}

\section{Multi-Qubit Calibration}

After single-qubit calibration, we implement multi-qubit calibration to correct the crosstalk, align the timing, and find the control parameters of qubits to achieve the target Hamiltonian with high precision. The reference frequency is fixed at $4.534~\mathrm{GHz}$, and the on-site potential manipulated in the experiment is the difference between the qubit frequency and this reference frequency.

\subsection{Z Crosstalk Correction}
The Z crosstalk results from the fact that the low-frequency Z bias signal is not completely localized to a single qubit. The individual Z bias signal of each qubit has a spatial distribution across the chip, but the strength decays with the distance between qubits. Assuming that the Z pulse amplitude (Zpa) of the $j$-th qubit $Q_j$ is $z_j$, and the vertical distance between its Z control line and the $i$-th qubit $Q_i$ is $r_{i,j}$, then the magnetic induction strength felt by $Q_i$ from the Z line of the $Q_j$ can be expressed as $B_{{i}\leftarrow{j}}\propto{z_j/r_{i,j}}$. The corresponding crosstalk flux thus is $\Phi_{{i}\leftarrow{j}}=B_{{i}\leftarrow{j}}S_{i}=c_{{i}\leftarrow{j}}z_j$, where $S_i$ represents the loop area of $Q_i$'s SQUID and $c_{{i}\leftarrow{j}}\propto{S_{i}/r_{i,j}}$ denotes the flux crosstalk per unit Zpa. To compensate for the crosstalk $\Phi_{{i}\leftarrow{j}}$, we apply $\Phi_{{i}\leftarrow{i}}=c_{{i}\leftarrow{i}}z_i$ on the Z line of $Q_i$ to satisfy
\begin{equation}
	\Phi_{{i}\leftarrow{i}}=-\Phi_{{i}\leftarrow{j}},
\end{equation}
thus the Z crosstalk coefficient of $Q_j$ to $Q_i$ can be calculated as
\begin{equation}
	M_{i,j}=\frac{c_{{i}\leftarrow{j}}}{c_{{i}\leftarrow{i}}}=-\frac{z_i}{z_j}.
\end{equation}
For each $z_j$, we scan the Zpa of $Q_i$ to find the corresponding $z_i$ for compensation. Hence, the Z crosstalk coefficient $M_{i,j}$ can be determined by linear fitting. As shown in Fig.~\ref{fig:zxtalk_spec}(a), we apply an XY drive pulse to the target qubit $Q_i$ while scanning its Zpa to compensate for the crosstalk from the Zpa of the source $Q_j$. If the crosstalk is compensated exactly, then we will measure a high probability of $Q_i$ in $\ket{1}$. Typical experimental data are shown in Fig.~\ref{fig:zxtalk_spec}(c)~and~\ref{fig:zxtalk_spec}(d). Note that the above Z crosstalk is linear and does not take into account the anti-cross effect of energy levels when qubits are close to each other. When measuring such linear crosstalk, it is preferable to adjust the step of the scanned source Zpa so that the frequency change corresponding to each scanning step is greater than twice the coupling strength. Thus, the impact of the anti-cross effect on the measured data will be reduced.

After measuring the Z crosstalk coefficient for each pair of qubits, we obtain the whole crosstalk matrix $\mathbf{M}$ (Fig.~\ref{fig:zxtalk_mat}). In the subsequent multi-qubit calibration and experiments, we will implement the following correction to the Z pulses of all the qubits:
\begin{equation}
	\mathbf{z}_{\text{corrected}}=\mathbf{M}^{-1}\mathbf{z},
\end{equation}
where $\mathbf{M}^{-1}$ is the inverse of $\mathbf{M}$, $\mathbf{z}=[z_1,z_2,\dots]^T$ denotes the vector of all the Zpas, and $\mathbf{z}_{\text{corrected}}$ represents the corrected vector. 


\begin{figure}[htbp]
	\centering
	\includegraphics[width=0.47\textwidth]{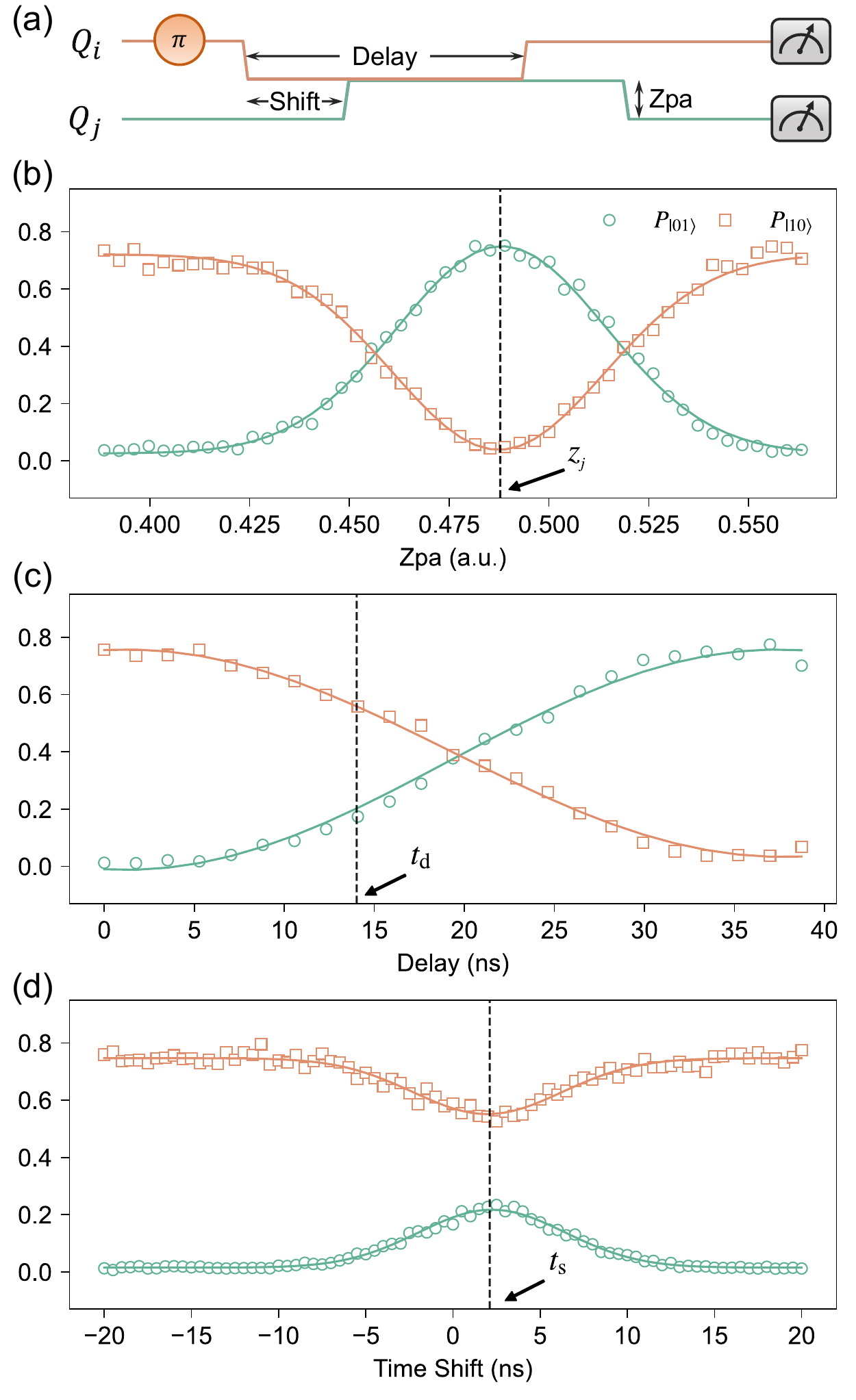}
	\caption{\textbf{Typical experimental data of timing calibration.} (a) Pulse sequences of two-qubit swap experiment for timing calibration. (b) Joint probabilities $P_{\ket{01}}$ and $P_{\ket{10}}$ as the function of the Zpa of $Q_j$. By fitting this data with the cosine function, we obtain the swap period $T$. Then we determine the delay $t_{\text{d}}$ slightly less than $T/4$. (c) Joint probabilities as the function of delay. (d) Joint probabilities as the function of time shift. The optimal time shift $t_{\text{s}}$ applied to $Q_j$ is calculated as the average of the Gaussian fitting results.}
	\label{fig:timing}
\end{figure}

\subsection{Timing Calibration}
In single-qubit timing calibration, we have calibrated the timing between each qubit's own XY and Z signals. However, different qubits have different control line lengths and this causes a timing shift between qubits. Here we introduce an efficient method to automatically calibrate the timing between qubits (Fig.~\ref{fig:timing}):
\begin{itemize}
	\item Find the Zpa corresponding to the two-qubit resonance: We initially prepare $\ket{10}$ by applying a $\pi$ pulse to $Q_i$, then vary the Zpa of $Q_j$ (the Zpa of $Q_i$ is fixed near the resonance point) and measure the joint probabilities $P_{\ket{01}}$ and $P_{\ket{10}}$, see Fig.~\ref{fig:timing}(b). For these probabilities to change significantly as the Zpa of $Q_j$ approaches the resonance point, we choose the delay to be approximately half of the swap period, i.e., $t_{\text{d}}\approx{T/2}$, where $T=\pi/g$ is the swap period and $g$ denotes the coupling strength of qubits. Here the coupling strength only needs to be given a rough value and does not need to be finely measured.
	\item Determine a sensitive delay time of the swap: With the Zpa of $Q_j$ corresponding to the resonance, we measure $P_{\ket{01}}$ and $P_{\ket{10}}$ as the function of delay, see Fig.~\ref{fig:timing}(c). By fitting this data with the cosine function, we can obtain the exact swap period $T$. The sensitive delay time is near where the derivative is at its maximum ($\approx T/4$). Here we choose a delay slightly less than $T/4$ for the next step.
	\item Optimize the relative time shift between two qubits: We measure $P_{\ket{01}}$ and $P_{\ket{10}}$ as the function of $Q_j$'s time shift and fit them with Gaussian functions, see Fig.~\ref{fig:timing}(d). The optimal time shift is at the peak (or dip) of the Gaussian function.
\end{itemize}

\subsection{On-Site Potential}
Our experiment requires a high-precision control of on-site disorder and Stark potential. Even though we have corrected the linear Z crosstalk and aligned the timing, the qubit (dressed-)frequency will still deviate from the target frequency when other qubits are simultaneously biased. This effect mainly comes from the interaction with other qubits (if the residual Z crosstalk is relatively negligible). Here we introduce a method to suppress this effect and calibrate qubit frequencies with a high precision.

\begin{figure*}[htbp]
	\centering
	\includegraphics[width=0.87\textwidth]{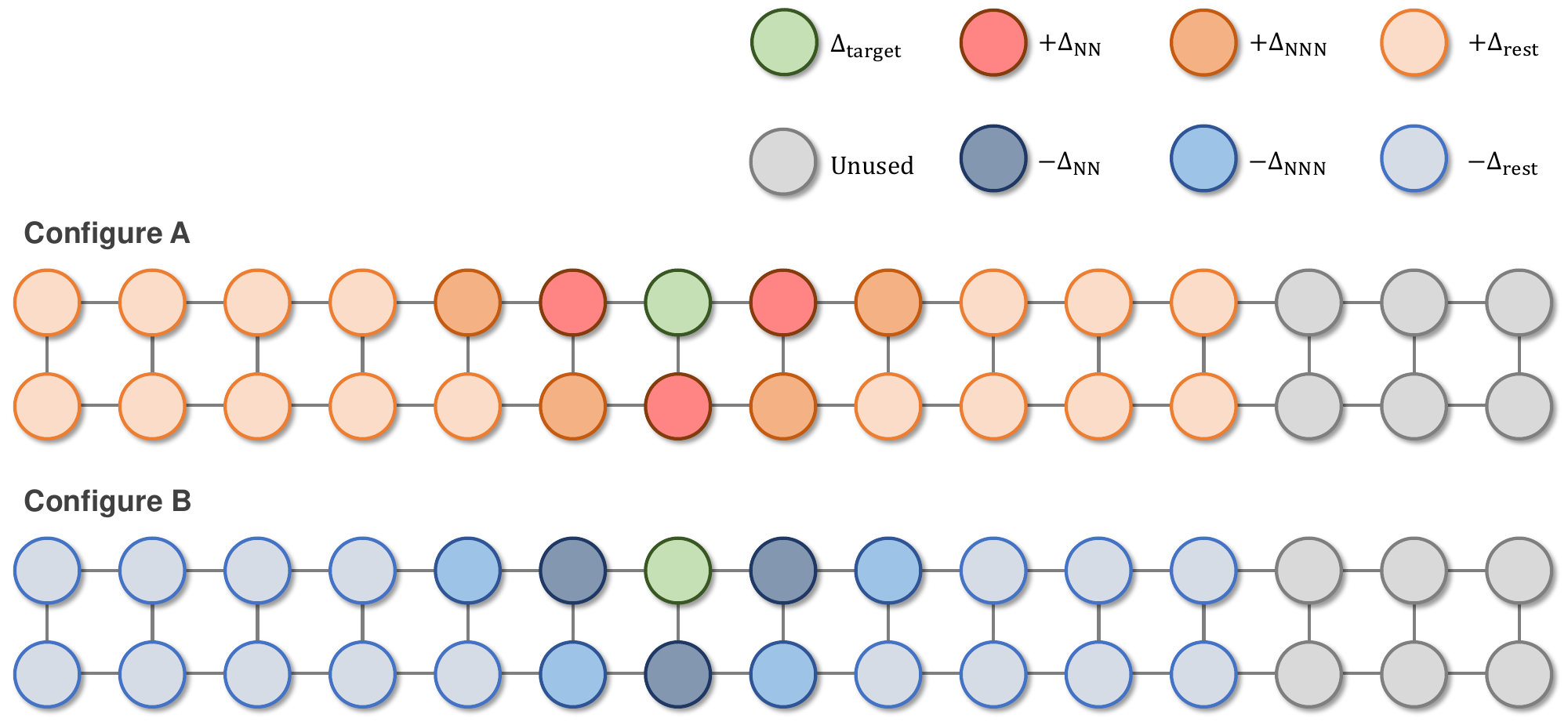}
	\caption{\textbf{Configurations for multi-qubit calibration.} As an example, here we take $Q_{7,\uparrow}$ as the target qubit. The frequency differences are $\Delta_{\text{NN}}/2\pi\approx160~\mathrm{MHz}$, $\Delta_{\text{NN}}/2\pi\approx50~\mathrm{MHz}$ and $\Delta_{\text{rest}}/2\pi\approx25~\mathrm{MHz}$. All unused qubits ($Q_{13,\uparrow}$, $Q_{13,\downarrow}$, $Q_{14,\uparrow}$, $Q_{14,\downarrow}$, $Q_{15,\uparrow}$, $Q_{15,\downarrow}$) are biased away from the reference frequency.}
	\label{fig:freq_cali}
\end{figure*}

\begin{figure}[htbp]
	\centering
	\includegraphics[width=0.47\textwidth]{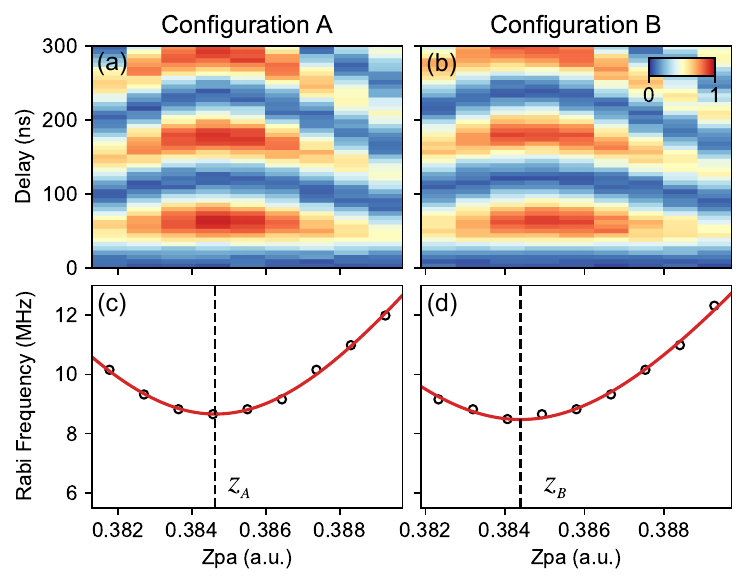}
	\caption{\textbf{Typical experimental data of calibrating on-site potential.} The results of Rabi oscillation experiments under configurations A (a) and B (b). The heatmap shows the probability of the target qubit in $\ket{1}$. (c) and (d) correspond to (a) and (b), respectively, representing the Rabi frequency as a function of Zpa.}
	\label{fig:rabilong}
\end{figure}

As shown in Fig.~\ref{fig:freq_cali}, there are two configurations for the calibration, namely A and B. Configuration A requires that the nearest-neighbor (NN) qubits, the next-nearest-neighbor (NNN) qubits, and the rest qubits biased from the target frequency of $+\Delta_{\text{NN}}$, $+\Delta_{\text{NNN}}$, and $+\Delta_{\text{rest}}$, respectively. Thus, the dressed frequency of the target qubit $Q_i$ can be approximately expressed as
\begin{equation}
	\tilde{\omega}_i^{\text{A}}\approx\omega_i+\sum_{j\neq{i}}\frac{g^2_{i,j}}{\Delta_j},
\end{equation}
where $\omega_i$ denotes the target frequency, $g_{i,j}$ is the coupling strength, and $\Delta_j=\omega_i-\omega_j$ is the frequency detuning. Similarly, for configuration B with the opposite detuning, we have
\begin{equation}
	\tilde{\omega}_i^{\text{B}}\approx\omega_i-\sum_{j\neq{i}}\frac{g^2_{i,j}}{\Delta_j}.
\end{equation}
We perform the Rabi oscillation experiments under these two configurations, and thus determine the corresponding Zpas of $\tilde{\omega}_i^{\text{A}}$ and $\tilde{\omega}_i^{\text{B}}$, i.e., $z_i^{\text{A}}$ and $z_i^{\text{B}}$ (Fig.~\ref{fig:rabilong}). 

\begin{figure*}[t]
	\centering
	\includegraphics[width=0.96\textwidth]{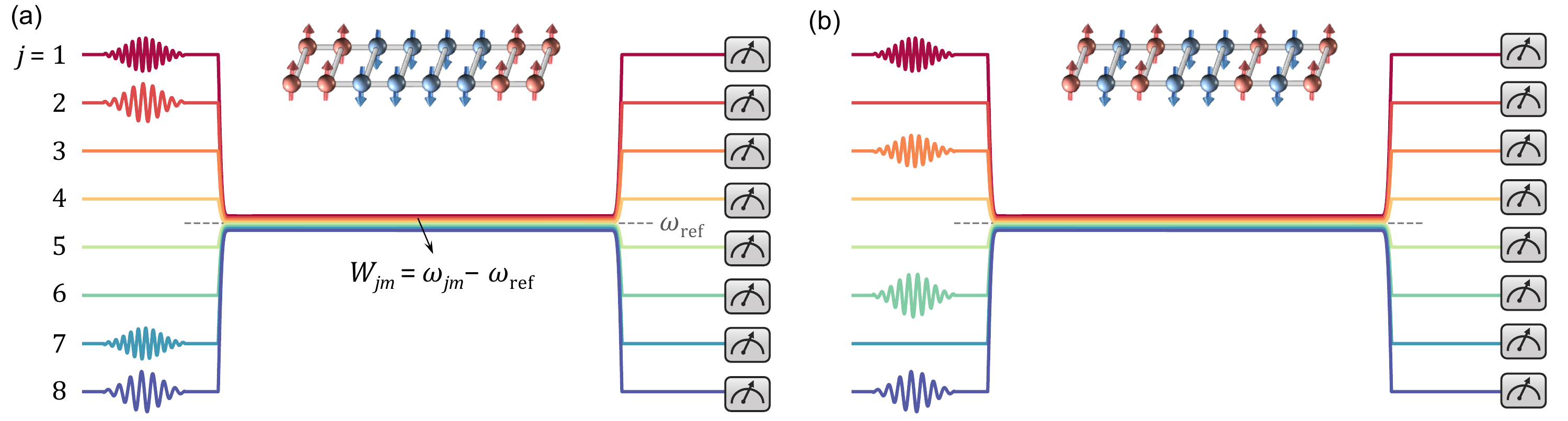}
	\caption{\textbf{Schematic of experimental pulse sequences for different initial states.} Here we take the case of $L=8$ as an example, where the initial states $\ket{\psi_{n_{\text{DW}}=2}}=\ket{\mathbb{11000011}}$ and $\ket{\psi_{n_{\text{DW}}=6}}=\ket{\mathbb{10100101}}$ correspond to (a) and (b), respectively. The on-site potential $W_{jm}$ is denoted as the difference between the qubit frequency $\omega_{jm}$ and the reference frequency $\omega_{\text{ref}}$. By controlling the Zpas of qubits, we manipulate $\omega_{jm}$ to realize the Stark (linear) potentials $W_{jm}/ \bar{J}_{\text{NN}}=-j\gamma$, or random potentials $W_{jm}/ \bar{J}_{\text{NN}}$ with a uniform distribution in $[-W, W]$.}
	\label{fig:pulse_seq}
\end{figure*}

To ensure a large adjustment range of Zpa, we usually set the target frequency below the maximum frequency of the qubit with the smallest maximum frequency. The relationship between the qubit's Zpa and frequency near such target frequencies is monotonic. Hence, the corresponding Zpa of the target $\omega_i$ can be approximately calculated as the average of $z_i^{\text{A}}$ and $z_i^{\text{B}}$:
\begin{equation}
	z_i=f(\omega_i)\approx\frac{1}{2}\left[f(\tilde{\omega}_i^{\text{A}})+f(\tilde{\omega}_i^{\text{B}})\right]=\frac{1}{2}(z_i^{\text{A}}+z_i^{\text{B}}),
\end{equation}
where $f$ is a monotonic function that represents the relationship between qubit's Zpa and frequency near the target frequency. For each qubit, we repeat the above calibration procedure and finally obtain the Zpas of all qubits corresponding to the target on-site potentials. The schematic of experimental pulse sequences is shown in Fig.~\ref{fig:pulse_seq}.





\section{Model and Hamiltonian}
\subsection{Jordan-Wigner Transformation for the $XX$-Ladder Model}
The main interacting term of the Hamiltonian describing our superconducting processor can be approximated as the $XX$-ladder model. The approximating ladder-XX model can be mapped to an interacting spinless fermion model, utilizing a generalized form of Jordan-Wigner transformation, described as
\begin{equation}
	\hat{S}_{j, 1}^{-}=\hat{c}_{j, 1} \exp \left[i \pi \sum_{l=0}^{j-1}\left(\hat{n}_{l, 1}+\hat{n}_{l, 2}\right)\right],
	\end{equation}
for the first row of chain, and 
\begin{equation}
	\hat{S}_{j, 2}^{-}=\hat{c}_{j, 2} \exp \left[i \pi \left[\sum_{l=0}^{j}\hat{n}_{l, 1}+\sum_{l=0}^{j-1}\hat{n}_{l, 2}\right]\right],
	\end{equation}
for the second row of chain, where $\hat{S}_{j,m}^{-}=(\hat{\sigma}_{j,m}^x-i \hat{\sigma}_{j,m}^y) / 2$ ($j=1,2,\ldots,L$ and $m=1,2$) and $\hat{n}_{j,m}=\hat{c}^\dagger_{j,m}\hat{c}_{j,m}$ is the spinless fermion number operator for site $(j,m)$. Therefore, the Hamiltonian in the spinless fermion representation reads
\begin{equation}
	\begin{aligned}
		\hat{H}=-2 J \sum_{j, \delta}[ & \hat{c}_{j,1} \exp \left(-i \pi \hat{\phi}_{j, j+\delta}^{(1)}\right) \hat{c}_{j+\delta,1}^{\dagger} \\
	+&\hat{c}_{j,1}  \exp \left(-i \pi \hat{\phi}_{j, j+\delta}^{(2)}\right) \hat{c}_{j+\delta,2}^{\dagger}] \\
	-&2 J \sum_j \hat{c}_{j, 1} \hat{c}_{j, 2}^{\dagger},
	\end{aligned}
	\end{equation}
where $\delta$ denotes the nearest neighbor, and $\hat{\phi}_{j, j+1}^{(1)}=\hat{n}_{j,2},\hat{\phi}_{j, j-1}^{(1)}=-\hat{n}_{j-1,2},\hat{\phi}_{j, j+1}^{(2)}=\hat{n}_{j,1},\hat{\phi}_{j, j-1}^{(2)}=-\hat{n}_{j-1,1}$. The many-body interaction in the Hamiltonian leads to the nonintegrability and many-body quantum chaos.

\subsection{Level Statistics}
\begin{figure}[htbp]
	\centering
	\includegraphics[width=0.9\linewidth]{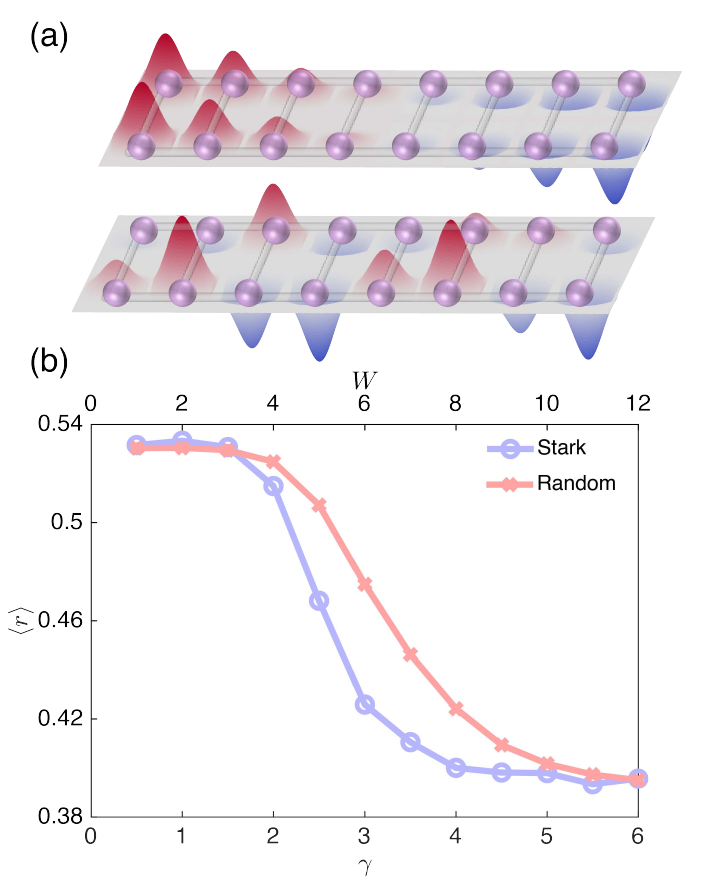}
	\caption{\textbf{Ergodicity to localization transition in Stark and random potentials.}
	(a) Schematic representation of the Stark potentials and random potentials sampled from a uniform distribution to study the transition from ergodicity to localization. All qubits are adjusted to corresponding frequencies deviated from the center frequency $\approx{4.534}~\mathrm{GHz}$. (b) The ratio of adjacent level spacings as the function of gradient $\gamma$ or the disorder strength $W$, with level statistics from the ergodic Wigner-Dyson distribution ($\langle r \rangle \simeq 0.531$) to the many-body localized Poisson distribution ($\langle r \rangle \simeq 0.386$). }
	\label{fig:LS}
\end{figure}

In this work, we study systems in both Stark potentials $W_{jm}/ \bar{J}_{NN}=-j\gamma$, and random potentials $W_{jm}/ \bar{J}_{NN}$ drawn from a uniform distribution in $[-W, W]$, as illustrated in the upper and lower panel of Fig.~\ref{fig:LS}(a), respectively. With the increasing of the gradient $\gamma$ or the disorder strength $W$, a transition from ergodicity to localization is expected in this typical nonintegrable model.

A common numerical way to characterize the ergodicity and MBL is the level statistics based on random matrix theory, quantified by the ratio of adjacent level spacings $r^{(n)}=\min \left(\delta^{(n)}, \delta^{(n+1)}\right) / \max \left(\delta^{(n)}, \delta^{(n+1)}\right)$ with $\delta^{(n)}=E_n-E_{n-1}>0$. As the gradient $\gamma$ (or the disorder strength $W$) increases, the level statistics changes from the Wigner-Dyson distribution with the average $\langle r \rangle \simeq 0.531$ in the ergodic phase, to the Poisson distribution with $\langle r \rangle \simeq 0.386$ in the many-body localized phase (see Fig.~\ref{fig:LS}(b)).

It is worth noting that, across all parameters depicted in Fig.~\ref{fig3} and Fig.~\ref{fig4}, the average ratio of adjacent level spacings $\langle r \rangle$ consistently remains close to $0.531$, signifying thermalization for at least the majority of spectrum.

\subsection{Participation entropy for Gaussian orthogonal ensemble and Gaussian unitary ensemble}
\subsubsection{Derivation}
Here, we give a direct derivation of the participation entropy predicted by the Gaussian orthogonal ensemble (GOE) and the Gaussian unitary ensemble (GUE)~\cite{Torres-Herrera2016,Boixo2018}.

For the GOE, the $N$ components of an eigenvector is real number, and its norm must be one. Therefore, 
\begin{equation}
    P_{\text{GOE}}(\mathbf{c})=C_{N} \delta\left(1-\sum_{n=1}^{N}c_{n}^{2}\right).
\end{equation}
For the function of random variants $f(\mathbf{x})$ subjects to distribution $P(f)$, we have $P(f)=\int{\mathbf{dx}\delta(P-f({\mathbf{x}}))p(\mathbf{x})}$. So the marginal distribution of a single component by
\begin{equation}
    P_{\text{GOE}}(p)=\int d c_{1} \cdots d c_{N} \delta\left(p-c_{i}^{2}\right) P_{\text{GOE}}(\mathbf{c}).
\end{equation}

Defining
\begin{equation}
    P_{\text{GOE}}(p;t)=\int d c_{1} \cdots d c_{N} \delta\left(p-c_{i}^{2}\right)C_{N} \delta\left(t-\sum_{n=1}^{N}c_{n}^{2}\right),
\end{equation}
and taking the Laplace transform, we have:
\begin{equation}
    \begin{aligned}
        &\int_{0}^{\infty} d t e^{-s t} P_{\text{GOE}}(p ; t)\\
		&=C_{N} \int d c_{i} \delta\left(p-c_{i}^{2}\right) e^{-s c_{i}^{2}}\left(\int d c e^{-s c^{2}}\right)^{N-1}\\
        &\propto \frac{e^{-s p}}{\sqrt{p}s^{(N-1)/2}}.
    \end{aligned}
\end{equation}
Then inverting the Laplace transform, we finally get
\begin{equation}
    \begin{aligned}
        &P_{\text{GOE}}(p ; t) \propto(t-p)^{(N-3)/2} \theta(t-p)/\sqrt{p}\\
        \Rightarrow &P_{\text{GOE}}(p)=P_{\text{GOE}}(p ; 1)=\frac{1}{\sqrt{\pi}} \frac{\Gamma(N / 2)}{\Gamma((N-1) / 2)} \frac{(1-p)^{(N-3) / 2}}{\sqrt{p}}.
    \end{aligned}
\end{equation}

Therefore, $P_{\text{GOE}}(p) \propto \exp{(-Np/2)}/\sqrt{p}$ for large $N \gg  1$, and
\begin{equation}
	\begin{aligned}
		&S^{\text{PE}}_{\text{GOE}}=-\sum_i^{N} {{p_i}\log {p_i}}\\
		&=-N\int_0^{\infty} P_{\text{GOE}}(p) p \log(p) dp\\
		&\propto -N\int_0^{\infty} \exp{(-Np/2)}\sqrt{p} \log(p) dp.
	\end{aligned}
\end{equation}
Finally, we get $S^{\text{PE}}_{\text{GOE}}=\log N -2 + \log(2) + \gamma_e \approx \log{0.482N}$, with $\gamma_e$ being Euler's constant.

Similarly, since the only constraint on $N$ components of an eigenvector for the GUE is that its norm must be one, we have
\begin{equation}
    P_{\text{GUE}}(\mathbf{c})=C_{N} \delta\left(1-\sum_{n=1}^{N}\left|c_{n}\right|^{2}\right).
\end{equation}

Defining
\begin{equation}
    P_{\text{GUE}}(p;t)=\int d^{2} c_{1} \cdots d^{2} c_{N} \delta\left(p-\left|c_{i}\right|^{2}\right)C_{N} \delta\left(t-\sum_{n=1}^{N}\left|c_{n}\right|^{2}\right),
\end{equation}
and taking the Laplace transform:
\begin{equation}
    \begin{aligned}
	&\int_{0}^{\infty} d t e^{-s t} P_{\text{GUE}}(p ; t)\\
	&=C_{N} \int d^{2} c_{i} \delta\left(p-\left|c_{i}\right|^{2}\right) e^{-s\left|c_{i}\right|^{2}}\left(\int d^{2} c e^{-s|c|^{2}}\right)^{N-1}\\
    &\propto \frac{e^{-s p}}{s^{N-1}},
    \end{aligned}
\end{equation}
then inverting the Laplace transform, we finally get
\begin{equation}
    \begin{aligned}
        &P_{\text{GUE}}(p ; t) \propto(t-p)^{N-2} \theta(t-p)\\
        \Rightarrow &P_{\text{GUE}}(p)=P_{\text{GUE}}(p ; 1)=(N-1)(1-p)^{N-2}.
    \end{aligned}
\end{equation}
Therefore, $P_{\text{GUE}}(p)=N\exp{(-Np)}$ for large $N \gg  1$, and
\begin{equation}
	\begin{aligned}
		&S^{\text{PE}}_{\text{GUE}}=-\sum_i^{N} {{p_i}\log {p_i}}\\
		&=-N\int_0^{\infty} p N e^{-N p} \log p d p\\
		&=\log N -1 + \gamma_e
	\end{aligned}
\end{equation}

\begin{figure}[htbp]
	\centering
	\includegraphics[width=0.96\linewidth]{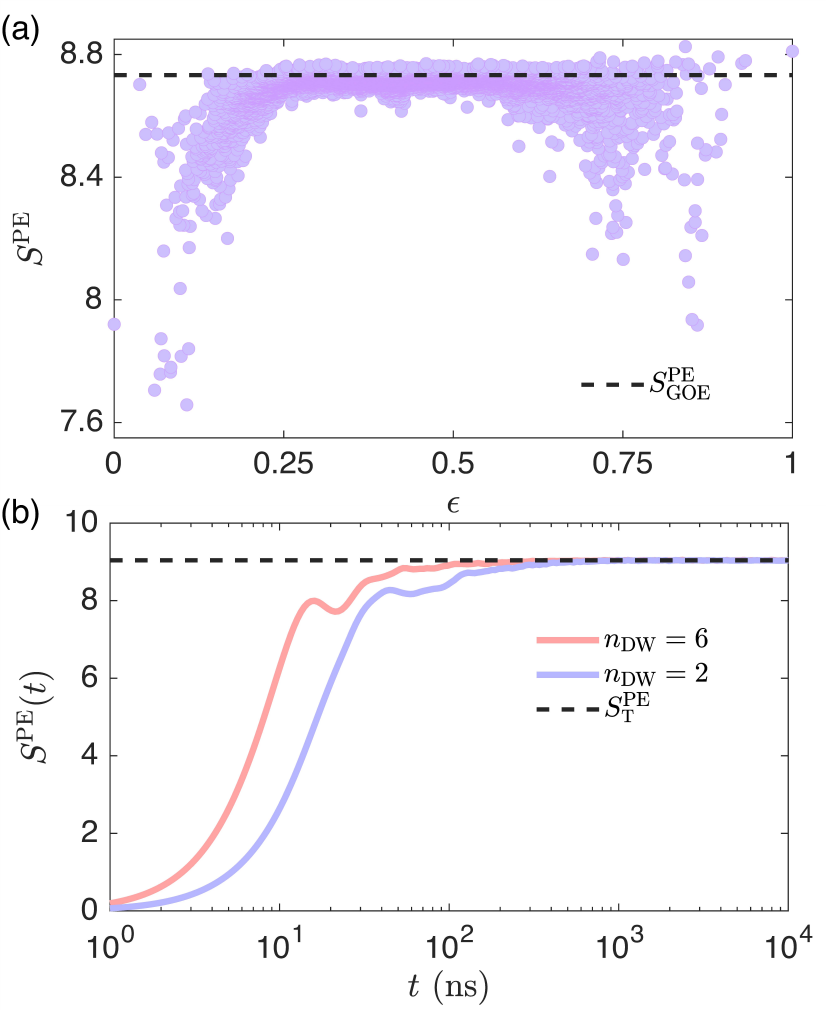}
	\caption{\textbf{Participation entropy for the $XX$-ladder model.} 
	(a)Participation entropy of the eigenstate against its normalized energy. The dashed line represent $S^{\text{PE}}_{\text{GOE}}\approx 8.73$. (b) Time-evolution of participation entropy for the initial states $\ket{\psi_{n_{\text{DW}}=6}}=\ket{\mathbb{10100101}}$ and $\ket{\psi_{n_{\text{DW}}=2}}=\ket{\mathbb{11000011}}$. The dashed line represent $S^{\text{PE}}_{\text{T}}\approx 9.04$.}
	\label{fig:SPE_Erg}
\end{figure}

\subsubsection{Numerical verification}
As stated before, the Hamiltonian [Eq.~(\ref{Ham1})] in the main text without the on-site potential is a typical nonintegrable model, for which the PE of highly excited eigenstates will be close to $S^{\text{PE}}_{\text{GOE}}=\log \mathcal{N} -2 + \log(2) + \gamma_e$. 

At the same time, under ergodic dynamics, the time-evolved states for generic initial conditions will approach complex random pure states after sufficient evolution time. The only constraint is that their norm must be one, consistent with the constraint in eigenvectors for the GUE.
Therefore, if the system thermalizes, the PE of the time-evolved states will approach $S^{\text{PE}}_{\text{T}}=S^{\text{PE}}_{\text{GUE}}=\log \mathcal{N} -1 + \gamma_e$.

Here, we numerically calculate the PE of all eigenstates for the $XX$-ladder model with the system size $8 \times 2$. The PE is plotted against the normalized energy $\epsilon=\frac{E-E_\text{min}}{E_\text{max}-E_\text{min}}$ for each eigenstate in Fig.~\ref{fig:SPE_Erg}(a). The PE of highly excited eigenstates ($\epsilon \approx 0.5$) is very close to $S^{\text{PE}}_{\text{GOE}}\approx 8.73$. We also plot the time-evolution of PE for initial states $\ket{\psi_0}=\ket{\psi_{n_{\text{DW}}=6}}$ and $\ket{\psi_{n_{\text{DW}}=2}}$ in Fig.~\ref{fig:SPE_Erg}(b). We can see that in a few hundred nanoseconds, the PE closely approximates $S^{\text{PE}}_{\text{T}}=S^{\text{PE}}_{\text{GUE}} \approx 9.04$ for both initial states.

\subsection{Statistical Properties of Eigenstates in Stark Systems}
\begin{figure*}[htbp]
	\centering
	\includegraphics[width=0.9\linewidth]{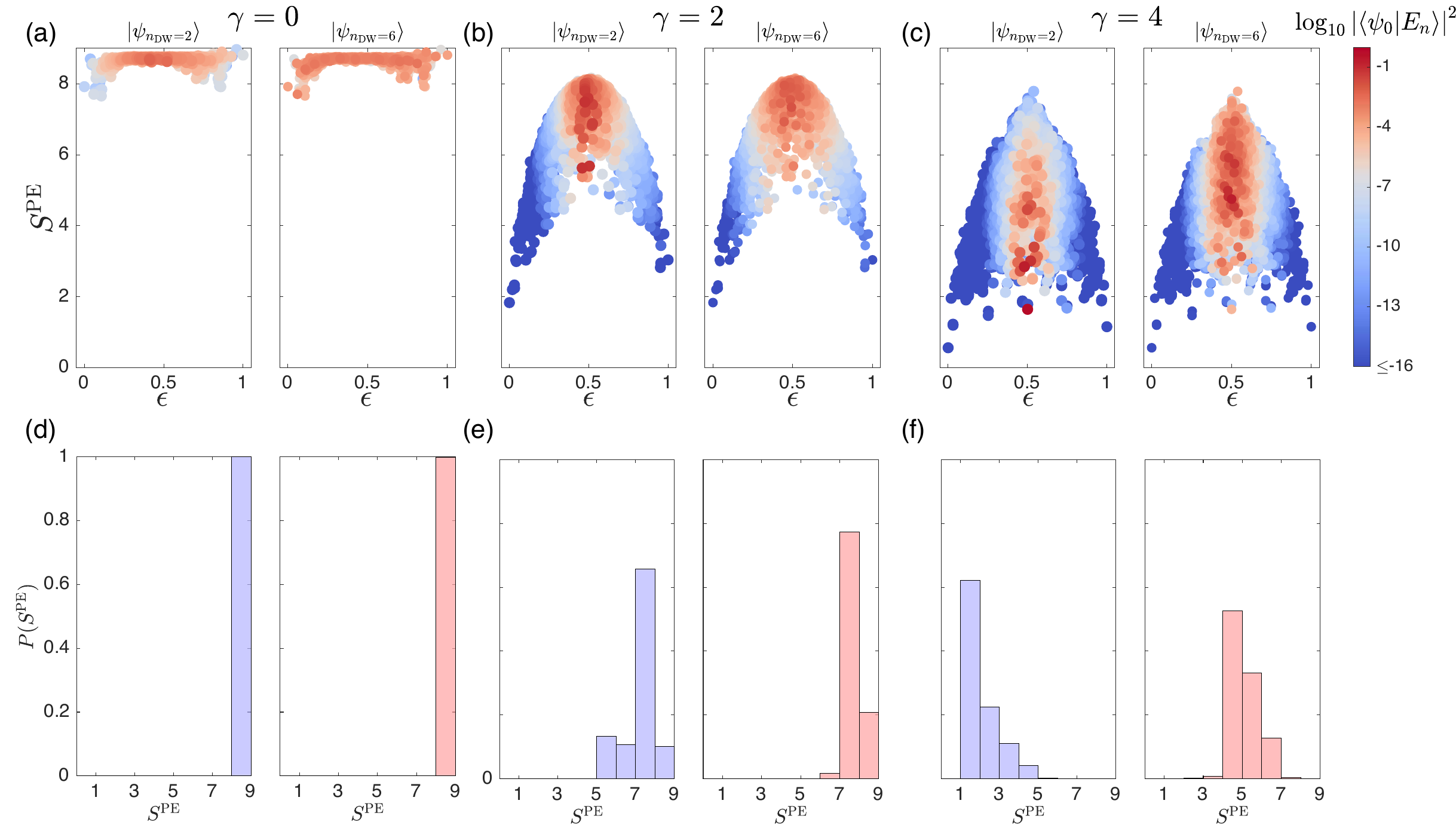}
	\caption{\textbf{Participation entropy of Eigenstates in Stark Systems.} 
	(a--c) Participation entropy of eigenstates for a Stark potential with the gradient $\gamma=0,\ 2$ and $4$. The colorbar indicates the logarithm of the eigenstate occupation numbers $\log_{10}{\left| {\langle \psi_0|E_n\rangle } \right|^2}$ for the initial states $\ket{\psi_{n_{\text{DW}}=2}}$ and $\ket{\psi_{n_{\text{DW}}=6}}$. (d--f) Histogram of the participation entropy at $\gamma=0,\ 2$ and $4$ in the diagonal ensembles corresponding to the initial states $\ket{\psi_{n_{\text{DW}}=2}}$ and $\ket{\psi_{n_{\text{DW}}=6}}$.}
	\label{fig:Eigenstates_PE}
\end{figure*}

\begin{figure*}[htbp]
	\centering
	\includegraphics[width=0.9\linewidth]{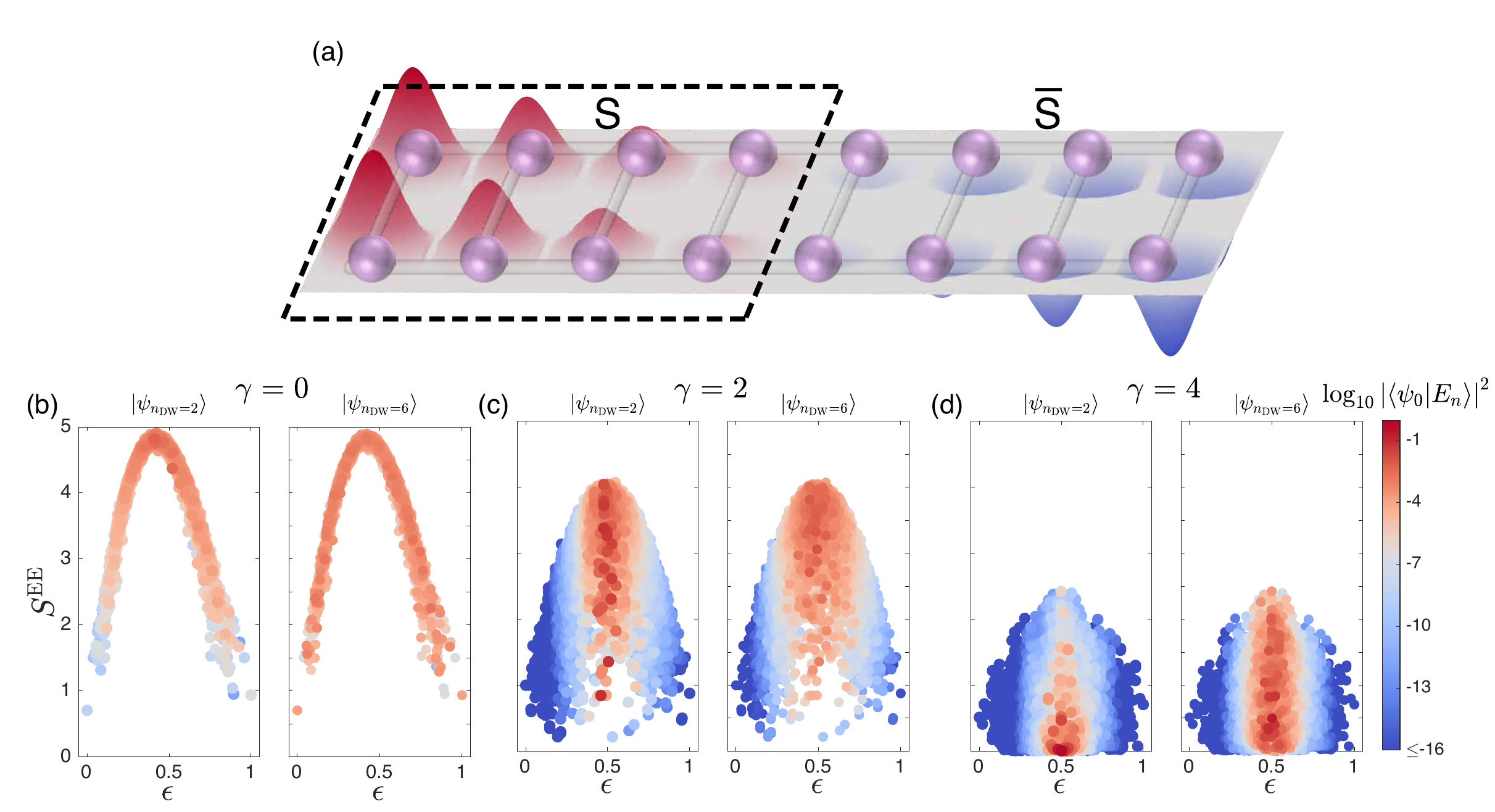}
	\caption{\textbf{Entanglement entropy of eigenstates in Stark Systems.} 
	(a) Half-chain entanglement cut (region S) of the ladder. (b--d) Entanglement entropy of eigenstates for a Stark potential with the gradient $\gamma=0,\ 2$ and $4$. The colorbar indicates the logarithm of the eigenstate occupation numbers $\log_{10}{\left| {\langle \psi_0|E_n\rangle } \right|^2}$ for the initial states $\ket{\psi_{n_{\text{DW}}=2}}$ and $\ket{\psi_{n_{\text{DW}}=6}}$.}
	\label{fig:Eigenstates_EE}
\end{figure*}

\begin{figure*}[htbp]
	\centering
	\includegraphics[width=0.9\linewidth]{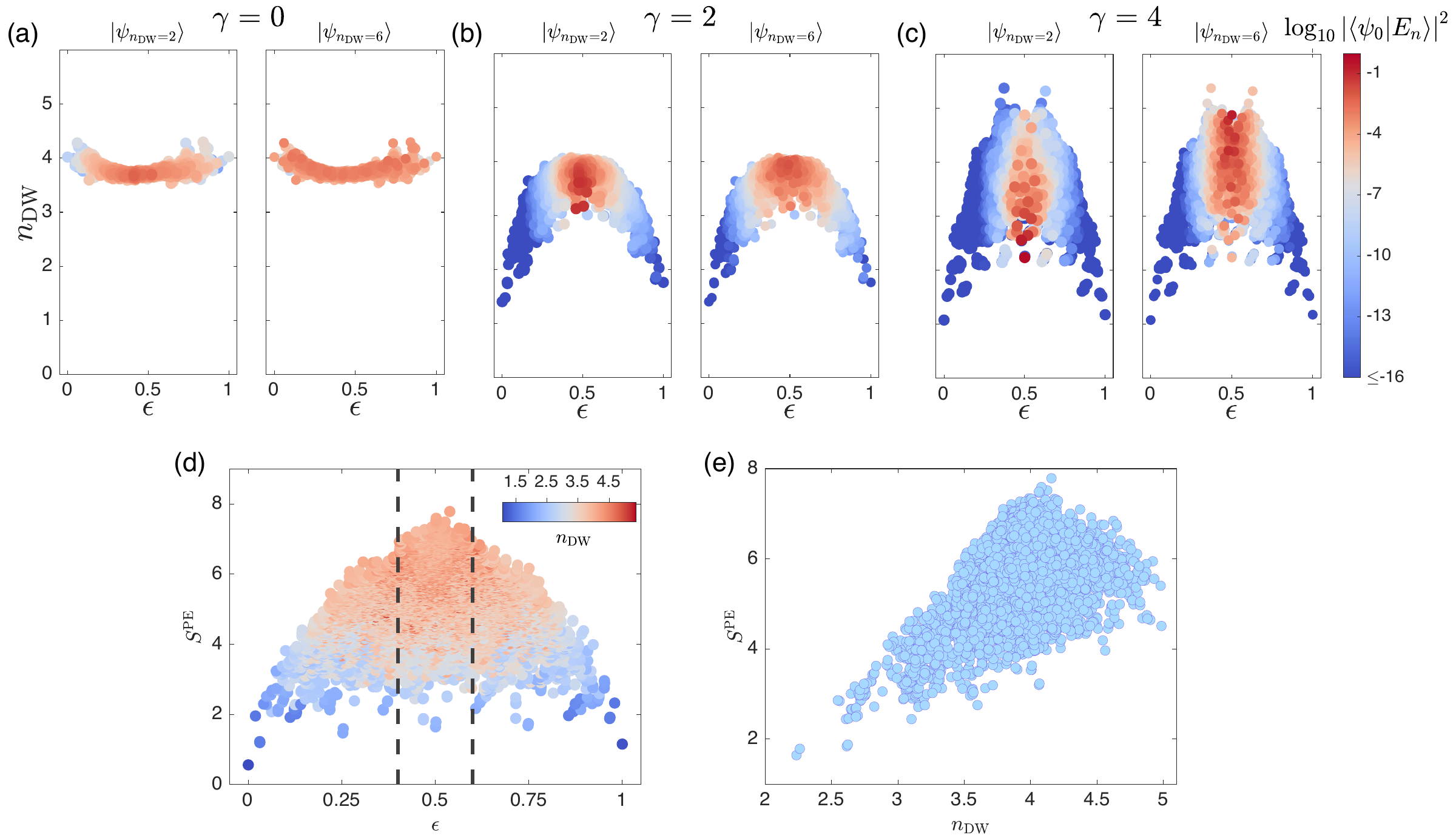}
	\caption{\textbf{Domain wall number of eigenstates in Stark Systems.} 
	(a--c) Domain wall number of eigenstates for a Stark potential with the gradient $\gamma=0,\ 2$ and $4$. The colorbar indicates the logarithm of the eigenstate occupation numbers $\log_{10}{\left| {\langle \psi_0|E_n\rangle } \right|^2}$ for the initial states $\ket{\psi_{n_{\text{DW}}=2}}$ and $\ket{\psi_{n_{\text{DW}}=6}}$. (d) Participation entropy of eigenstates at $\gamma=4$ with the colorbar representing the domain wall number $n_{\text{dw}}$. (e) Participation entropy as a function of $n_{\text{DW}}$ for eigenstates with $\epsilon\in[0.4,0.6]$, corresponding to the region between the dashed lines in (d).}
	\label{fig:Eigenstates_nDW}
\end{figure*}

\subsubsection{Participation entropy}
To quantify the localization in Hilbert space of spin configurations, we can calculate the PE of all eigenstates with exact diagonalization, defined as
\begin{equation}
	S^{\text{PE}}(n) = -\sum_i^{\mathcal{N}} {{p_i(n)}\log {p_i(n)}},
\end{equation}
where $p_i(n) = {\left| {\langle i|E_n\rangle } \right|^2}$, with $\{\ket{i}\}$ being spin configuration basis, and $\mathcal{N}$ is the dimension of Hilbert space {($\mathcal{N}= {16 \choose 8}= 12870$ for the $Q = L$ sector with $L=8$)}.

In Fig.~\ref{fig:Eigenstates_PE}(a--c), the PE of eigenstates is displayed for the Hamiltonian [Eq.~(\ref{Ham1})] with a Stark potential at $\gamma = 0, 2$ and $4$. The colorbars represent the eigenstate occupation numbers $\left| {\langle \psi_0|E_n\rangle } \right|^2$ (i.e., the weights that eigenstates $\ket{E_n}$ have in the diagonal ensemble corresponding to $\ket{\psi_0}$). At $\gamma = 0$, the PE $S^{\text{PE}} \approx S^{\text{PE}}_{\text{GOE}}$ for eigenstates with $\epsilon \approx 0.5$  (same as in Fig.~\ref{fig:SPE_Erg}(a)). As $\gamma$ increases, we can observe that the PE of eigenstates decreases overall, with the range of fluctuations increasing. 

We also displayed the histogram of the PE in the diagonal ensembles for $\ket{\psi_0}=\ket{\psi_{n_{\text{DW}}=2}}$ and $\ket{\psi_{n_{\text{DW}}=6}}$ in Fig.~\ref{fig:Eigenstates_PE}(d--f). For nonzero $\gamma$, the histogram of the PE in the diagonal ensemble for $\ket{\psi_{n_{\text{DW}}=2}}$ tends to distribute more at lower values of PE, compared to the diagonal ensemble for $\ket{\psi_{n_{\text{DW}}=6}}$. From the distribution of the PE in the diagonal ensemble, we can infer that in quench dynamics, the PE at long times also differs for different initial states (as illustrated in Fig.~\ref{fig4} of the main text).

\subsubsection{Entanglement entropy}
Another quantity that is often used to characterize MBL is the entanglement entropy (EE)~\cite{Nandkishore2015,Abanin2019}, defined as
\begin{equation}
	S^{\text{EE}}(n) = -\text{Tr}[\hat\rho_S(n)\log {\hat\rho_S(n)}],
\end{equation}
where $\hat\rho_S(n)=\text{Tr}_{\bar S}[\ket{E_n}\bra{E_n}]$ is the reduced density matrix of the subsystem S (with $\bar S$ being the complement of $S$) for the eigenstate $\ket{E_n}$. As shown in Fig.~\ref{fig:Eigenstates_EE}(a), the subsystem S considered here is the left half of the ladder.
In Fig.~\ref{fig:Eigenstates_EE}(b--d), we display the EE of eigenstates for the Hamiltonian [Eq.~(\ref{Ham1})] with a Stark potential at $\gamma = 0, 2$ and $4$. The colorbars represent the eigenstate occupation numbers $\left| {\langle \psi_0|E_n\rangle } \right|^2$ in the diagonal ensemble. Similar to the PE, the EE of eigenstates decreases overall as $\gamma$ increases, and substantial fluctuations in the EE are observed across the eigenstates for nonzero $\gamma$, indicating violations of ETH. At the same time, compared to $\ket{\psi_0}=\ket{\psi_{n_{\text{DW}}=6}}$, eigenstates with lower EE exhibit larger eigenstate occupation numbers in the diagonal ensemble for $\ket{\psi_{n_{\text{DW}}=2}}$, which implies that the system also exhibits initial-state dependent dynamics in terms of the EE.

\subsubsection{Domain wall number}
Besides the PE and the EE, violations of ETH can also be diagnosed from observables like domain wall number, defined as
\begin{equation}
	n_{\text{DW}}=\sum_{j=1}^{L-1} (1-\langle \hat{\bar{\sigma}}_{j}^{z} \hat{\bar{\sigma}}_{j+1}^{z} \rangle)/2,
\end{equation}
with $\hat{\bar{\sigma}}_{j}^{z}=(\po{j,1}{z}+\po{j,2}{z})/2$. In Fig.~\ref{fig:Eigenstates_nDW}(a--c), we display the domain wall number of the eigenstates, with colorbars being the logarithm of the eigenstate occupation numbers. At $\gamma=0$, according to the ETH, the domain wall numbers $n_{\text{DW}}$ for the majority of eigenstates are close to the ETH prediction of $n_{\text{DW}}^{\text{ETH}}\approx3.73$ for the ladder model. As $\gamma$ increases, the $n_{\text{DW}}$ of eigenstates fluctuates, with a number of eigenstates significantly deviating from the ETH prediction. For nonzero $\gamma$, eigenstate occupation numbers predominantly distribute on eigenstates with relatively smaller $n_{\text{DW}}$ in the diagonal ensemble for $\ket{\psi_{n_{\text{DW}}=2}}$, but they mainly distribute on eigenstates with larger $n_{\text{DW}}$ for $\ket{\psi_{n_{\text{DW}}=6}}$.

In Fig.~\ref{fig:Eigenstates_nDW}(d), we plot the PE of eigenstates at $\gamma=4$, where the colorbar represents $n_{\text{dw}}$. It is evident that eigenstates with higher PE tend to correspond to larger values of $n_{\text{DW}}$. To elucidate this relationship more clearly, in Fig.~\ref{fig:Eigenstates_nDW}(e), the PE is plotted against $n_{\text{DW}}$ for eigenstates in the middle of the spectrum, $\epsilon\in[0.4,0.6]$. This illustrates a discernible positive correlation between the PE and $n_{\text{DW}}$.

\begin{figure}[htbp]
	\includegraphics[width=1\linewidth]{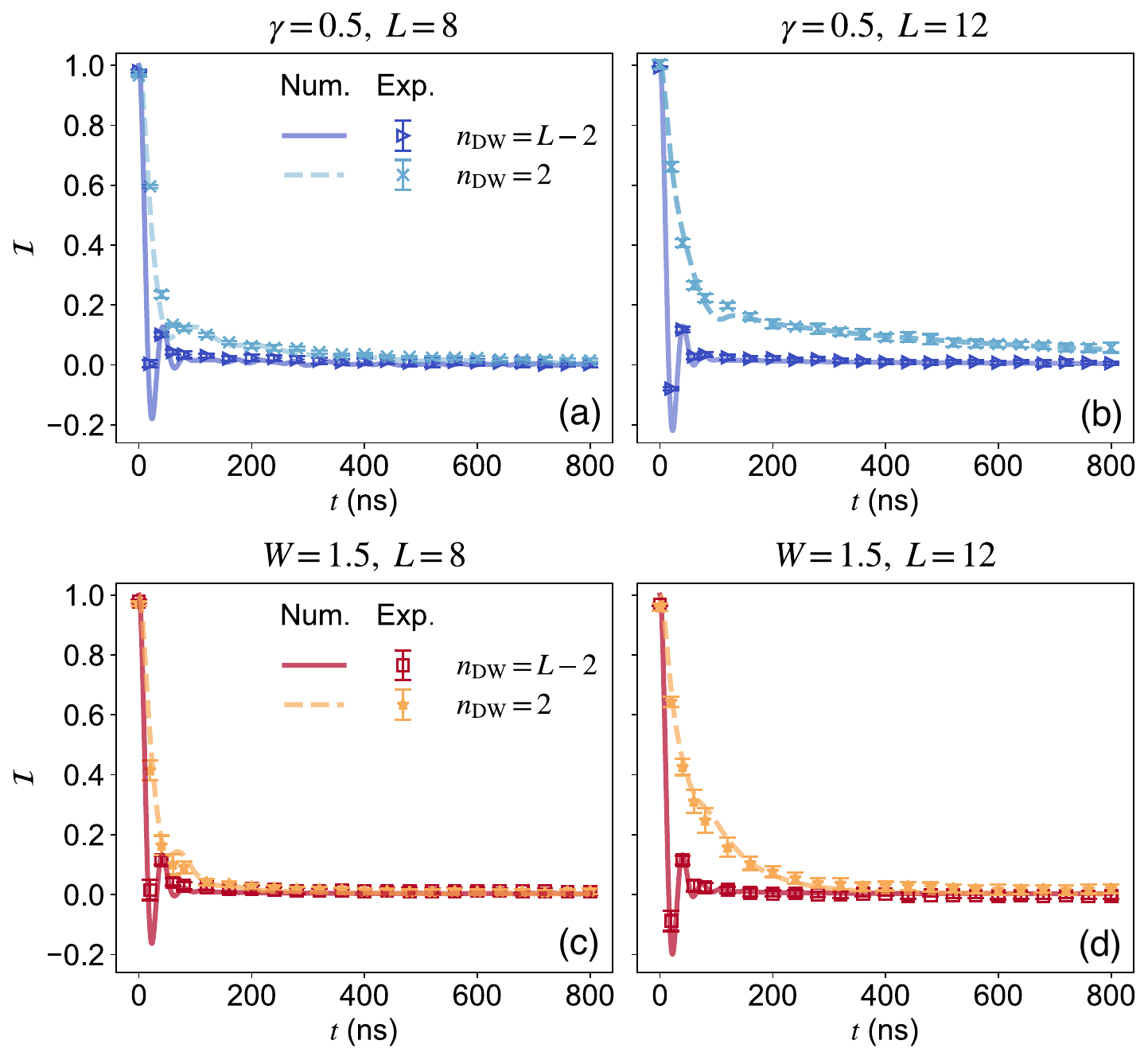}\\
	\caption{\textbf{Time evolution of imbalance for different initial states in different systems:} (a) Stark system with the gradient $\gamma=0.5$ for $L=8$, and (b) Stark system with $\gamma=0.5$ for $L=12$; (c) disordered system with the disorder strength $W=1.5$ for $L=8$, and (d) disordered system with $W=1.5$ for $L=12$. In Stark systems (a, b), points are experimental data, each averaged over $5000$ repetitions with error bars indicating the standard deviation. In disordered systems (c, d), the experimental data points are the average of 20 disorder realizations with $5000$ repetitions for each realization, and the error bar shows the standard error of the mean. Lines represent numerical simulation.}
    \label{fig:Imba_exp}
\end{figure}

\begin{figure*}[htbp]
	\includegraphics[width=0.85\linewidth]{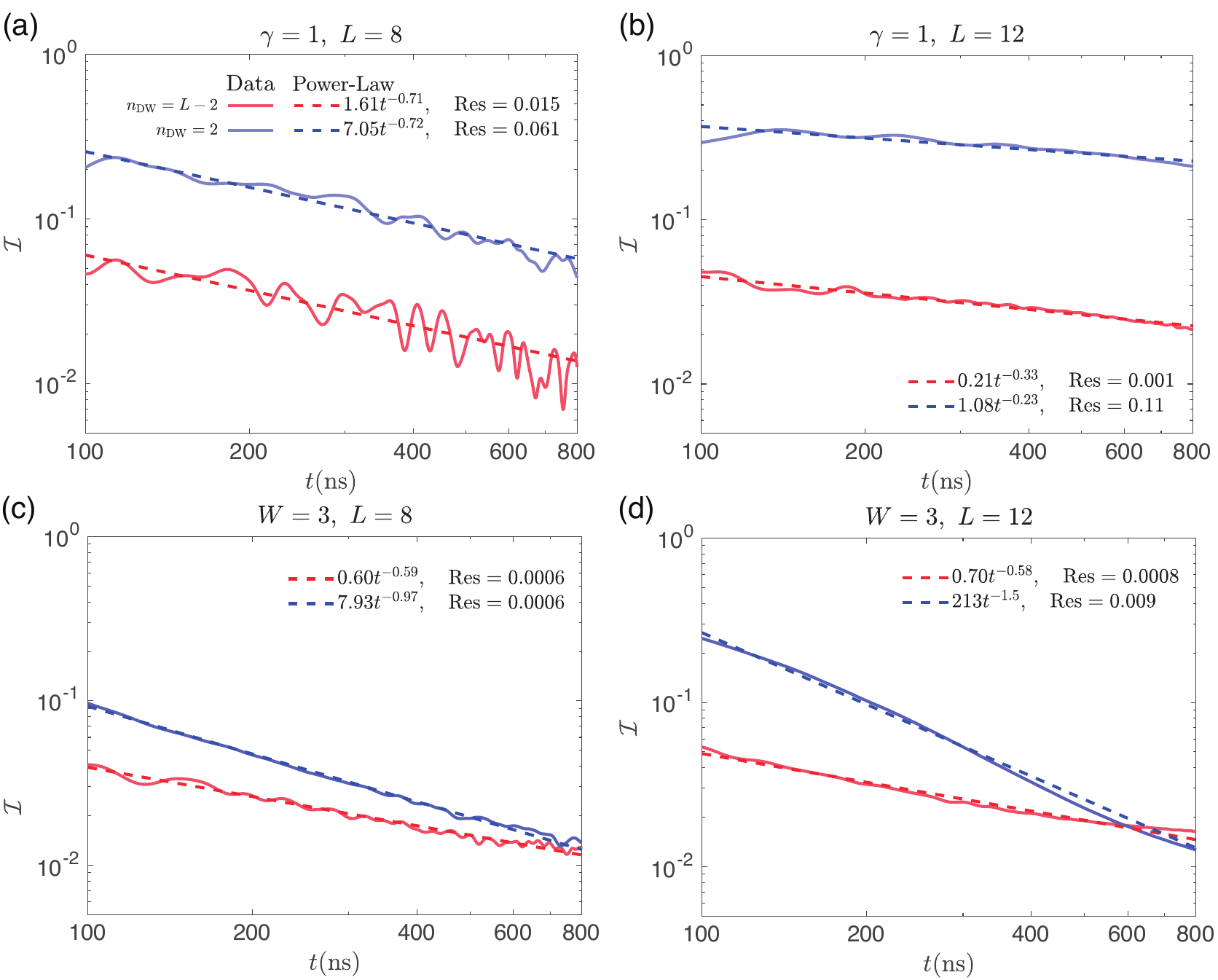}\\
	\caption{\textbf{Time evolution of imbalance and fitting with power-law functions.} (a, b) The dynamics of imbalance in Stark systems with the gradient $\gamma=1$ for (a) $L=8$, and (b) $L=12$. (c, d) The dynamics of imbalance in disordered systems with the disorder strength $W=3$ for (c) $L=8$, and (d) $L=12$. Time evolution of imbalance is fitted from $100$ ns to $800$ ns with power-law function $f(t)\sim at^{-\xi} $.}
    \label{fig:Imba_fit}
\end{figure*}
\section{Supplementary Results on Dynamics}
\subsection{Experimental data of imbalance for smaller \texorpdfstring{$\gamma$}{} and \texorpdfstring{$W$}{}}

In the main text, we present the results of imbalance for the gradient $\gamma=1$ in Stark systems, and for the disorder strength $W=3$ in disordered systems. Here, we show in Fig.~\ref{fig:Imba_exp} additional experimental data of imbalance for smaller $\gamma=0.5$ and $W=1.5$. 

In Stark systems, when $L = 8$, the imbalance relaxes to zero for both initial states (see Fig.~\ref{fig:Imba_exp}(a)). However, as the system length increases to $L = 12$, the relaxation of imbalance for $\ket{\psi_0} = \ket{\psi_{n_{\text{DW}} = 2}}$ significantly slows down, compared to $\ket{\psi_0} = \ket{\psi_{n_{\text{DW}} = L-2}}$ (see Fig.~\ref{fig:Imba_exp}(b)). In contrast, in disordered systems, both initial states exhibit a quick relaxation of imbalance to zero, irrespective of system size, as shown in Fig.~\ref{fig:Imba_exp}(c) and (d). This behavior is consistent with the results for $\gamma = 1$ and $W = 3$ discussed in the main text.

\begin{figure*}[htbp]
	\centering
	\includegraphics[width=0.86\linewidth]{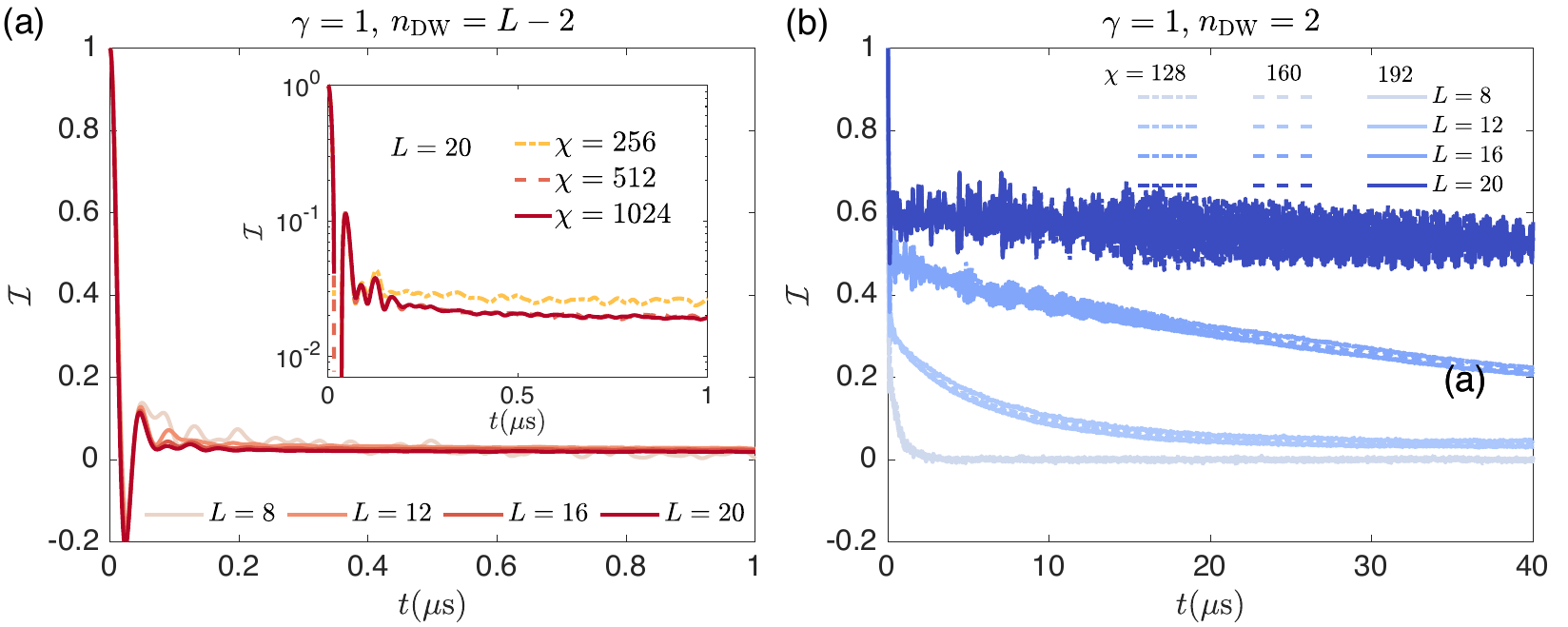}\\
	\caption{\textbf{Time evolution of imbalance in Stark system at $\gamma=1$.} (a) Time evolution of imbalance for $\ket{\psi_0}=\ket{\psi_{n_{\rm{DW}}=L-2}}$ with different system lengths. The maximum bond dimensions $\chi=512$ is employed in TDVP algorithm. The inset shows the results for $L=20$ with various bond dimensions. (b) Time evolution of imbalance for $\ket{\psi_0}=\ket{\psi_{n_{\rm{DW}}=2}}$ with different system lengths. The results of different maximum bond dimensions are displayed.}
	\label{fig:Imba_scaling}
\end{figure*}
\begin{figure*}[htbp]
	\centering
	\includegraphics[width=1\linewidth]{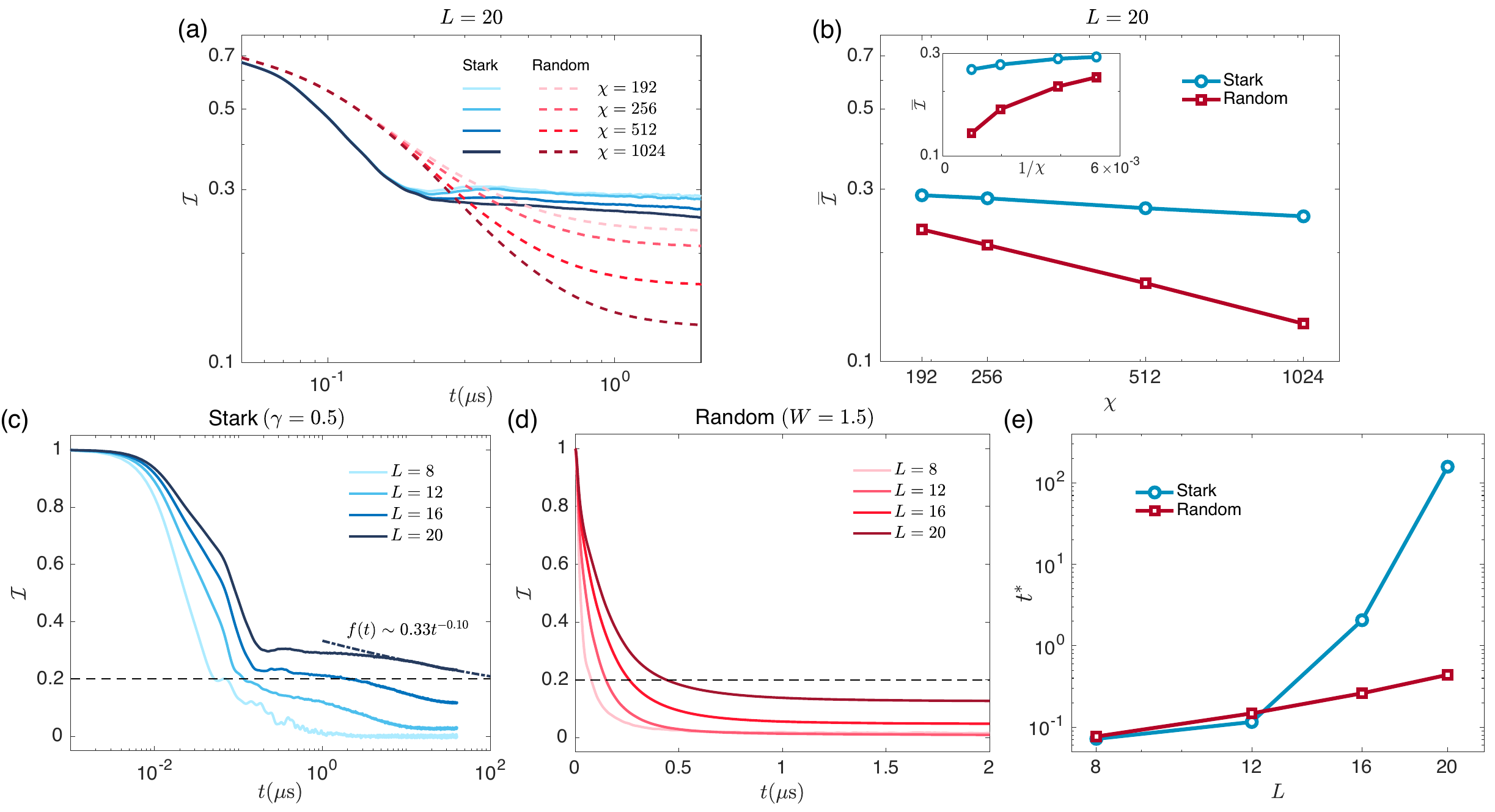}\\
	\caption{\textbf{Time evolution of imbalance for $\ket{\psi_0}=\ket{\psi_{n_{\rm{DW}}=2}}$ in Stark and disordered systems.} (a) Time evolution of imbalance for $L=20$ with varying maximum bond dimensions in the Stark system with $\gamma=0.5$ and in the disordered system with $W=1.5$. (b) Time-averaged imbalance near $2$ $\mu$s as a function of maximum bond dimensions $\chi$ in the Stark and disordered systems. Inset: $\bar{\mathcal{I}}$ as a function of $1/\chi$. (c) Time evolution of imbalance in Stark systems for different system lengths with $\chi=192$. For $L=20$, fitting of the imbalance from $10$ to $40$ $\mu$s with the power-law function $f(t)\sim at^{\xi}$ is plotted. (d) Time evolution of imbalance in disordered systems for different system lengths with $\chi=1024$. The dashed lines in (c) and (d) denote $\mathcal{I}^*=0.2$. (e) The relaxation time estimated by the imbalance decaying to $\mathcal{I}^*=0.2$ in Stark and disordered systems.}
	\label{fig:Imba_scaling2}
\end{figure*}

\begin{figure*}[htbp]
	\centering
	\includegraphics[width=0.96\linewidth]{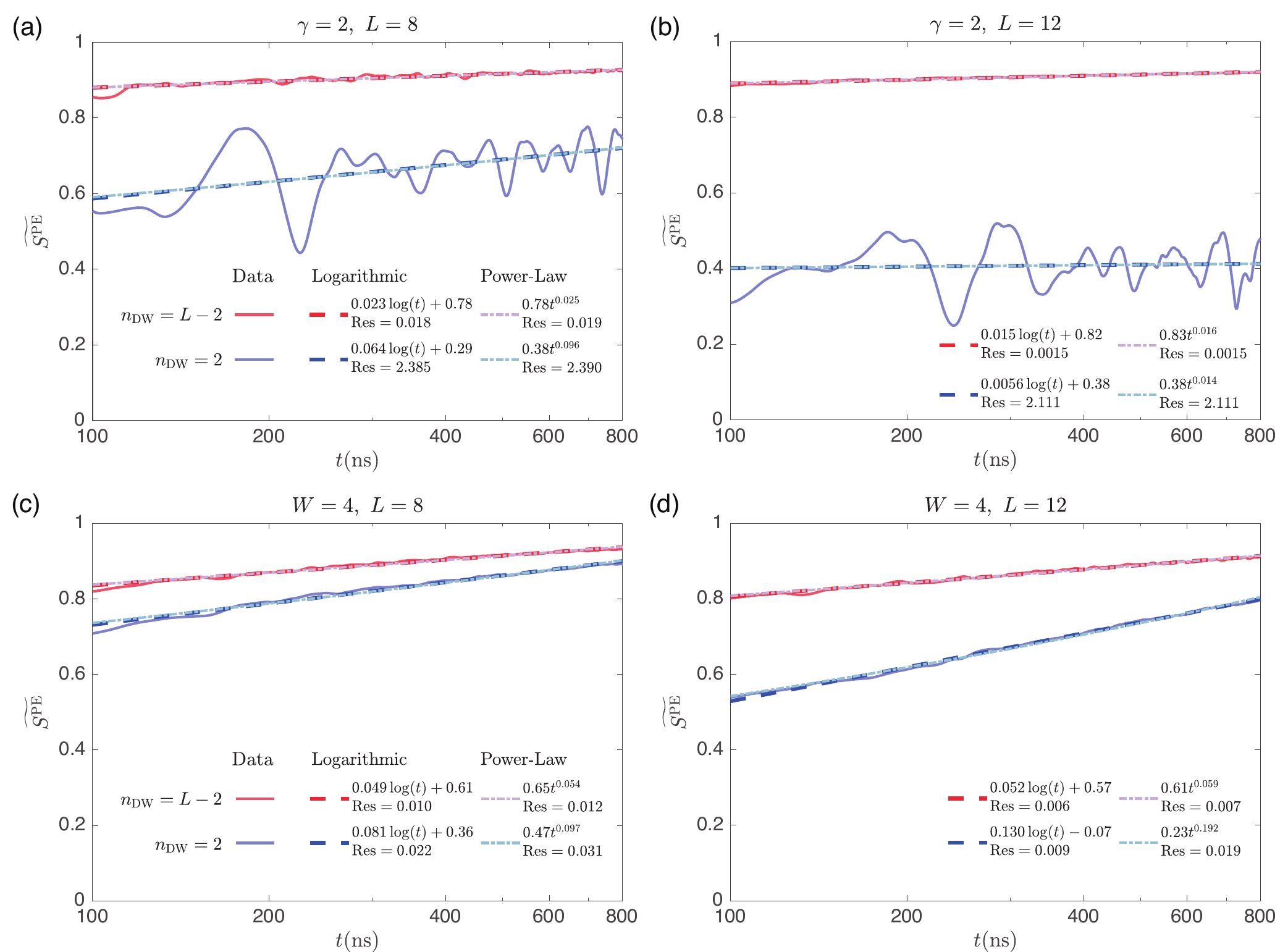}\\
	\caption{\textbf{Time evolution of participation entropy and fitting with logarithmic and power-law functions.} (a, b) The dynamical PE in Stark systems with the gradient $\gamma=2$ for (a) $L=8$, and (b) $L=12$. (c, d) The dynamical PE in disordered systems with the disorder strength $W=4$ for (c) $L=8$, and (d) $L=12$. Time evolution of PE is fitted from $100$ ns to $800$ ns with logarithmic function $f(t)\sim a\log(t)+b$, and power-law function $f(t)\sim at^b $.}
	\label{fig:SPEt_fit}
\end{figure*}

\begin{figure*}[htbp]
	\centering
	\includegraphics[width=0.93\linewidth]{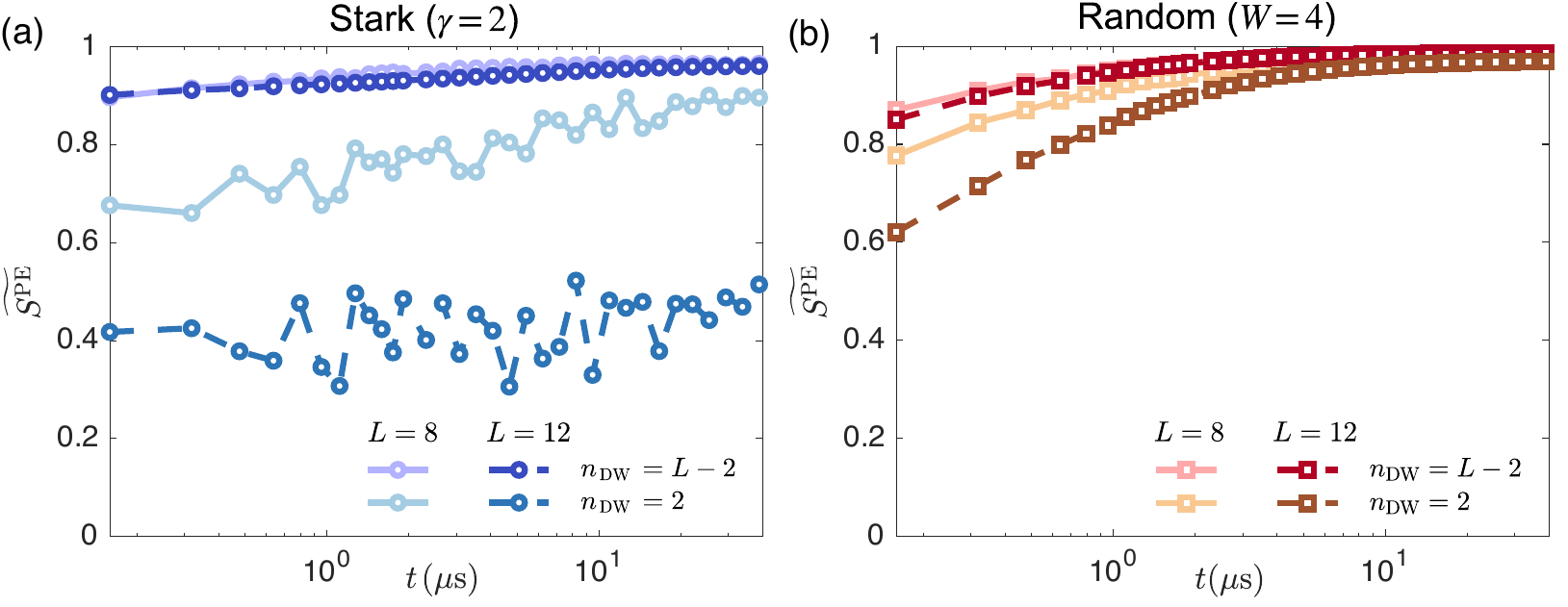}\\
	\caption{\textbf{Time evolution of participation entropy for longer times.} Dynamical participation entropy for times up to $40$ $\mu$s in  (a) the Stark system with $\gamma=2$, and (b) the disordered system with $W=4$, for different initial states $\ket{\psi_0}=\ket{\psi_{n_{\text{DW}}=L-2}}$ and $\ket{\psi_{n_{\text{DW}}=2}}$ with the system lengths $L=8$ and $L=12$.}
	\label{fig:SPEt_longtime}
\end{figure*}

\subsection{Fitting the decay of imbalance}

To quantitatively describe the decay of imbalance in the intermediate time, we fit the time evolution of imbalance from $100$ ns to $800$ ns with a power-law function $f(t)\sim at^{-\xi} $. As shown in Fig.~\ref{fig:Imba_fit}, the power-law function describes the decay of imbalance well.

We observe that for $\ket{\psi_0}=\ket{\psi_{n_{\rm{DW}}=2}}$, as the system length increases from $L=8$ to $L=12$, the exponent $\xi$ decreases from $\xi=0.72$ to $\xi=0.23$ in Stark systems (see Fig.~\ref{fig:Imba_fit}(a) and (b)), while it increases from $\xi=0.97$ to $\xi=1.5$ in disordered systems (see Fig.~\ref{fig:Imba_fit}(c) and (d)). Besides, in disordered systems, although the imbalance for $\ket{\psi_0}=\ket{\psi_{n_{\rm{DW}}=2}}$ decays to a lower value (say, $\mathcal{I}=0.1$) later compared to $\ket{\psi_0}=\ket{\psi_{n_{\rm{DW}}=L-2}}$, the latter has a higher value in terms of the exponent of the imbalance decay in the intermediate time.

\subsection{Finite-size and finite-time analysis}
As is well known, finite-size and finite-time effects pose challenges in observation of MBL in disordered systems~\cite{Abanin2021,Sierant2022}. Regarding HSF, to better elucidate the finite-size and finite-time effects in Stark systems, it is necessary to perform calculations over larger system sizes and timescales.  For these calculations, we use time-dependent variational principle (TDVP) algorithm based on matrix product states (MPS), where the wave function $\ket{\psi(t)}$ is determined by 
\begin{equation}
	\frac{d \ket{\psi(t)} }{d t} = -i \mathcal{P}_{\rm{MPS}}\mathcal{H}\ket{\psi(t)},
\end{equation}
where $\mathcal{P}_{\rm{MPS}}$ projects the time evolution onto the variational manifold of the MPS. Here, we employ a hybrid one-site-two-site approach, i.e., we expand the bond dimension with two-site approach at the initial time steps, and switch to the one-site algorithm once the maximum bond dimension is reached. For this part of calculation, we consider a ladder model with only nearest-neighbor hopping fixed at $J_{NN}/2\pi=7$ MHz, instead of using the actual parameters of the experimental device. The results for the gradient $\gamma=1$ in Stark systems are shown in Fig.~\ref{fig:Imba_scaling}.

In Fig.~\ref{fig:Imba_scaling}(a), we display the dynamics of imbalance for the initial state $\ket{\psi_0}=\ket{\psi_{n_{\rm{DW}}=L-2}}$ in Stark systems with sizes $L=8$, $12$, $16$, and $20$  (corresponding to $16$, $24$, $32$, and $40$ qubits). For the initial state $\ket{\psi_0}=\ket{\psi_{n_{\rm{DW}}=L-2}}$, we observe rapid relaxation of imbalance, which approaches zero within $100$ns, regardless of system size. The inset in Fig.~\ref{fig:Imba_scaling}(a) shows the dynamics of imbalance for system size $L=20$ with maximum bond dimensions $\chi= 256$, $512$, and $1024$, demonstrating that $\chi=512$ is sufficient for convergence of the calculation. In contrast, for $\ket{\psi_0}=\ket{\psi_{n_{\rm{DW}}=2}}$, as shown in Fig.~\ref{fig:Imba_scaling}(b), with timescales far exceeding the experimental time of up to $40$ $\mu$s, the slow relaxation becomes more distinctive, reflected by the increase in $\mathcal{I}(t)$ as $L$ increases. When $L = 20$, the decay of imbalance is barely discernible.

These distinct behaviors observed between the dynamics for different initial states with increasing system size provide an important evidence for HSF, which leads to differing localization mechanisms in Stark systems compared to disordered systems. As a comparison, we further calculate the dynamics for $\ket{\psi_0}=\ket{\psi_{n_{\rm{DW}}=2}}$ with $\gamma=0.5$ in Stark systems, and $W=1.5$ in disordered systems, ensuring that the systems satisfy (weak) ETH. The results are plotted in Fig.~\ref{fig:Imba_scaling2}.

Since the systems are in the ergodic phase ($\langle r \rangle \sim0.53$, as shown in Fig.~\ref{fig:LS}), we should carefully verify the convergence of the calculations based on MPS. In Fig.~\ref{fig:Imba_scaling2}(a), we display the time evolution of imbalance $\mathcal{I}$ up to $t=2$ $\mu$s for $L=20$ (40 qubits), and the time average of imbalance $\bar{\mathcal{I}}$ near $2$ $\mu$s is depicted in Fig.~\ref{fig:Imba_scaling2}(b). It can be observed that for the Stark system, an increase in maximum bond dimension only slightly alters $\bar{\mathcal{I}}$. However, for the disordered system, the calculation of $\bar{\mathcal{I}}$ within $t=2$ $\mu$s fails to exhibit the signature of convergence with a maximum bond dimension $\chi$ of up to $1024$, suggesting that $\bar{\mathcal{I}}$ will likely further decline towards zero with increasing $\chi$. This is also a circumstantial evidence that the strong ETH is satisfied in the weakly disordered systems, but not in a weak Stark potential.

For the Stark system, we calculated the evolution of imbalance for a longer time up to $40$ $\mu$s ($\bar{J}_\text{NN}t \sim  2\times 10^3$) for different system lengths with $\chi=192$. Similar to Fig.~\ref{fig:Imba_scaling}(b) at $\gamma=1$, the system at $\gamma=0.5$ exhibits divergent relaxation times and larger late-time values as $L$ increases. We can define a threshold $\mathcal{I}^*=0.2$, as a measure of the relaxation time $t^*$. We plot $t^*$ in Fig.~\ref{fig:Imba_scaling2}(e), where for $L=20$, since the imbalance does not decline below $\mathcal{I}=0.2$ within $40$ $\mu$s, we perform a power-law fitting and extrapolate to obtain an estimated value of $t^*$. As a comparison, we performed the same calculation for the disordered system with $\chi=1024$. Considering that the calculation fails in converging, this actually provide an overestimated value for $t^*$. It can be observed that the relaxation time $t^*$ in Stark systems increases much faster with the system length, compared to disordered systems even for the overestimated values.

\subsection{Growth of participation entropy}
In the main text, we demonstrate the dynamics of PE, wherein $S^{\text{PE}}$ for $\ket{\psi_0}=\ket{\psi_{n_{\rm{DW}}=L-2}}$ rapidly increases to a high value within a short time under various conditions. However, for $\ket{\psi_0}=\ket{\psi_{n_{\rm{DW}}=2}}$, in Stark systems, $S^{\text{PE}}$ exhibits long-term oscillations, while in disordered systems, it displays continuous growth.

To quantitatively describe the growth of PE in the intermediate and late time, in this subsection, we try to fit the time evolution of PE from $100$ ns to $800$ ns with two fitting schemes:(1) logarithmic form $f(t)\sim a\log(t)+b$; (2) a power-law form $f(t)\sim at^b $, since the growth form of PE is not definitively established. As shown in Fig.~\ref{fig:SPEt_fit}, although the logarithmic fitting has slightly lower residuals in some cases, it is actually difficult to definitively  determine which fitting scheme is superior within the current timescale. We observe that for $\ket{\psi_0}=\ket{\psi_{n_{\rm{DW}}=2}}$, as the system length increases from $L=8$ to $L=12$, the parameter $a$ (which indicates the growth rate of PE) for logarithmic fitting decreases from $a=0.064$ to $a=0.0056$ in Stark systems (see Fig.~\ref{fig:SPEt_fit}(a) and (b)), while it increases from $a=0.081$ to $a=0.130$ in disordered systems (see Fig.~\ref{fig:SPEt_fit}(c) and (d)). In short, as the system length increases, the growth rate for PE decreases in Stark systems but increases in disordered systems.

Additionally, it is worth noting that in disordered systems, after $S^{\text{PE}}$ for $\ket{\psi_0}=\ket{\psi_{n_{\rm{DW}}=L-2}}$ rapidly grows to a higher value in the short term, its intermediate and late-time growth rate of PE, for instance with $L=12$, $a=0.052$ is lower than $a=0.130$ for $\ket{\psi_0}=\ket{\psi_{n_{\rm{DW}}=2}}$, which exhibits such a logarithmic growth from the very beginning of the quench dynamics. This different behavior in relaxation is related to the different connectivity for initial states with different $n_\text{DW}$ in the Hilbert space of spin configurations, for the Hamiltonian [Eq.~(\ref{Ham1})] in the main text with nearest and next-nearest-neighboring coupling~\cite{Prasad2022}.

In fact, besides the growth rate, we should also focus on the saturation behavior of $S^{\text{PE}}$, which characterizes the available Hilbert space where the initial state can propagate and cover eventually. Therefore, we numerically calculate the time evolution of PE for times up to $40$ $\mu$s ($\bar{J}_\text{NN}t \sim  2\times 10^3$) with Krylov space method. The results are displayed in the Fig.~\ref{fig:SPEt_longtime}. It can be seen that in the Stark system (see Fig.~\ref{fig:SPEt_longtime}(a)), $S^{\text{PE}}$ for $\ket{\psi_0}=\ket{\psi_{n_{\rm{DW}}=2}}$ exhibits significantly slower growth; its oscillatory behavior persists up to $40$ $\mu$s without noticeable increase for $L=12$. In contrast, in the disordered system (see Fig.~\ref{fig:SPEt_longtime}(b)), both initial states approach $S^{\text{PE}}\sim 1$ within a few microseconds.

\subsection{Local participation entropy}

Although the efficient readout of our superconducting processor enables the measurement of PE with the system length $L=8$ (16 qubits), it remains a challenging experimental task for systems with $L=12$ (24 qubits) or larger. To address this issue, we introduce the concept of local participation entropy, defined as the PE for a subsystem $\mathcal{S}$:
\begin{equation}
    S^{\mathrm{PE}}[\mathcal{S}]=\sum_{i}p_i^\mathcal{S}\log p_i^\mathcal{S},
\end{equation}
where $p_i^\mathcal{S}$ represents the probabilities of a time-evolved state in spin configuation basis $\ket{i}^\mathcal{S}\in\{ \ket{\boldsymbol{i_{Q}}}|i_Q\in\{0,1\} ,Q \in \mathcal{S}\}$ for the subsystem $\mathcal{S}$. In this work, we consider the subsystem $\mathcal{S}$ which includes consecutive $l \times 2$ qubits, $\mathcal{S}(j;l)=\{Q_{j,m},Q_{j+1,m},\ldots,Q_{j+l-1,m}\}$ with $m=1,2$.

\begin{figure}[b]
	\centering
	\includegraphics[width=1\linewidth]{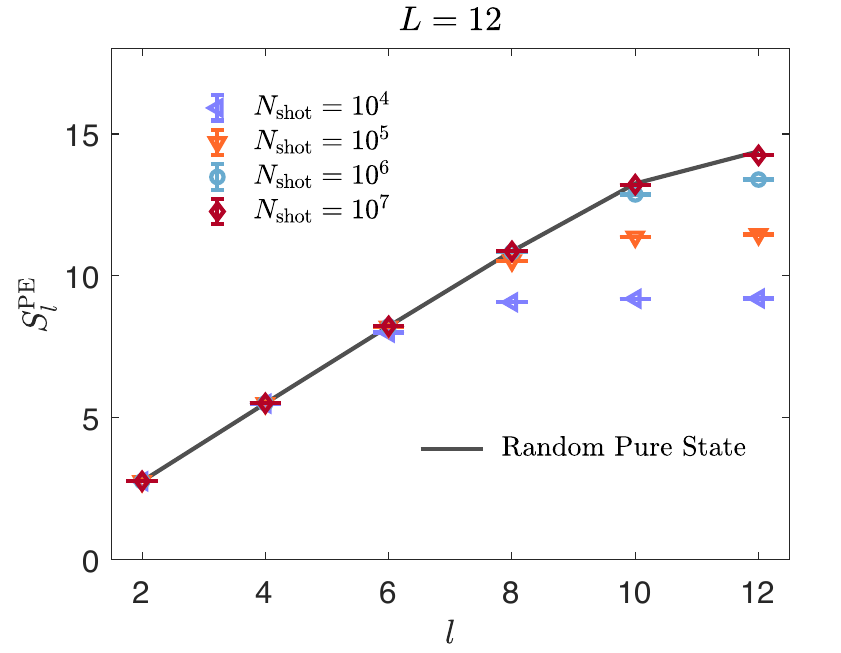}
	\caption{\textbf{Local participation entropy for random pure states with varying numbers of single-shot measurements.}
	The black line represents the local participation entropy directly calculated from the probabilities for random pure states with $L=12$, and the data points are calculated using different numbers of samples drawn from the probability distribution. The results are averaged from 5 different realizations of random pure states, and the error bar denotes the standard deviation.}
	\label{fig:numshots}
\end{figure}
\begin{figure}[htbp]
	\centering
	\includegraphics[width=1\linewidth]{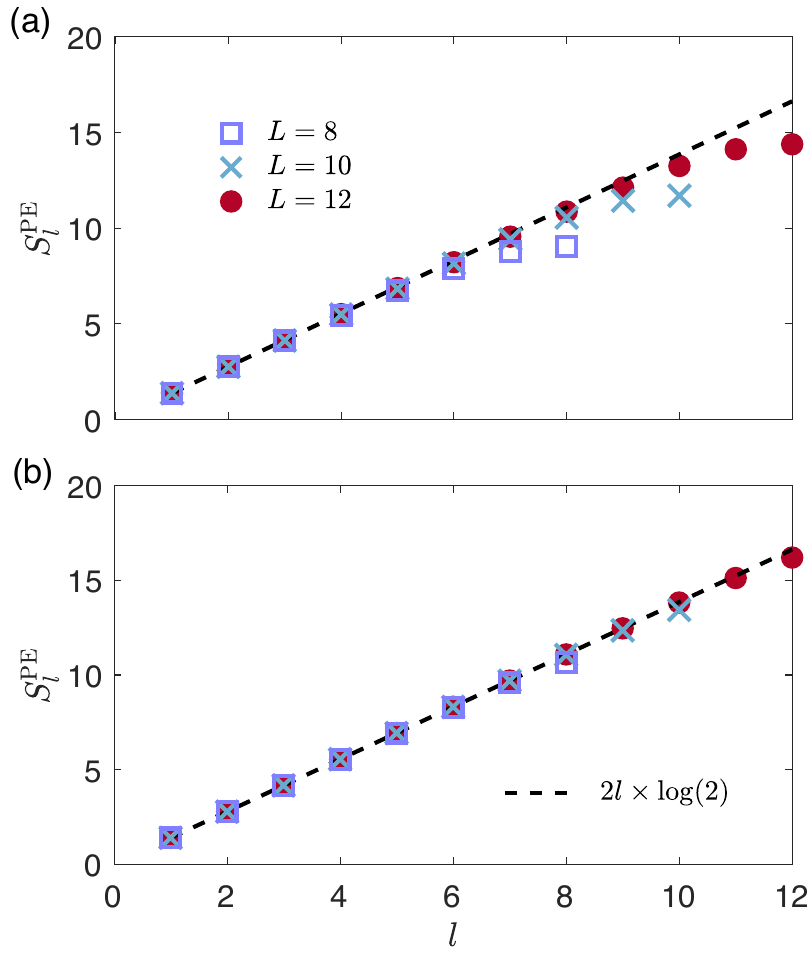}\\
	\caption{\textbf{Local participation entropy for random pure states with different system lengths.}
	(a) Local participation entropy of random pure states for systems with total $U(1)$ charge conservation in the $Q = L$ symmetry sector, with dimension $\mathcal{N}=\binom{2L}{L}$. (b) Local participation entropy of random pure states for systems without total $U(1)$ charge conservation in the full Hilbert space, with dimension $\mathcal{N}=2^{2L}$.} 
	\label{fig:pe_l_rp}
\end{figure}

\begin{figure*}[htbp]
	\centering
	\includegraphics[width=1\linewidth]{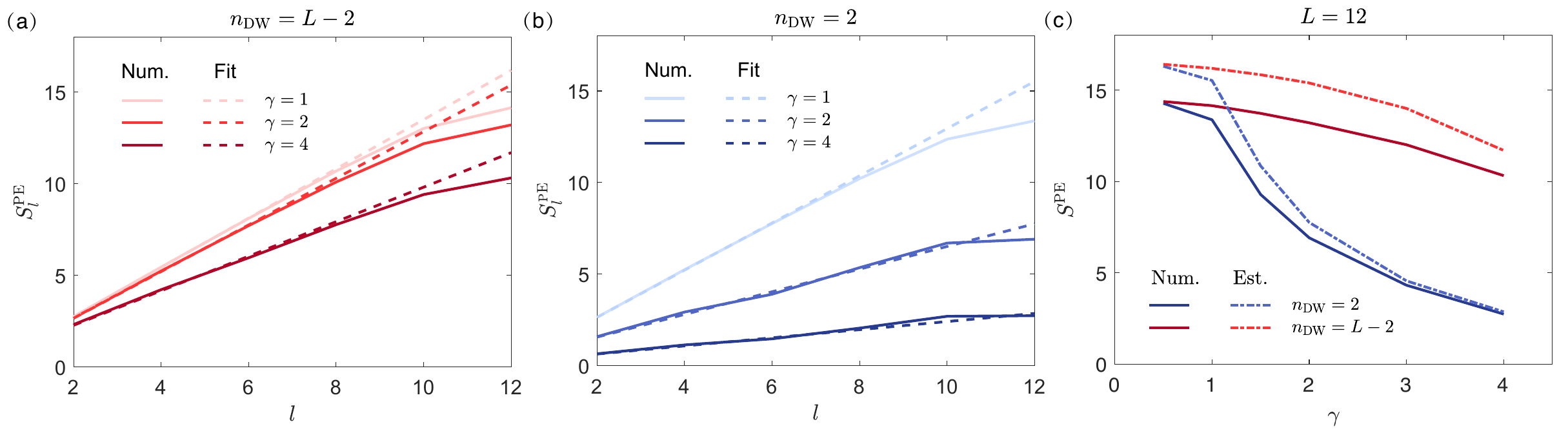}
	\caption{\textbf{Late-time local participation entropy with different gradients for $L=12$.}
	Late-time local participation entropy in Stark systems with $\gamma=1,\ 2$ and $4$ for (a) $\ket{\psi_{0}} = \ket{\psi_{n_{\mathrm{DW}}=L-2}}$, and (b) $\ket{\psi_{0}} = \ket{\psi_{n_{\mathrm{DW}}=2}}$. The solid lines are numerical results calculated from the time-evolved states at $t=800$ ns, and the dashed lines denote the fit of the data from $l=2$ to $l=6$ with a linear function. (c) The solid lines are the directly calculated participation entropies, while the dash-dotted lines denote the participation entropies estimated by fitting local participation entropy.}
	\label{fig:pe_l_fit}
\end{figure*}
Here, we first numerically examine the spatially averaged local PE $S^{\mathrm{PE}}_l \equiv \langle S^{\mathrm{PE}}[\mathcal{S}(j; l)]\rangle_j$ for complex random pure states with $L=12$, with different numbers of single-shot measurements $N_{\rm{shot}}$. The results are displayed in Fig.~\ref{fig:numshots}. For the subsystem length $l = 12$,  we can find that even $10^7$ signle-shot measurements are not enough to achieve an accurate value of $S^{\mathrm{PE}}_l$. However, for $l \leq 6$, approximately $10^5$ shots are sufficient to obtain relatively accurate values. This clearly shows that local PE can be efficiently measured in quantum processors for subsystems of moderate size.

\added{We note that the curve of local PE for random pure states exhibits a downward bending as the subsystem length $l$ approaches the entire system length $L$ (see Fig.~\ref{fig:numshots}). We attribute this downward bending primarily to the conservation of total $U(1)$ charge, which restricts the dimensions of available Hilbert space for the subsystem. This effect becomes more pronounced as the subsystem length approaches the entire system length. 
In Fig.~\ref{fig:pe_l_rp}, we present the local PE of random pure states for systems conserving total $U(1)$ charge in the $Q = L$ symmetry sector (Fig.~\ref{fig:pe_l_rp}(a)), and for systems without total $U(1)$ charge conservation in the full Hilbert space (Fig.~\ref{fig:pe_l_rp}(b)). The latter is more closely aligned with the linear function $\sim 2l\times\log(2)$ over a wider range of $l$ compared to the former. This provides strong evidence that the downward bending is due to the conservation of total $U(1)$ charge.}

\added{Additionally, as shown in Fig.~\ref{fig:pe_l_rp}(a), for the random pure states in the $Q = L$ symmetry sector, the curves of local PE for larger values of $L$ approach the linear function more closely,  The ratio between the PE for random pure states and the estimated PE from linear fitting is $r \equiv \frac{S^{\mathrm{PE}}}{S^{\mathrm{PE}}_\mathrm{est}}\sim\frac{\log\binom{2L}{L}-1+\gamma_e}{2L\log(2)}$, which approaches $1-\frac{\log(\sqrt{\pi L})}{2L\log(2)}$ for sufficient large $L$, converging to $1$ as $L$ increases. This implies a diminishing deviation between the actual value of PE and the estimated value from linear fitting with increasing system sizes.}

In the main text, we have shown that the late-time local PE in the Stark system with $\gamma=2$ exhibits approximately linear growth with the subsystem length $l$, with different slopes for initial states $\ket{\psi_{0}} = \ket{\psi_{n_{\mathrm{DW}}=L-2}}$ and $\ket{\psi_{n_{\mathrm{DW}}=2}}$. In this subsection, we extend our analysis by computing the late-time local PE for different $\gamma$ for the system length $L=12$, as shown in Fig.~\ref{fig:pe_l_fit}. For both initial states, $S^{\mathrm{PE}}_l$ exhibits linear growth for small subsystem lengths $l$ across different values of $\gamma$. We therefore fit the data of local PE for $l=2$ to $l=6$ with a linear function, and extrapolate to estimate the upper bound for the PE $S^{\mathrm{PE}}_{\mathrm{est}}$. A comparison between the directly calculated PE and the estimated upper bound of PE for different $\gamma$ is shown in Fig.~\ref{fig:pe_l_fit}(c). We observe that $S^{\mathrm{PE}}_{\mathrm{est}}$ follows the same trend as $S^{\mathrm{PE}}$, and the differences in $S^{\mathrm{PE}}_{\mathrm{est}}$ between different initial states reveals the presence of the HSF in the Stark systems.

Overall, for a wide range of parameters, measuring local PE allows us to effectively estimate the upper bound of the PE, requiring relatively fewer experimental measurements compared to directly measuring the PE of the entire system. This suggests that the approach can serve as a scalable observable for future experimental studies of HSF with larger system sizes.

\section{Numerical Krylov Subspace}
\begin{figure*}[ht]
	\centering
	\includegraphics[width=0.85\linewidth]{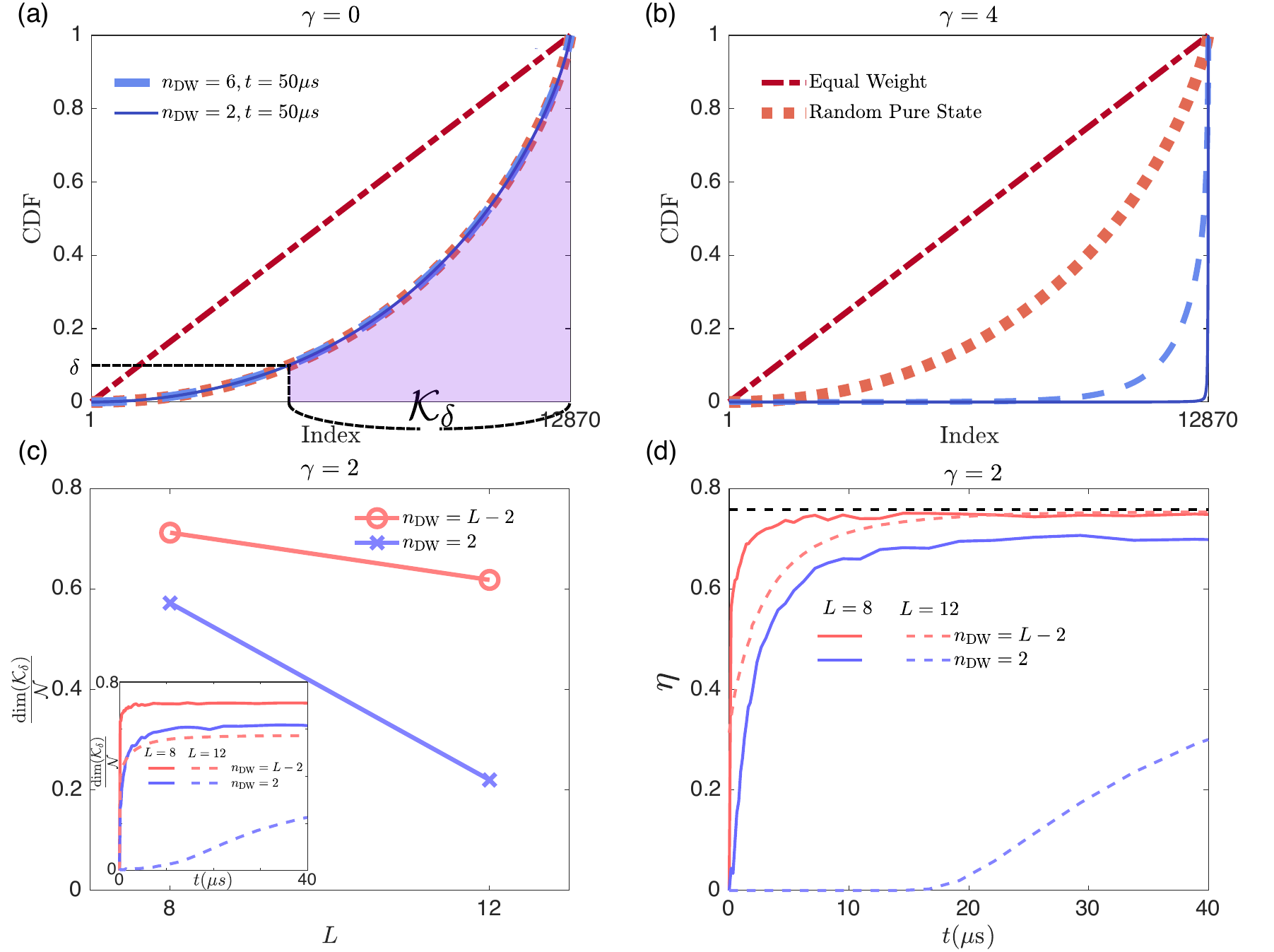}
	\caption{\textbf{Numerical Krylov subspace in Stark systems.} 
	(a), (b) The cumulative distribution function of the time-evolved states for $\ket{\psi_0}=\ket{\psi_{n_{\text{DW}}=2}}$ and $\ket{\psi_{n_{\text{DW}}=6}}$ at $\gamma=0$, and $4$. The cumulative distribution function for the state with equal weight on all the bases, and for a complex random pure state are also provided for comparison. The shaded region in (a) corresponds to the bases which construct the numerical Krylov subspace. (c) The dimension of the numerical Krylov subspace for the initial states $\ket{\psi_{n_{\text{DW}}=2}}$ and $\ket{\psi_{n_{\text{DW}}=L-2}}$ with $L=8$ and $12$ at $\gamma=2$. The inset shows the time evolution of the dimension of $\mathcal{K}_\delta(t)$ from $t=0$ to $40 \mu$s. (d) The overlap between numerical Krylov subspaces $\eta\equiv\frac{{\text{dim}(\mathcal{K}_\delta(\psi_0)\cap\mathcal{K}_\delta(\overline{\psi_0}))}}{{\text{dim}(\mathcal{K}_\delta(\psi_0)\cup\mathcal{K}_\delta(\overline{\psi_0}))}}$ for $\ket{\psi_0}$ and $\ket{\overline{\psi_0}}$, with $\ket{\psi_0}=\ket{\psi_{n_{\text{DW}}=2}}$ and $\ket{\psi_{n_{\text{DW}}=L-2}}$, respectively, and $\ket{\overline{\psi_0}}=\prod_{j,m}\po{jm}{x}\ket{\psi_0}$. The dashed line denotes the typical value of the overlap between the subspaces corresponding to two random pure states.}
	\label{fig:num_Krylov}
\end{figure*}

\begin{figure*}[htbp]
	\centering
	\includegraphics[width=1\linewidth]{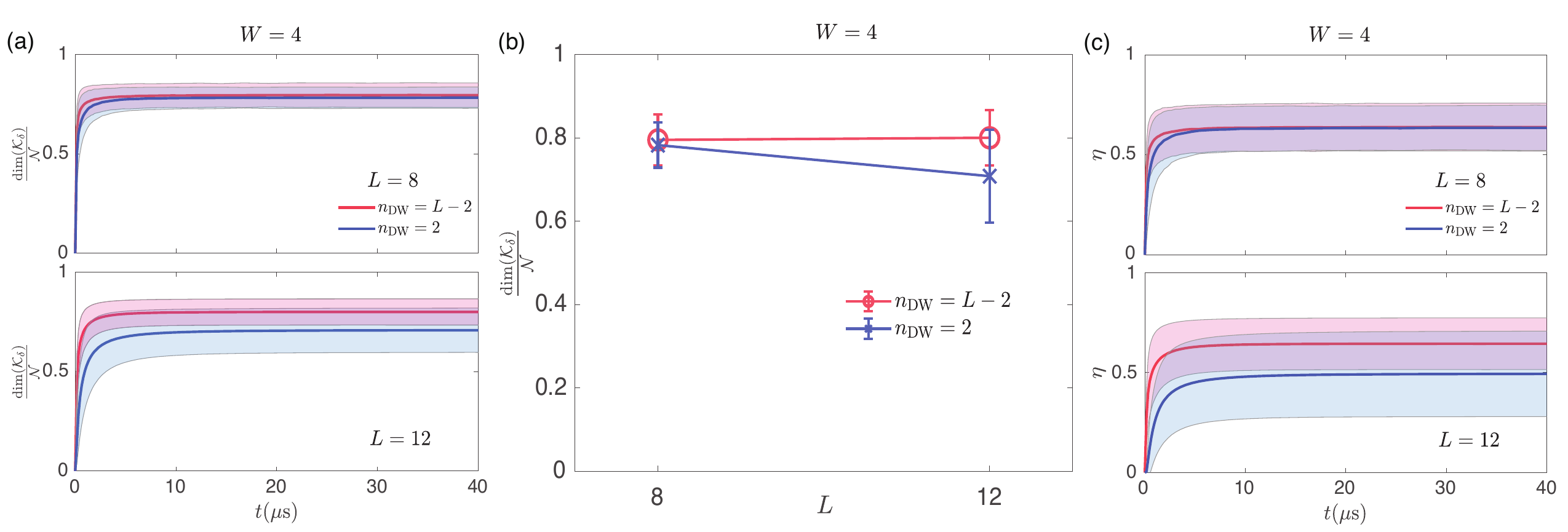}
	\caption{\textbf{Numerical Krylov subspace in disordered systems.} 
	(a) Time evolution of the dimension of $\mathcal{K}_\delta(t)$ from $t=0$ to $40 \mu$s. (b) The dimension of the numerical Krylov subspace as a function of the system length, for the initial states $\ket{\psi_{n_{\text{DW}}=2}}$ and $\ket{\psi_{n_{\text{DW}}=L-2}}$ at disorder strength $W=4$. (c) The overlap $\eta$ between numerical Krylov subspaces for $\ket{\psi_0}$ and $\ket{\overline{\psi_0}}$, with $\ket{\psi_0}=\ket{\psi_{n_{\text{DW}}=2}}$ and $\ket{\psi_{n_{\text{DW}}=L-2}}$, respectively, and $\ket{\overline{\psi_0}}=\prod_{j,m}\po{jm}{x}\ket{\psi_0}$.}
	\label{fig:num_Krylov_dis}
\end{figure*}

To illustrate the construction of the numerical Krylov subspace, we firstly introduce the cumulative distribution function (CDF) for a time-evolved state $\ket{\psi(t)}$. The calculation of CDF involves the following steps:
\begin{itemize}
	\item [i)] Calculate the probabilities of spin configuration basis $p_i=\left| {\langle \psi(t)|i\rangle } \right|^2$ (i.e., the moduli squared of the wave function coefficients expressed in spin configuration basis $\{\ket{i}\}$);
	\item [ii)] Sort the probabilities in non-decreasing order, denoted as $p_{i^\prime}$, with $\{i^\prime\}$ representing the sorted indices;
	\item [iii)] Compute the CDF as the sum of the sorted probabilities to the index $n$, given by $\text{CDF}(n)=\sum_{i^\prime=1}^n p_{i^\prime}$.
\end{itemize}

The CDF defined here illustrates the bases with the probabilities arranged from low to high, contributing to the state, with the curvature representing the degree of unevenness in the distribution. One can expect the CDF would be a straight line for the state with equal weight on all the bases $\{\ket{i}\}$, while if the state is a single configuration, the CDF would be $0$ for indices $n \leq \mathcal{N}-1$ of the sorted basis, and $1$ at index $n = \mathcal{N}$. 

In Fig.~\ref{fig:num_Krylov}(a) and (b), we displays the CDF of the time-evolved states for $\ket{\psi_0}=\ket{\psi_{n_{\text{DW}}=2}}$ and $\ket{\psi_{n_{\text{DW}}=6}}$ at $\gamma=0$, and $4$. For both initial states, the curve of CDF for the time-evolved states lies slightly beneath that for the state with equal weight on all the bases,  nearly coinciding with the curve for a complex random pure state at $\gamma=0$. At $\gamma=4$, the CDF further curve downward for both initial states. Notably, for $\ket{\psi_0}=\ket{\psi_{n_{\text{DW}}=2}}$, the curve is very close to that of a single configuration state, suggesting that the time-evolved state predominantly exhibits high probabilities on only a limited number of bases. 

Based on the CDF, we can define the subset $\mathcal{K}_\delta(t)$ for an initial state as the complement of the cumulative portion of the basis corresponding to the cumulative probabilities $\delta$. This subset satisfies $\sum_{i\in\mathcal{K}_\delta(t)}\left| {\langle \psi(t)|i\rangle } \right|^2 = 1-\delta$, as depicted by the shaded region in Fig.~\ref{fig:num_Krylov}(a).
Therefore, the numerical Krylov subspace $\mathcal{K}_\delta$ can be further defined as largest subset satisfying $\mathop{\arg\max}\limits_{\mathcal{K}_\delta(t),\ t\leq\tau}\{\text{dim}(\mathcal{K}_\delta(t))\}$ within the evolution time $\tau$~\cite{Scherg2021}.

Here, we calculate $\mathcal{K}_\delta$ for $\delta=0.01$ and $\tau=40 \mu$s, at $\gamma=2$. The dimension of the numerical Krylov subspace is plotted in Fig.~\ref{fig:num_Krylov}(c) for the initial states $\ket{\psi_{n_{\text{DW}}=2}}$ and $\ket{\psi_{n_{\text{DW}}=L-2}}$ with $L=8$ and $12$, with the inset showing the dimension of $\mathcal{K}_\delta(t)$ from $t=0$ to $40 \mu$s. We observe that the dimension $\text{dim}\mathcal{K}_\delta(t)$ quickly saturates for $\ket{\psi_{n_{\text{DW}}=L-2}}$, regardless of the system size; while for $\ket{\psi_{n_{\text{DW}}=2}}$, the growth of $\text{dim}(\mathcal{K}_\delta(t))$ slows considerably with increasing system length $L$. Correspondingly, the dimension of the numerical Krylov Subspace for $\ket{\psi_{n_{\text{DW}}=2}}$ decreases much more dramatically with increasing $L$, compared to $\ket{\psi_{n_{\text{DW}}=6}}$. Thus, the numerical Krylov subspaces $\mathcal{K}_\delta$ exhibit different scaling behaviors for the two initial states, which leads to the initial-state dependent dynamics of the PE as shown in Fig.~\ref{fig4} in the main text.

We have shown that initial states with different $n_{\text{DW}}$ belong to different fragments, therefore exhibiting distinct dynamics. However, it is worth noting that the reverse is not necessarily valid: initial states sharing the same $n_{\text{DW}}$ may also each reside within nearly disconnected fragments. Here, we consider a pair of initial states $\{\ket{\psi_0}, \ket{\overline{\psi_0}\}}$, which have the maximum Hamming distance $L \times 2$ between each other~\cite{Yao2023}, with $\ket{\overline{\psi_0}}=\prod_{j,m}\po{jm}{x}\ket{\psi_0}$. For example, $\{\ket{\psi_{n_{\text{DW}}=2}}=\ket{\mathbb{11000011}},\ \ket{\overline{\psi}_{n_{\text{DW}}=2}}=\ket{\mathbb{00111100}}\}$ and $\{\ket{\psi_{n_{\text{DW}}=L-2}}=\ket{\mathbb{101000101}},\ \ket{\overline{\psi}_{n_{\text{DW}}=L-2}}=\ket{\mathbb{01011010}}\}$ for $L=8$. From symmetry considerations, one can expect that the initial states $\{\ket{\psi_0}, \ket{\overline{\psi_0}\}}$ exhibit the same dynamics in terms of PE, EE, imbalance and other quantities. However, this does not necessarily mean these initial states belong to the same fragment. To explore this aspect, we introduce the quantity (commonly referred to as the Jaccard coefficient in statistics): 
\begin{equation}
	\eta\equiv\frac{{\text{dim}(\mathcal{K}_\delta(\psi_0)\cap\mathcal{K}_\delta(\overline{\psi_0}))}}{{\text{dim}(\mathcal{K}_\delta(\psi_0)\cup\mathcal{K}_\delta(\overline{\psi_0}))}},
\end{equation}
which measures the overlap between numerical Krylov subspaces for $\ket{\psi_0}$ and $\ket{\overline{\psi_0}}$. 

The results are presented in Fig.~\ref{fig:num_Krylov}(d). At $\gamma=2$, long-time $\eta$ between $\ket{\psi_{n_{\text{DW}}=L-2}}$ and $\ket{\overline{\psi}_{n_{\text{DW}}=L-2}}$ is close to $\eta$ between two random pure states (indicated by dashed lines in Fig.~\ref{fig:num_Krylov}(d)) regardless of the system size, which implies that they belong to the same Krylov subspace. In contrast, $\eta$ between $\ket{\psi_{n_{\text{DW}}=2}}$ and $\ket{\overline{\psi}_{n_{\text{DW}}=2}}$ fails to reach the value between two random pure states even after a long time, and this discrepancy becomes more pronounced with increasing system size. Specially, for $L=12$, $\eta$ remains almost zero for a long time, and then slowly increase when $t \gtrsim 20 \mu$s ($\bar{J}_\text{NN}t \gtrsim  10^3$), which provides evidence that $\ket{\psi_{n_{\text{DW}}=2}}$ and $\ket{\overline{\psi}_{n_{\text{DW}}=2}}$ belong to different fragments, dynamically disconnected from each other. Note that the increase of $\eta$ after $t \gtrsim 20 \mu$s stems from the coupling of different fragments by higher-order terms in the perturbative expansion of Hamiltonian [Eq.~(\ref{Ham1})], which break the conservation of the emergent dipole moment and eventually leads to the destruction of fragmentation.

For comparison, we perform the calculation involving the effective Krylov subspaces for disordered systems, similar to the approach used in Fig.~\ref{fig:num_Krylov}(c), (d). The numerical results are plotted in Fig.~\ref{fig:num_Krylov_dis}. In Fig.~\ref{fig:num_Krylov_dis}(a), we show the averaged dimensions $\rm{dim}\mathcal{K}_\delta(t)$ as a function of time over 20 disorder realizations at $W=4$, with the shaded area representing the standard deviation. We observe that $\rm{dim}\mathcal{K}_\delta(t)$ quickly saturates for both initial states, regardless of the system size. For $\tau=40\mu \rm{s}$, the dimension of the numerical Krylov subspace $\rm{dim}\mathcal{K}_\delta$ is shown in Fig.~\ref{fig:num_Krylov}(b), which exhibits minor variation with $L$ for both initial states, compared to the Stark systems. Furthermore, as shown in Fig.~\ref{fig:num_Krylov}(c), we also calculated $\eta$ for disordered systems, which increases rapidly in the very beginning even for $\ket{\psi_0}=\ket{\psi_{n_{\rm{DW}}=2}}$, contrasting with Stark systems where $\eta$ remains nearly zero for approximately a thousand tunneling times.

\section{Time Evolution with Krylov Space Method}
In dealing with large systems, the full diagonalization becomes unfeasible. In our study involving up to 24 qubits, we employ the Krylov space method, which offers a more efficient approach for computing the time evolution of the system.

The unitary evolution of a isolated system governed by the Schr\"{o}dinger equation can be expressed as 
\begin{equation}
	\ket{\psi(t+\tau)}=\exp{(-i\tau \hat{H}/\hbar)}\ket{\psi(t)},
	\label{eq:time_evo}
\end{equation}
where $\ket{\psi(t)}$ denotes the state at time $t$. The Krylov subspace of dimension $m$ is spanned by the successive powers of $H$ applied to $\ket{\psi(t)}$
\begin{equation}
	\mathcal{K}_m(\mathbf{H},\mathbf{v}_0)=\text{Span}\{\mathbf{v}_0,\mathbf{H}\mathbf{v}_0, \mathbf{H}^2\mathbf{v}_0,\ldots, \mathbf{H}^{m-1}\mathbf{v}_0\},
\end{equation}
with $\mathbf{H}$ and $\mathbf{v}_0$ being the matrix form of the Hamiltonian and the state, respectively. For Hermitian matrix $\mathbf{H}$, via the Lanczos algorithm, an orthogonal transformation matrix $\mathbf{K}_m=(\mathbf{v}_0\ \mathbf{v}_1\ \mathbf{v}_2\ \ldots\ \mathbf{v}_{m-1})$ can be constructed, which turns $\mathbf{H}$ into a tridiagonal $m \times m$ matrix form,
\begin{equation}
	\mathbf{H}_{m}=\mathbf{K}_m^{\dagger} \mathbf{HK}_m=\left(\begin{array}{ccccc}
	\alpha_0 & \beta_1 & & & 0 \\
	\beta_1 & \alpha_1 & \beta_2 & & \\
	& \beta_2 & \alpha_2 & \ddots & \\
	& & \ddots & \ddots & \beta_{m-1} \\
	0 & & & \beta_{m-1} & \alpha_{m-1}
	\end{array}\right).
	\end{equation}
Here, $\alpha_j$, $\beta_j$ and $\mathbf{v_j}$ are determined by the Lanczos algorithm, which is an iterative procedure
\begin{equation}
	\begin{aligned}
	& \alpha_i=\mathbf{v}_i \cdot \mathbf{H} \mathbf{v}_i, \\
	& \beta_i \mathbf{v}_{i+1}=\mathbf{H} \mathbf{v}_i-\alpha_i \mathbf{v}_i-\beta_{i-1} \mathbf{v}_{i-1} .
	\end{aligned}
	\end{equation}

For a short time step $\tau$, the time evolution in~\ref{eq:time_evo} can be approximately calculated as
\begin{equation}
	\ket{\psi(t+\tau)} \approx \mathbf{K}_m \exp{(-i\tau \mathbf{H}_m/\hbar)}\mathbf{K}_m^\dagger \mathbf{v}_0.
	\label{eq:time_evo_Krylov}
\end{equation}
Given that the eigenvalues of $\mathbf{H}_m$ are the most crucial $m$ eigenvalues governing the dynamics at the current time $t$, the Krylov space method can yields highly accurate results. In this work, we adaptively select the time steps while ensuring convergence with $m=15$ Krylov vectors.

\section{Post-Selection of Measurement Results}

\begin{figure}[htbp]
	\centering
	\includegraphics[width=0.96\linewidth]{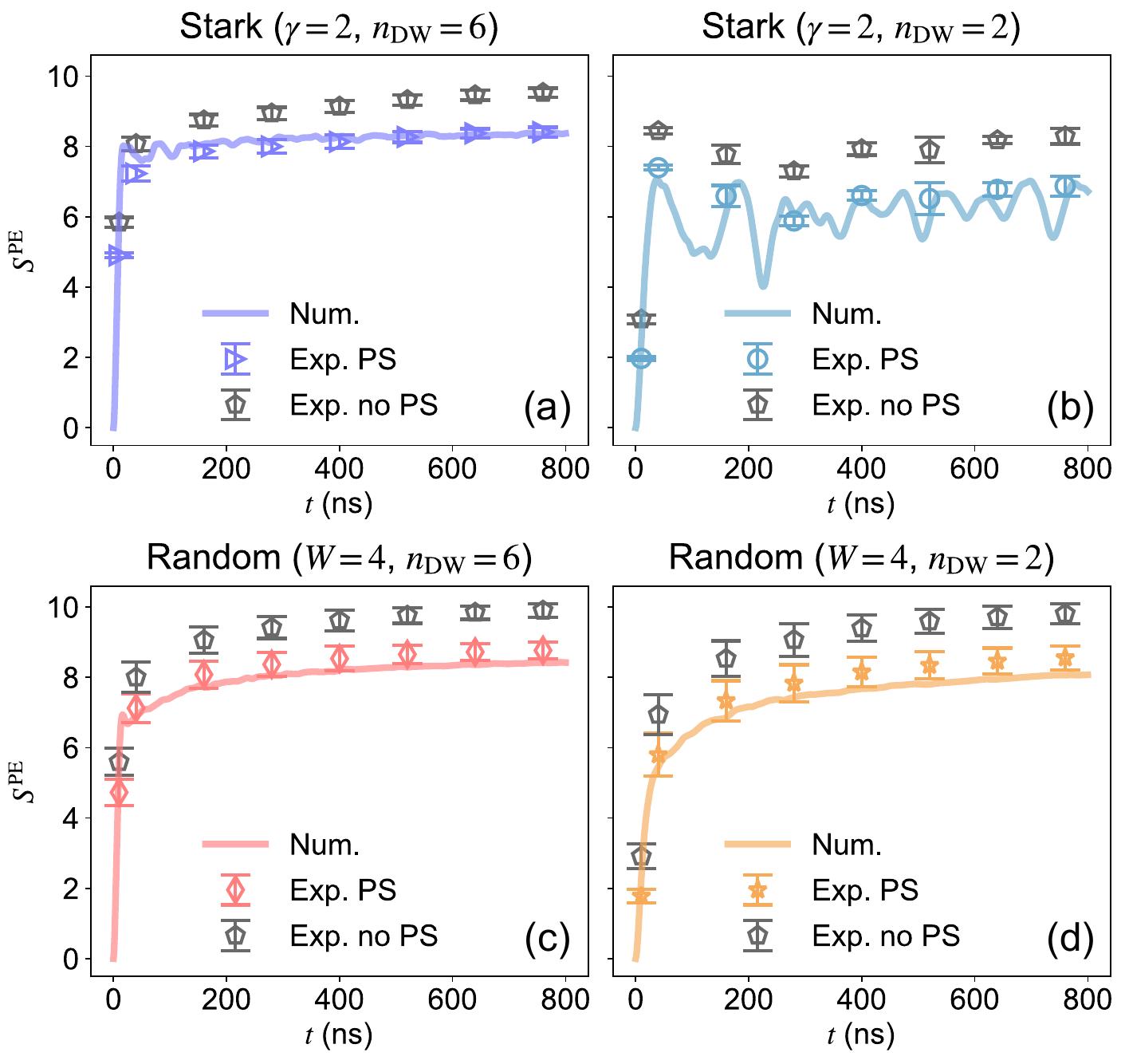}
	\caption{\textbf{Effect of Post-Selection on Experimental Results.}
		The experimental data for the participation entropy (PE) with and without (grey data point) post-selection (PS) within the $Q = L$ sector, in comparison with the numerical results (solid lines), for specific initial states in both Stark ($\gamma=2$) and random  ($W=4$) potentials.}
	\label{fig:postselect}
\end{figure}

The Hamiltonian [Eq.~(\ref{Ham1})] in the main text conserves the total $U(1)$ charge $\hat{Q} \equiv \sum_{j,m}{\po{jm}{+}\po{jm}{-}}$ during the time evolution. Despite the experimental evolution time being up to $800$ ns, significantly less than the qubit energy relaxation times $T_1 \sim 33.2$ $\mu$s listed in Supplementary Tab.~\ref{tab:paras}, the leakage out of the $Q = L$ sector could still lead to an overestimation of the participation entropy (PE). To address this, we perform post-selection on the measured probabilities within the $Q = L$ sector for the data presented in Fig.~\ref{fig4} in the main text.

Comparisons between the experimental results with and without post-selection, for $\psi_{n_{\text{DW}}=2}$ and $\psi_{n_{\text{DW}}=L-2}$ in a Stark and random potential, are depicted in Fig.~\ref{fig:postselect}. Notably, the inclusion of post-selection demonstrates a significantly improved agreement between the experimental and numerical results.
\bibliography{refe_sm}
\clearpage